\documentclass[amsmath,amssymb,floatfix]{revtex4}
\input epsf
\newcommand {\cen}[1]{\begin{center} #1 \end{center}}
\newcommand {\eq}{\begin{equation}}
\newcommand {\qe}{\end{equation}}

\newcommand {\bfk} {{\bf k}}
\newcommand {\bfq} {{\bf q}}

\newcommand {\bfp} {{\bf p}}
\newcommand {\bfv} {{\bf v}}
\newcommand {\bfa} {{\bf a}}
\newcommand {\bfb} {{\bf b}}

\newcommand {\bft} {{\bf t}}
\newcommand {\bfm} {{\bf m}}
\newcommand {\bfn} {{\bf n}}
\newcommand {\bfl} {{\bf l}}
\newcommand {\bfH} {{\bf H}}

\newcommand {\h}{\frac{1}{2}}
\newcommand {\calf} {{\cal F}}
\newcommand {\calM} {{\cal M}}
\newcommand {\calK} {K(w,t)}
\newcommand {\bfs}{\mbox {\boldmath $\sigma$}}

\newcommand {\slq}{\slash{\!\!\!q}}
\newcommand {\slp}{\slash{\!\!\!p}}
\newcommand {\slk}{\slash{\!\!\!k}}
\newcommand {\qt}{\widetilde{q}}
\newcommand {\kt}{\widetilde{k}}
\newcommand {\qb}{w}

\newcommand {\qmm}{q_0^{max}}

\newcommand {\pr}{Phys. Rev. }
\newcommand {\nucp}{Nucl. Phys.}
\def\ecart{\noalign{\medskip}}

\usepackage{epsfig}
\begin{document}
\hspace{5.5in}{\Large LPNHE 2006-04}  \\
\cen{\large \bf
     Two-Pion Exchange in Proton-Proton Scattering}

\vspace{.5in}

\cen{\bf \large  W. R. Gibbs}

\cen{New Mexico State University, Las Cruces NM 88003}

\cen{and}

\cen{\bf \large B. Loiseau}
\cen{LPNHE-Laboratoire de Physique Nucl\'eaire et de Hautes 
\'Energies, Groupe Th\'eorie,\\  
IN2P3-CNRS, Universit\'es P. \& M. Curie et Denis Diderot,
 4 Pl. Jussieu, 75252 Paris, France}

The contribution of the box and crossed two-pion-exchange diagrams to
proton-proton scattering at 90$^{\circ}_{c.m.}$ is calculated in the
laboratory momentum range up to 12 GeV/c. Relativistic form factors
related to the nucleon and pion size and representing the pion source
distribution based on the quark structure of the hadronic core are
included at each vertex of the pion-nucleon interaction. These form
factors depend on the four-momenta of the exchanged pions and scattering
nucleons. Feynman-diagram amplitudes calculated without form factors are
checked against those derived from dispersion relations. In this
comparison, one notices that a very short-range part of the crossed
diagram, neglected in dispersion-relation calculations of the
two-pion-exchange nucleon-nucleon potential, gives a sizable
contribution. In the Feynman-diagram calculation with form factors the
agreement with measured spin-separated cross sections, as well as
amplitudes in the lower part of the energy range considered, is much
better for pion-nucleon pseudo-vector vis \`a vis pseudo-scalar
coupling.  While strengths of the box and crossed diagrams are
comparable for laboratory momenta below 2 GeV/c, the crossed diagram
dominates for larger momenta, largely due to the kinematics of the
crossed diagram allowing a smaller momentum transfer in the nucleon
center of mass. An important contribution arises from the
principal-value part of the integrals which is non-zero when form
factors are included. It seems that the importance of the exchange of
color singlets may extend higher in energy than expected.

\newpage

\section{Introduction}

The nucleon-nucleon interaction at intermediate energy (up to 12 GeV/c
laboratory momentum, P$_{Lab}$) has been the focus of a number of
experimental \cite{lin,bhatia,crosbie,court,crabb,fallon,
akerlof,kammerud, abe} and theoretical \cite{brod1,farrar,
sivers1,sivers2} studies. The momentum dependence of the spin transfer
shows a very interesting behavior. The spin correlation observable, 
$C_{NN}$, for proton-proton scattering at $90^{\circ}_{c.m.}$, has values
near unity at low energies, decreases to a constant value of around 0.07
from 4 GeV/c to 8 GeV/c and then increases to values around 0.5.

\begin{figure}[htb]
\epsfig{file=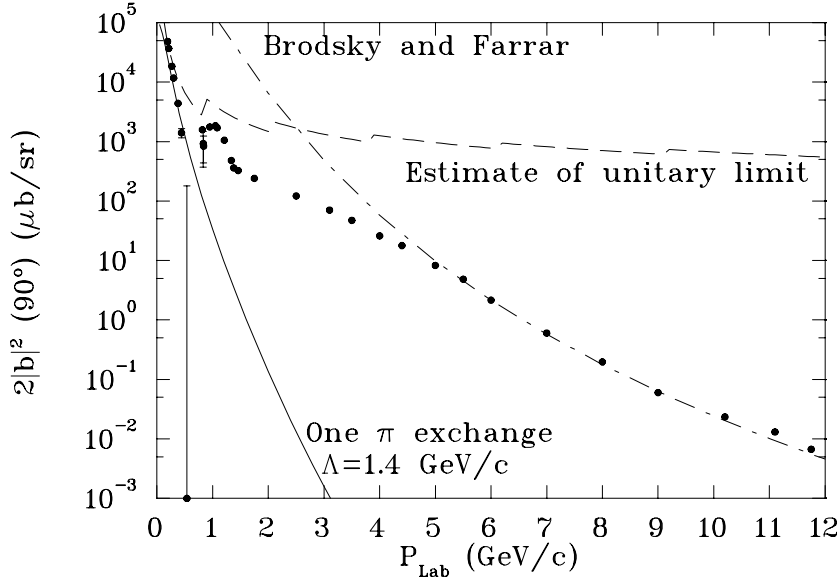,angle=90,height=3.in}
\caption{Values of $2|b|^2$ (see Ref. \cite{Bystricky78} and
appendix A)  at $90^{\circ}_{c.m.}$, extracted as discussed in the
text, compared with one-pion exchange and an estimated unitary
limit. The dash-dot curve gives the energy dependence
predicted by Brodsky and Farrar \cite{bf} normalized at 6
GeV/c.}
\label{b2ope}
\end{figure}

While one-pion exchange has been shown to be a very important contributor
to the NN interaction at low energies \cite{rosa,fg,ballot} and at higher
energies at small momentum transfer \cite{gl}, it alone predicts a
constant value of 1/3 for $C_{NN}$. Not only is this value in disagreement
with the data, but it is not expected that single pion exchange will still
be important at this higher momentum transfer (see Fig. \ref{b2ope} and
comments below).

Brodsky \cite{brod1} and Farrar \cite{farrar} have shown that the simple
quark-exchange mechanism also gives 1/3 for $C_{NN}$. No calculations of
absolute cross sections with quark-exchange models exist to our knowledge,
although predictions do exist \cite{bf} for the energy dependence of the
cross section. These predictions for the slope are in quite good
agreement, not only with proton-proton scattering, but with other
scattering processes at high energies.

While it is natural to think that this behavior could be an indicator of
the nature of the elastic scattering process, this idea of the
identification of a mechanism has languished for many years for lack of
candidate theories.  We present here the calculation of the contribution
of two-pion exchange in this energy region.  We find that this mechanism
predicts the right size for the cross section, hence provides a candidate
theory for the dominant mechanism in this intermediate energy region.

The interest in two-pion exchange as a contributor to the nucleon-nucleon
interaction is very old, beginning just after the discovery of the pion
\cite{tns,tmo,bw,so,mach}. The seminal work of Partovi and Lomon
\cite{partovi} was one of the first to create a viable potential based on
this idea.  Since these works were aimed at obtaining the
two-pion-exchange contribution to the NN potential, non-relativistic
approximations were made.

The Paris potential group \cite{Cottingham73,Lacombe81,
Richard75} has worked extensively with this contribution within a
dispersion relation approach.  Again, the interest was primarily
to obtain a non-relativistic potential for the nucleon-nucleon
interaction.

Studies of the relation of chiral symmetry to the two-pion-exchange potential have 
been made recently \cite{friarcoon,
robilotta} which led to more recent studies by Rentmester et al.
\cite{rent} on the long range part of the two-pion-exchange
potential. Ref. \cite{rent} concludes that
there is strong evidence for the existence and importance of
two-pion exchange.

Seeing this significant body of study at low energy, it is
natural to ask about the two-pion interaction at higher energies.  
We will investigate its role by calculating the lowest order
Feynman graphs \cite{feynman} for this process in the range P$_{Lab}$
from 0 to 12 GeV/c.

Before beginning the calculations, some discussion is needed as to the 
general approach. A common methodology has been to proceed by ranges 
\cite{tns,tmo}. In this point of view one says that the one- and 
two-pion exchange are valid (and give essentially all of the 
potential) beyond a certain inter-nucleon separation. In order to 
confront the data it is then necessary to construct the potential at 
shorter ranges, perhaps with phenomenology as in Ref. \cite{Lacombe81}.
 We take the point of view that the pion exchanges are between quarks and hence
(to some approximation) the interaction calculated remains valid even at short
distances. There is no reason {\it a priori} to exclude pion exchange at short
distances provided that the distribution of quarks within the nucleon is
properly taken into account. Even when the relative coordinate between the {\it
centers} is zero the range of the exchange between the quarks will be of the
order of the size of the nucleon. In fact, there are a number of studies
\cite{qpiq} of baryon spectra based on the exchange of mesons between quarks
within the interior of a single nucleon.

Of course, we do not expect that one- and two-pion exchanges represent 
the entire interaction, other contributions are certainly present. We 
calculate the two-pion contribution with a view to examining to what 
extent the scattering properties can be explained with this mechanism 
alone.

Since a common view is that nucleons and pions have an intrinsic
finite size, the present approach includes the form
factor resulting from this distribution in the Feynman integrals.
There is a long history of treating form factors of complex systems
in nuclear physics.  It is often regarded as the amplitude for
the probability of being able to transfer the three-momentum represented
by its argument to the object while leaving it intact, 
hence is related to the probability of an {\it elastic} scattering or 
absorption.
  
The inclusion of such a form factor is not simple, however. The full
Feynman integrals are in four dimensions, requiring a form factor with a
dependence on all 4 variables. Many form factors deal only with
three-momentum, although the problem of adding a fourth variable has been
addressed by a number of groups (see section \ref{form} for a discussion).  
We approach the problem of the form factor from the point of view of
invariants and analyticity with guidance from the argument that the
function should represent a boost into the center of mass frame of the
nucleon in the on-shell limit.  We will argue that a selection can be made
on this dependence according to the view of the physical origin of the
size of the system, i.e. if it arises from the interaction with virtual
mesons (or more generally the sea) or with the ``permanent'' quark core
constituents. We assume here that the pion exchanges take place between
quarks contained within the core of the nucleon.

We will treat the cases of both pure pseudo-scalar (PS) and pure
pseudo-vector (PV) coupling even though some authors have found evidence
for a mixture of the couplings \cite{mix1,mix2,kon}. Some of the integrals
involved in the PV calculation are divergent and, while the PS integrals
converge, the resulting cross sections are orders of magnitude too large at
the higher energies if no form factor is included.  The introduction of
form factors solves divergence problems which arise in the calculation but,
more importantly, it takes into account the confinement core of the nucleon
explicitly introducing an interaction range of this size.
 
To represent the proton-proton amplitude, we use the Saclay
\cite{Bystricky78} decomposition into components called $a$, $b$, $c$, $d$
and $e$.  These independent amplitudes have individual characteristics
similar to common non-relativistic representations (see appendix A in Ref.  
\cite{gl}) and hence provide a natural extension of our familiar concepts
of the spin dependence of the amplitude at low energy. At $90^{\circ}$
center of mass scattering angle, some simplification occurs since $a\equiv
0$ and $b\equiv -c$ so that only 3 independent amplitudes are needed.

For data comparison we combine the values of $C_{NN}$
\cite{lin,bhatia,crosbie} with the measured cross sections at $90^{\circ}$
\cite{akerlof,kammerud, abe} to obtain the value of the absolute square of
the Saclay $b$ amplitude and the sum of the absolute squares of the $d$
and $e$ amplitudes (see Appendix A).  These experimental measurements will
constitute the primary data with which we will compare.  Figure
\ref{b2ope} shows the values of $2|b|^2$ compared with a calculation of
the corresponding values from one-pion exchange with a dipole form factor
with a range of $\Lambda=1.4$ GeV/c (see Section \ref{form} for a
discussion of this quantity).  The one-pion-exchange mechanism is clearly
important below 1 GeV/c but the predicted cross section decreases rapidly
above that.

There are other problems with the comparison with data of a simple 
perturbative calculation of the type that we make here.  The results 
will not, in general, be unitary.  At low energies it is not 
reasonable to compare with data without a unitary theory, which is 
the reason that a potential is normally calculated and followed by a 
solution of the Schr\"{o}dinger equation or a relativistic 
generalization.  We have made an estimate of what might be expected 
for a maximum cross section at 90$^{\circ}$ based on unitarity by 
setting each partial-wave S-matrix element to --1 and performing the 
partial-wave sum up to $k_{c.m.}R$ where $R$ was taken as 1 fm. From 
this estimate we see that above about 1.5 GeV/c the experimental 
cross section is sufficiently below this limit that unitarity can be 
expected to play a minor role. The introduction of a form factor also 
leads to questions about causality \cite{bjd,polyzou,nussenzveig} 
which we discuss in section \ref{form}.

Section II gives the general method of calculation with the introduction of
the pseudo-scalar and pseudo-vector operators for the box and crossed
kinematics. Section III develops the method for the numerical treatment of
the singularities of the propagators. Section IV introduces the form
factors used and discuses their physical basis. Section V presents the
dispersion relation calculation.  Section VI gives the basic results of the
study while section VII gives a summary of the work and states some
conclusions. Appendix A gives the projection of the spin-dependent
amplitudes onto the Saclay amplitudes, appendix B outlines the general
method used to treat the singularities numerically and appendix C gives an
interpretation of the variable used in the form factor in terms of a
four-dimensional cross product.

\section{Method of Calculation}

The differential cross section in the center of mass (c.m.) is written in
terms of the matrix element, $\calM$, as

\eq
\sigma(\theta)=\left(\frac{m^2}{4\pi E}\right)^2|\calM|^2.
\qe
Here $E$ is the energy of one proton in the c.m. and $m$ is
the proton mass. Spin sums are implicit. Since we consider
both the box and crossed diagrams, $\calM$ will be the sum of the
two. We will study only the cases of pure pseudo-scalar or
pure pseudo-vector coupling.

\subsection{Pseudo-scalar Coupling}

\subsubsection{Box Diagram}

\begin{figure}
\hspace*{.8in}
\begin{picture}(500,180)
\put(5,170){$k_1=(E,\bfk)$}
\put(275,170){$k_1'=(E,\bfk')$}
\put(120,170){$p=(E+q_0,\bfk+\bfq)$}
\put(3,30){$k_2=(E,-\bfk)$}
\put(275,30){$k_2'=(E,-\bfk')$}
\put(110,30){$p'=(E-q_0,-\bfk-\bfq)$}
\put(30,110){$q=(q_0,\bfq)$}
\put(265,110){$q'=(q_0,\bfk-\bfk'+\bfq)$}
\put(5,160){\line(1,0){380}}
\put(5,161){\line(1,0){380}}
\put(5,50){\line(1,0){380}}
\put(5,49){\line(1,0){380}}
% first pion
\put(100,50){\line(0,1){10}}
\put(100,70){\line(0,1){10}}
\put(100,90){\line(0,1){10}}
\put(100,110){\line(0,1){10}}
\put(100,130){\line(0,1){10}}
\put(100,150){\line(0,1){10}}
% first arrow up
%\put(95,128){\line(1,2){5}}
\put(95,129){\line(1,2){5}}
\put(95,130){\line(1,2){5}}
%\put(105,129){\line(-1,2){5}}
\put(105,130){\line(-1,2){5}}
\put(105,131){\line(-1,2){5}}
% second arrow up
%\put(95,68){\line(1,2){5}}
\put(95,69){\line(1,2){5}}
\put(95,70){\line(1,2){5}}
%\put(105,69){\line(-1,2){5}}
\put(105,70){\line(-1,2){5}}
\put(105,71){\line(-1,2){5}}
% second pion
\put(260,50){\line(0,1){10}}
\put(260,70){\line(0,1){10}}
\put(260,90){\line(0,1){10}}
\put(260,110){\line(0,1){10}}
\put(260,130){\line(0,1){10}}
\put(260,150){\line(0,1){10}}
% first arrow down
%\put(255,78){\line(1,-2){5}}
\put(255,77){\line(1,-2){5}}
\put(255,76){\line(1,-2){5}}
\put(260,66){\line(1,2){5}}
\put(260,67){\line(1,2){5}}
%\put(260,68){\line(1,2){5}}
% second arrow down
%\put(255,138){\line(1,-2){5}}
\put(255,137){\line(1,-2){5}}
\put(255,136){\line(1,-2){5}}
\put(260,126){\line(1,2){5}}
\put(260,127){\line(1,2){5}}
%\put(260,128){\line(1,2){5}}
\end{picture}
\caption{Kinematics for the box diagram. Note that in our
calculation, starting from Eq. \protect{(\ref{def})} form factors are used
at each vertex 
even though we have not indicated their presence here. }\label{boite}
\end{figure}
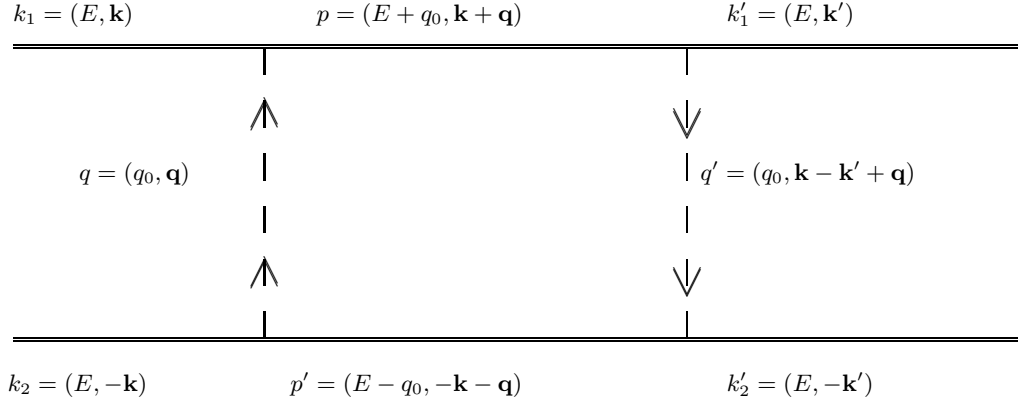

We may write the Feynman diagram for the matrix element with
the box kinematics (see Fig.~ \ref{boite}) as

$$ \calM_{PS}\delta(k_1+k_2-k'_1-k'_2)=$$
$$
\frac{g^4}{(2\pi)^4}\int \frac{dq dq' dp dp' 
\left[\bar{u}(k'_1)\gamma_5(\slash{\!\!\!p}+m)\gamma_5
u(k_1)\right]_1\left[\bar{u}(k'_2)\gamma_5(\slash{\!\!\!p'}+m)\gamma_5
u(k_2)\right]_2}
{(p^2-m^2+i\epsilon)(p'^2-m^2+i\epsilon)(q^2-\mu^2+i\epsilon)
(q'^2-\mu^2+i\epsilon)}\calf(k_1,k_2,k_1',k_2'q,q')$$
\eq \times \delta (q-p+k_1) \delta (q'-p+k'_1)
\delta (-q-p'+k_2) \delta (-q'-p'+k'_2).\label{def}
\qe
Here $g$ is the pseudo-scalar coupling constant with the normalization 
$\frac{g^2}{4\pi}= 13.75$ and $\calf(k_1,k_2,k_1',k_2',q,q')$ is the 
form factor derived from the intrinsic size of the interacting system.  
Since $\calf$ is assumed to have no spin dependence and has no poles 
on the real axes, it plays no direct role in the treatment of 
singularities or spin reduction. We suppress the arguments of $\calf$ 
in the equations for the remainder of this section and for the next 
section.
Using the first and third delta function to eliminate the integral
over $p$ and $p'$ and the (one dimensional) relationship
$ \delta(x)\delta(y)=2\delta(x+y)\delta(x-y) $,
the remaining two delta functions can be written as
$$ 2^4\delta(k_1+k_2-k'_1-k'_2)\delta(2q-2q'+k_1-k_2+k'_2-k'_1). $$
The first delta function factors out of the integral and cancels
the one on the left-hand-side of Eq. (\ref{def}).

We work in the center-of-mass system where
\eq 
\bfk_1=-\bfk_2\equiv\bfk,\ \ E_1=E_2,\ \ \bfk'_1=-\bfk'_2\equiv\bfk'
,\ \ E'_1=E'_2, 
\qe
since the external lines are on shell and
\eq E_1=E'_2\equiv E \qe
by the conservation of energy.  With these relations, the remaining
delta function under the integral sign becomes
$$ 2^4\delta(2q-2q'+2k-2k')=\delta(q-q'+k-k'). $$
For the box, (see Fig. \ref{boite}), we have 
\eq
\bfp=\bfk+\bfq=\bfk'+\bfq',\ \ \ \bfp'=-\bfk-\bfq=-\bfk'-\bfq',\ \ \ 
\bfp'=-\bfp,
\qe
and
\eq 
q_0=q'_0;\ \  p_0=E+q_0;\ \ p'_0=E-q_0 .
\qe
Thus, the expression to be evaluated reduces to a
4-dimensional integral

\eq
\calM_{PS}=\frac{g^4}{(2\pi)^4}\int \frac{dq
\left[\bar{u}(E,\bfk')\gamma_5(\slash{\!\!\!p}+m)\gamma_5u(E,\bfk)\right]^1
\left[\bar{u}(E,-\bfk')\gamma_5(\slash{\!\!\!p'}+m)\gamma_5u(E,-\bfk)\right]^2}
{(p^2-m^2+i\epsilon)(p'^2-m^2+i\epsilon)(q^2-\mu^2+i\epsilon) 
(q'^2-\mu^2+i\epsilon)}\calf .
\qe
We can reduce the spinor algebra of the operators on the two lines
using the general relationship (for on-shell spinors)

\eq \bar{u}(E_{v'},\bfv')\gamma_5u(E_v,\bfv)=\chi^{\dagger}
\frac{\sqrt{(E_{v'}+m)(E_v+m)}}{2m}
\bfs\cdot \left[\frac{\bfv}{E_v+m}-\frac{\bfv '}{E_{v'}+m}
\right]\chi, 
\qe
where $\chi$ is a Pauli spinor and we have used
\eq
u(E_v,\bfv)=\sqrt{\frac{E_v+m}{2m}}\left(
\begin{array}{c}
1\\
\frac{\bfs\cdot\bfv}{E_v+m}
\end{array}\right)\chi .
\qe
Here $E_v=\sqrt{\bfv^2+m^2}$ and $\bfs$ is the Pauli matrix.

The expression can be written as an operator in spin space in the form
\eq \calM_{PS}=
\frac{g^4}{(2\pi)^4}\int \frac{dq\  \Theta_1(p_0,\bfp)
\Theta_2(p_0',\bfp')\calf}
{(p^2-m^2+i\epsilon)(p'^2-m^2+i\epsilon)(q^2-\mu^2+i\epsilon)
(q'^2-\mu^2+i\epsilon)}.
\label{gen}
\qe
With
\eq \bar{u}_{r'}(E,\bfk')\gamma_5(\slash{\!\!\!p}+m)\gamma_5u_r(E,\bfk)=
\bar{u}_{r'}(E,\bfk')(-\gamma_0p_0+m+\gamma\cdot\bfp)u_r(E,\bfk),
\qe 
$\Theta_1$ can be written as an operator between Pauli spinors 
$\chi_{r'}$ and $\chi_r$  where the index r corresponds to the 
spin of the proton, 

$$
2m\Theta_1=
-p_0(E+m)\left(1+\frac{\bfs_1\cdot\bfk'\bfs_1\cdot\bfk}{(E+m)^2}\right)
+m(E+m)\left(1-\frac{\bfs_1\cdot\bfk'\bfs_1\cdot\bfk}{(E+m)^2}\right)
$$

$$
+(E+m)\left(\begin{array}{cc}
1&-\frac{\bfs_1\cdot\bfk'}{E+m}\end{array}\right)
\left( \begin{array}{cc}
0&\bfs_1\cdot\bfp\\
-\bfs_1\cdot\bfp&0\\
\end{array}\right)
\left( \begin{array}{c}
1\\
\frac{\bfs_1\cdot\bfk}{E+m}\\
\end{array}\right),
$$

$$2m\Theta_1
=E^+(m-p_0)-\frac{\bfs_1\cdot\bfk'\bfs_1\cdot
\bfk}{E^+}(m+p_0)+\bfs_1\cdot\bfp\bfs_1\cdot\bfk
+\bfs_1\cdot\bfk'\bfs_1\cdot\bfp,
$$

\eq
2m\Theta_1
=E^+(m-p_0)-\frac{m+p_0}{E^+}
\left(\bfk\cdot\bfk'+i\bfs_1\cdot\bfk'\times\bfk\right)
+\bfp\cdot(\bfk+\bfk')+i\bfs_1\cdot\bfp \times (\bfk-\bfk'),
\qe
where $E^+=E+m$. These expressions allow us to identify $G$ and 
$\bfH$ in the operator,
\eq
2m\Theta_1=G+i\bfs_1\cdot \bfH , 
\qe
\eq
G=E^+(m-p_0)-\frac{m+p_0}{E^+}\bfk\cdot\bfk'
+\bfp\cdot (\bfk+\bfk'),
\label{gdef}
\qe
\eq
\bfH=-\left[\frac{m+p_0}{E^+}\bfk'\times\bfk-
\bfp\times (\bfk-\bfk')\right] \label{hdef}.
\qe

\begin{figure}
\hspace*{.8in}
\begin{picture}(500,180)
\put(120,170){$p=(E-q_0,\bfk'-\bfq)$}
\put(118,30){$p'=(E-q_0,-\bfk-\bfq)$}
\put(10,170){$k_1=(E,\bfk)$}
\put(280,170){$k_1'=(E,\bfk')$}
\put(235,90){$q'=(q_0,\bfk-\bfk'+\bfq)$}
\put(90,90){$q=(q_0,\bfq)$}
\put(10,30){$k_2=(E,-\bfk)$}
\put(280,30){$k_2'=(E,-\bfk')$}
\put(5,160){\line(1,0){380}}
\put(5,159){\line(1,0){380}}
\put(5,50){\line(1,0){380}}
\put(5,49){\line(1,0){380}}
% first pion
\put(100,50){\line(3,2){10}}
\put(115,60){\line(3,2){10}}
\put(130,70){\line(3,2){10}}
\put(145,80){\line(3,2){10}}
\put(160,90){\line(3,2){10}}
\put(175,100){\line(3,2){10}}
\put(190,110){\line(3,2){10}}
\put(205,120){\line(3,2){10}}
\put(220,130){\line(3,2){10}}
\put(235,140){\line(3,2){10}}
\put(250,150){\line(3,2){15}}
%first arrow up
\put(139,70){\line(2,5){4}}
\put(140,70){\line(2,5){4}}
\put(134,79){\line(1,0){9}}
\put(134,80){\line(1,0){9}}
% second arrow up
\put(216,120){\line(2,5){4}}
\put(217,120){\line(2,5){4}}
\put(209,129){\line(1,0){9}}
\put(209,130){\line(1,0){9}}
% first arrow down
\put(216,90){\line(2,-5){4}}
\put(217,90){\line(2,-5){4}}
\put(210,79){\line(1,0){10}}
\put(210,80){\line(1,0){10}}
% second arrow down
\put(141,140){\line(2,-5){4}}
\put(142,140){\line(2,-5){4}}
\put(135,129){\line(1,0){10}}
\put(135,130){\line(1,0){10}}
% second pion
\put(100,160){\line(3,-2){10}}
\put(115,150){\line(3,-2){10}}
\put(130,140){\line(3,-2){10}}
\put(145,130){\line(3,-2){10}}
\put(160,120){\line(3,-2){10}}
\put(175,110){\line(3,-2){10}}
\put(190,100){\line(3,-2){10}}
\put(205,90){\line(3,-2){10}}
\put(220,80){\line(3,-2){10}}
\put(235,70){\line(3,-2){10}}
\put(250,60){\line(3,-2){15}}
\end{picture}
\caption{Crossed diagram kinematics. Form factors are to be included
here as in Fig. \protect{\ref{boite}}.}\label{croise}
\end{figure}
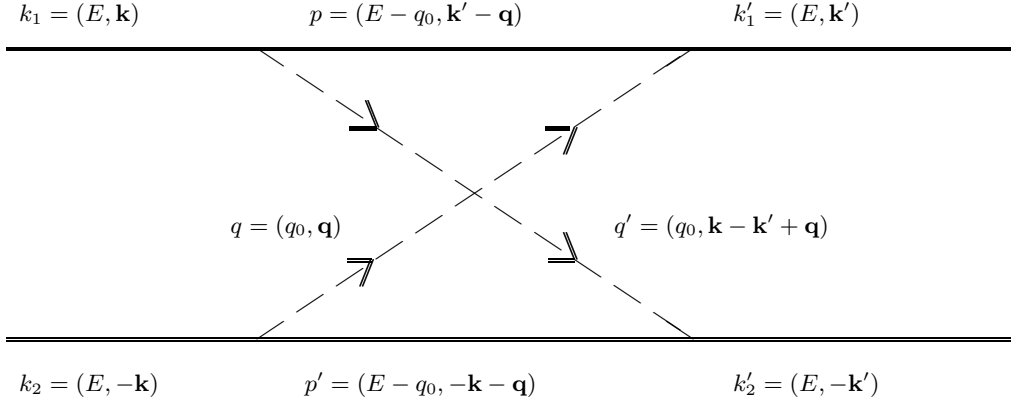

For the lower line where
\eq
2m\Theta_2=G'+i\bfs_2\cdot \bfH',
\qe
\eq
G'=E^+(m-p'_0)-\frac{m+p'_0}{E^+}\bfk\cdot\bfk'
-\bfp'\cdot (\bfk+\bfk');\ \ 
\bfH'=-\left[\frac{m+p'_0}{E^+}\bfk'\times\bfk+
\bfp'\times (\bfk-\bfk')\right].
\qe

\subsubsection{Crossed Diagram}

The expressions for the matrix element in Eq. (\ref{gen}), and for 
$\Theta_1$ and $\Theta_2$, for the two-pion crossed diagram are the same 
as in the case of the box except that now the relations between momenta 
are those corresponding to the diagram in Fig. \ref{croise},

\eq \bfp+\bfq '=\bfk ,\ \ \bfp '+\bfq=-\bfk,\ \  \bfp+\bfq=\bfk',\ \
\bfp'+\bfq'=-\bfk', \qe

\eq p_0=p_0';\ \ q_0=q_0';\ \ q_0=E-p_0. \qe
Since only neutral pions can be exchanged in the case of the box diagram 
and both neutral and charged pions can be exchanged in the crossed 
diagram, a factor of 5 multiplies the crossed diagram result. This isospin
factor is explicitly given in the dispersion relation approach in Eqs.
(\ref{Tbox}) and (\ref{Tcro}).

\subsection{Pseudo-vector Coupling}
 
In this case the interaction is given by
$\frac{f}{\mu}\bar{\psi}\gamma_{\mu}\gamma_5\tau\cdot
\partial^{\mu}\phi_{\pi}\psi $ and for one nucleon 
propagator, $P_{PV}$, the expectation value leads to
\eq
P_{PV}(p^2-m^2+i\epsilon)=-\frac{f^2}{\mu^2}\bar{u}(k')
\slq'\gamma_5(\slp+m)\slq\gamma_5u(k).
\qe
Using $k+q=p=k'+q'$
and the Dirac equation $(\slk-m)u(k)=0$ we can write
\eq
P_{PV}(p^2-m^2+i\epsilon)=\frac{f^2}{\mu^2}\bar{u}(k')
\gamma_5(\slp+m)(\slp+m)(\slp+m)\gamma_5u(k).
\qe
By regrouping the terms
\eq
P_{PV}(p^2-m^2+i\epsilon)=\frac{f^2}{\mu^2}\bar{u}(k')
\gamma_5(\slp+m)[(\slp-m)+2m][(\slp-m)+2m]\gamma_5u(k),
\qe
we find
\eq
P_{PV}(p^2-m^2+i\epsilon)=\frac{f^2}{\mu^2}\bar{u}(k')
\gamma_5(p^2-m^2)(\slp+3m)\gamma_5u(k)
+\frac{f^24m^2}{\mu^2}\bar{u}(k')\gamma_5(\slp+m)\gamma_5u(k).
\qe
The last term can be seen to be the pseudo-scalar expression
considered in the last section for one nucleon propagator.  Hence,
the operator can be separated into a term which corresponds to a
contact term and one which is identical to the pseudo-scalar
expression developed before.
\eq
P_{PV}=\frac{f^2}{\mu^2}\bar{u}(k')
\gamma_5(\slp+3m)\gamma_5u(k)
+g^2\frac{\bar{u}(k')\gamma_5(\slp+m)\gamma_5u(k)}
{p^2-m^2}\equiv P_C+P_{PS},
\qe
where we have used the standard correspondence 
$g^2=\frac{f^24m^2}{\mu^2} $.
We can write 
\eq
P_C=\frac{g^2}{4m^2}\left[
2m\bar{u}_{r'}(k')u_r(k)+\chi_{r'}^{\dagger}\Theta_1(p_0,\bfp)
\chi_r\right].
\qe

It is useful to define an operator analogous to the
$\Theta_1(p_0,\bfp)$ as before for the contact term for the upper
line for the case of the box diagram
\eq \chi_{r'}^{\dagger}C_1(p_0,\bfp)\chi_r=
2m\bar{u}_{r'}(k')u_r(k)+\chi_{r'}^{\dagger}\Theta_1(p_0,\bfp)
\chi_r.
\qe
We can also define operators corresponding to the contact
terms which are very similar to those defined before in Eqs.
(\ref{gdef}) and (\ref{hdef}).

\eq
G_C=G+2m\left(E^+-\frac{\bfk '\cdot\bfk}{E^+} \right);\ \
G_C'=G'+2m\left(E^+-\frac{\bfk '\cdot\bfk}{E^+} \right),
\label{gcdef}\qe
\eq
\bfH_C=\bfH-2m \frac{\bfk '\times\bfk}{E^+};\ \ 
\bfH_C'=\bfH'-2m \frac{\bfk '\times\bfk}{E^+},
\label{hcdef}   
\qe
with
\eq
2mC_1(p_0,\bfp)=G_C+i\bfs_1\cdot\bfH_C;\ 
\  2mC_2(p_0,\bfp)=G_C'+i\bfs_2\cdot\bfH_C'.
\qe
The same generalization to the crossed diagram can be made as in the
previous section and we do not repeat it here.
We can write for either the box or crossed diagram 
$$
\calM_{PV}=
\frac{g^4}{16m^4}\frac{1}{(2\pi)^4}\int dq\frac{C_1(p_0,\bfp)
C_2(p'_0,\bfp')\calf}{(q_0^2-\omega^2)(q_0^2-\omega'^2)}
+\frac{g^4}{4m^2}\frac{1}{(2\pi)^4}\int dq\frac{C_1(p_0,\bfp)
\Theta_2(p'_0,\bfp')\calf}{(p'^2-m^2)(q_0^2-\omega^2)(q_0^2-\omega'^2)}
$$
\eq
+\frac{g^4}{4m^2}\frac{1}{(2\pi)^4}\int dq\frac{\Theta_1(p_0,\bfp)
C_2(p'_0,\bfp')\calf}{(p^2-m^2)(q_0^2-\omega^2)(q_0^2-\omega'^2)}
 +\calM_{PS}. \label{4terms}
\qe
The first term will be referred to as the bubble diagram and
the second and third (numerically equal) as the triangle.

\section{Treatment of Singularities}

In principle, each of the propagators contribute two $\delta$-functions
and two principal-value integrals. However, for one fermion on shell
neither pion can be on shell and for one pion on shell neither fermion can
be on shell. This fact considerably reduces the number of terms which
contribute. For numerical calculation, the poles and their order must be
identified and separated into $\delta$-function and principal value parts.
The general technique for doing this is outlined in Appendix B. We cannot
use the classical methods of reduction \cite{feynman} because of the
existence of the form factor at each vertex. In the following we treat the
case for a general scattering angle $\chi$ and assume that the incident
beam direction, parallel to $\bfk$, is along the z-axis and the final
proton momentum, $\bfk'$, lies in the z-x plane.

\subsection{Contact-Contact Term (Bubble Diagram)}

For the first term in Eq. (\ref{4terms}) for either the
crossed or box configuration we have

\eq \calM_{CC}=
\frac{g^4}{16m^4}\frac{1}{(2\pi)^4}\int d\bfq
\int_{-\infty}^{\infty} dq_0\frac{C_1(p_0,\bfp)
C_2(p'_0,\bfp)\calf}{(q_0^2-\omega^2+i\epsilon)(q_0^2-\omega'^2+i\epsilon)}
\qe
\eq
=\frac{g^4}{16m^4}\frac{1}{(2\pi)^4}\int d\Omega\int_0^{\infty}
d\qt \int_{-\infty}^{\infty} dq_0\frac{O(q_0,\bfq)\calf}
{(q_0^2-\omega^2+i\epsilon)(q_0^2-\omega'^2+i\epsilon)},
\qe
where $\qt=|\bfq|$ and the volume factor $\qt^2$ has been 
included in $O(q_0,\bfq)$. Taking large finite limits for the integrals

\eq \calM_{CC}
=-\frac{g^4}{16m^4}\frac{1}{(2\pi)^4}\int d\Omega\int_0^{\qt_{max}}
\frac{d\qt}{\omega^2-\omega'^2} 
\int_{-\qmm}^{\qmm} dq_0 O(q_0,\bfq)
\left[\frac{1}{q_0^2-\omega^2+i\epsilon}
-\frac{1}{q_0^2-\omega'^2+i\epsilon}\right]\calf.
\label{127}\qe
If we consider the contribution of one pole in $q_0$ ($q_0=\omega$
for example) then a counting of powers of $\qt$ in the numerator
and denominator shows that, for large $\qt$, the integrand is proportional
to $\qt$ so that the integral diverges as $\qt_{max}^2$ if the form 
factor is set to a constant. The integral can now be evaluated by
the method given in Appendix B.  

For the first term of the integrand of the solid angle we have
\eq
\int_0^{\qt_{max}}\frac{d\qt}{\omega^2-\omega'^2+i\epsilon}
\int_{-\qmm}^{\qmm} dq_0 \frac{O(q_0,\bfq)\calf}                          
{q_0^2-\omega^2+i\epsilon}
\qe
\eq
=\int_0^{\qt_{max}}\!\!\!\!\!\! 
\frac{d\qt}{\omega^2-\omega'^2+i\epsilon}
\int_{-\qmm}^{\qmm} dq_0 \left[\frac{O(q_0,\bfq)-T(q_0,\bfq)}  
{q_0^2-\omega^2+i\epsilon}+\frac{T(q_0,\bfq)}
{q_0^2-\omega^2+i\epsilon}\right]\calf
=\int_0^{\qt_{max}}\!\!\!\!\!\!\frac{d\qt}{\omega^2-\omega'^2+i\epsilon}
F(\qt,\Omega),
\qe
where
$$
F(\qt,\Omega)=\int_{-\qmm}^{\qmm}dq_0\frac{O(q_0,\bfq)-T(q_0,\bfq)}
{q_0^2-\omega^2}\calf
$$
\eq
+\calf\frac{[O(\omega,\bfq)+O(-\omega,\bfq)]}
{2\omega}\ln {\frac{\qmm-\omega}{\qmm+\omega}}
-\frac{i\pi}{2\omega} [O(\omega,\bfq)+O(-\omega,\bfq)]\calf
\qe
and
\eq
T(q_0,\bfq)=\left[\frac{q_0+\omega}{2\omega}O(\omega,\bfq)
-\frac{q_0-\omega}{2\omega}O(-\omega,\bfq)\right]\calf.
\qe
We have included the $+i\epsilon$ in the denominator of the
external factor $1/(\omega^2-\omega'^2)$ but, in fact, there is
no singularity in the combined integral so this choice is
arbitrary. The remaining indicated integral is computed
numerically. We now need to treat the singularity in the
$\qt$ integral in a similar manner. Since
\eq
\omega^2-\omega'^2=\bfq^2-\bfq'^2=-2k^2(1-\cos\chi)
-2\bfq\cdot(\bfk-\bfk')=-2kX(\qt-\qb)
\qe
with
\eq
X=x(1-\cos\chi)-\sqrt{1-x^2}\sin\chi\cos\phi
\stackrel{\chi=90^{\circ}}{\longrightarrow} 
\cos\theta-\sin\theta\cos\phi;\ \ {\rm and}\ \ 
\qb\equiv -k(1-\cos\chi)/X. \label{wdef}
\qe
Here $\theta$ and $\phi$ are the angles of $\bfq$ in polar coordinates
and $x=\cos\theta$. We can now write
\eq
\int_0^{\qt_{max}}\frac{d\qt}{\omega^2-\omega'^2+i\epsilon}F(\qt,\Omega)
=-\frac{1}{2kX}\int_0^{\qt_{max}}\frac{d\qt F(\qt,\Omega)}
{\qt-\qb+i\epsilon}
\qe
\eq
=-\frac{1}{2kX}\int_0^{\qt_{max}}\frac{d\qt[ F(\qt,\Omega)
-F(\qb,\Omega)]}{\qt-\qb}-\frac{1}{2kX}
\ln (\qt_{max}/\qb-1)+\frac{i\pi}{2kX}F(\qb,\Omega).
\qe
The imaginary parts will cancel when the two terms of 
Eq. (\ref{127}) are combined.

\subsection{Contact-PS and PS-Contact Terms (Triangle Diagram)}

In this case there is one nucleon propagator in addition to the two
pion propagator and the  second and third terms in Eq. (\ref{4terms}) can 
be written in the form
\eq
-\frac{g^4}{4m^2}\frac{1}{(2\pi)^4}\int d\Omega\int_0^{\qt_{max}}
\frac{d\qt}{\omega^2-\omega'^2+i\epsilon}
\int_{-\qmm}^{\qmm} \frac{dq_0 O(q_0,\bfq)}
{p_0^2-E_p^2+i\epsilon}
\left[\frac{1}{q_0^2-\omega^2+i\epsilon}-
\frac{1}{q_0^2-\omega'^2+i\epsilon}\right]\calf.
\qe
A count of the powers of $\qt$ at a pole of $q_0$ shows that the integrand
varies as $1/\qt$ for large $\qt$ so the integral diverges (in the absence
of a form factor) logarithmically in $\qt_{max}$. Again there are no
double poles and the first method of Appendix B can be used to calculate
the 4 poles in $q_0$.

For the crossed-pion configuration we define, for the first
term in the above equation
\eq
P_1=E-E_p;\ \ P_2=E+E_p;\ \ P_3=\omega;\ \ P_4=-\omega,
\qe
while for the second we have,
\eq
P_1=E-E_p;\ \ P_2=E+E_p;\ \ P_3=\omega';\ \ P_4=-\omega'.
\qe
With the definition
\eq
G(q_0,\bfq)=\sum_{i=1}^4\frac{\prod_{j\ne i}(q_0-P_j)}
{\prod_{j\ne i}(P_i-P_j)}O(P_i,\bfq)\calf
\qe
the integral can be written as
\eq
-\frac{g^4}{4m^2}\frac{1}{(2\pi)^4}\int d\Omega\int_0^{\qt_{max}}
\frac{d\qt}{\omega^2-\omega'^2+i\epsilon} 
\left\{
\int_{-\qmm}^{\qmm} dq_0\left[\frac{[O(q_0,\bfq)-G(q_0,\bfq)]}
{\prod_{i=1}^4(q_0-P_i+i\epsilon)}+\frac{G(q_0,\bfq)}
{\prod_{i=1}^4(q_0-P_i+i\epsilon)}\right]\calf\right\}.
\qe
The last integral in the above becomes
\eq
\int_{-\qmm}^{\qmm} dq_0\frac{G(q_0,\bfq)\calf}
{\prod_{i=1}^4(q_0-P_i+i\epsilon)}
=\sum_{i=1}^4\int_{-\qmm}^{\qmm} dq_0\frac{O(P_i,\bfq)\calf}
{(q_0-P_i+i\epsilon)\prod_{j\ne i}(P_i-P_j)} 
\qe
\eq
=\sum_{i=1}^4 \frac{O(P_i,\bfq)\calf}{\prod_{j\ne i}(P_i-P_j)}
\ln{\left(\frac{q_0^{max}-P_i}{q_0^{max}+P_i}\right)}
+i\pi\sum_{i=1}^4 \frac{O(P_i,\bfq)\calf}
{\prod_{j\ne i}(P_i-P_j)}.
\qe
   
\subsection{PS-PS Term}

The pseudo-scalar term is given by Eq. (\ref{gen}). There are now two more 
factors of $\qt$ in the denominator so the integral is convergent. For 
this case, i.e. when there are two nucleon poles, the box and crossed 
diagrams must be treated separately.

\subsubsection{Crossed Diagram}

For the crossed-pion configuration, where $p_0=p'_0=E-q_0$,
there is formally a second order pole when $E_p=E_{p'}$, so we write
$$
\frac{1}{({p}_0^2-E_{p}^2+i\epsilon)
(p_0^2-E_{p'}^2+i\epsilon)
(q_0^2-\omega^2+i\epsilon)(q_0^2-\omega'^2+i\epsilon)}
$$
\eq
=\frac{1}{E_p^2-E_{p'}^2}\left[\frac{1}{p_0^2-E_p^2+i\epsilon}
-\frac{1}{p_0^2-E_{p'}^2+i\epsilon}\right]
\frac{1}{\omega^2-\omega'^2}
\left[\frac{1}{q_0^2-\omega^2+i\epsilon}
-\frac{1}{q_0^2-\omega'^2+i\epsilon}\right].
\label{fourterms}\qe

The integration can now be done with the methods of Appendix B
with four poles in the $q_0$ integration.

The first factor can be evaluated using
\eq
E_p^2-E_{p'}^2=\bfp^2-\bfp'^2=(\bfk'-\bfq)^2-(\bfq+\bfk)^2
=-2\qt |\bfk|f(x),
\qe
where
\eq
f(x)=x(1+\cos\chi)+\sqrt{1-x^2}\cos\phi\sin\chi
\stackrel{\chi=90^{\circ}}{\longrightarrow} 
(x+\sqrt{1-x^2}\cos\phi).
\qe
From this expression we see that the second pole has 
been transformed to the $x$ integration. The 
singularity falls at
\eq
x_0=-\frac{\sin\chi\cos\phi}{\sqrt{(1+\cos\chi)^2+\sin^2\chi
\cos^2\phi}}\stackrel{\chi=90^{\circ}}{\longrightarrow} 
-\frac{\cos\phi}{\sqrt{1+\cos^2\phi}}	
\qe
and gives a delta-function contribution of the type
$\delta[f(x)]$ with
\eq 
f'(x_0)=1+\cos\chi+\frac{\sin^2\chi\cos^2\phi}{1+\cos\chi}
\stackrel{\chi=90^{\circ}}{\longrightarrow} 1+\cos^2\phi.
\qe
Thus, for the first of the four terms in Eq. (\ref{fourterms}) we
can write
\eq
-\frac{g^4}{2k(2\pi)^4}\int d\phi \int_{-1}^1 \frac{dx}{f(x)}
\int_0^{\qt_{max}}\frac{d\qt}{\qt(\omega^2-\omega'^2)} 
\int_{-\qmm}^{\qmm} \frac{dq_0 O(q_0,\bfq)\calf}
{(p_0^2-E_p^2+i\epsilon)(q_0^2-\omega^2+i\epsilon)}
\qe

\eq =
-\frac{g^4}{2k(2\pi)^4}\int d\phi\int_{-1}^1 dx 
\frac{J(x,\phi)}{f(x)+i\epsilon}
\qe

\eq =
-\frac{g^4}{2k(2\pi)^4}\int d\phi \int_{-1}^1 dx \left[\frac{J(x,\phi)-
Q(x,\phi)}{f(x)+i\epsilon}+\frac{Q(x,\phi)}{f(x)+i\epsilon}\right],
\label{60}
\qe
where
\eq Q(x,\phi)\equiv 
\frac{f(x)}{f'(x_0)(x-x_0+i\epsilon)}J(x_0,\phi).
\qe
The first integral in Eq. (\ref{60}) now has no singularity
and can be performed numerically. The second becomes
\eq
\int_{-1}^1 dx\frac{Q(x,\phi)}{f(x)+i\epsilon}
=\frac{J(x_0,\phi)}{f'(x_0)}\int_{-1}^1 \frac{dx}{x-x_0+i\epsilon}
=\frac{J(x_0,\phi)}{f'(x_0)}\ln\left(\frac{1-x_0}{1+x_0}\right)
-i\pi\frac{J(x_0,\phi)}{f'(x_0)}.
\qe

\subsubsection{Box Diagram}

For the box diagram we have for the pole structure
$$
\frac{1}{(p_0^2-E_{p}^2+i\epsilon)
({p'}_0^2-E_{p}^2+i\epsilon)
(q_0^2-\omega^2+i\epsilon)(q_0^2-\omega'^2+i\epsilon)}.
$$
The pion pole part can be expanded as before
but the nucleon poles have a very different development since the
fourth integration variable is not the same in the two factors
but $E_p=E_{p'}$ leading to a second order pole.
We can write (with $p_0=E+q_0$ and $p'_0=E-q_0$)
$$
\frac{1}{[p_0^2-(E_p-i\epsilon)^2][{p'}_0^2-(E_p-i\epsilon)^2]}=
\frac{1}{[(E+q_0)^2-(E_p-i\epsilon)^2][(E-q_0)^2-(E_p-i\epsilon)^2]}
$$
\eq
=\frac{1}{[q_0^2-(E+E_p-i\epsilon)^2][q_0^2-(E-E_p+i\epsilon)^2]}.
\qe
The first factor can be handled by standard methods since there
is no second order pole.  The second factor, however, does 
contain a second order pole when $E_p=E$.  For this factor 
we write
$$
\frac{1}{q_0^2-(E-E_p+i\epsilon)^2}=\frac{E+E_p}{2(E^2-E_p^2+i\epsilon)}
\left[\frac{1}{q_0-E+E_p-i\epsilon}-\frac{1}{q_0+E-E_p+i\epsilon}\right].
$$
The integral can now be done with the methods of Appendix B with
five poles in the $q_0$ integral in the following manner.

We express
$$
E^2-E_p^2=k^2-\bfp^2=-\qt^2-2k\qt x=-\qt(\qt+2kx).
$$
Including the pion poles with this factor we have
$$
\frac{E+E_p}{2\qt(\qt+2kx+i\epsilon)}\frac{1}{2kX(\qt-\qb)}
\times
$$
\eq
\left[\frac{1}{q_0-E+E_p-i\epsilon}-\frac{1}{q_0+E-E_p+i\epsilon}\right]
\left[\frac{1}{q_0^2-\omega^2+i\epsilon}
-\frac{1}{q_0^2-\omega'^2+i\epsilon}\right],
\qe
where $w$ was defined in Eq. (\ref{wdef}).

We expand the factor
\eq
\frac{1}{(\qt+2kx)(\qt-\qb)}=\frac{1}{\qb+2kx}\left[
\frac{1}{\qt+2kx}+\frac{1}{\qb-\qt}\right] \label{twoterm}
\qe
and write
\eq
\frac{1}{\qb+2kx}=\frac{X}{k(1-2xX)}=\frac{X}{kf(x)},
\qe
where
\eq
f(x)=1-\cos\chi-2xc(x)\ \ 
{\rm and}\ \ \
c(x)=x-\sin\chi\sqrt{1-x^2}\cos\phi-x\cos\chi .
\qe

The function $f(x)$ has two zeros given by
\eq
x_{\pm}=\pm \sqrt{\frac{1\pm \frac{z}{\sqrt{1+z^2}}}{2}}
\ \ {\rm with}\ \ 
z=\frac{\sin\chi\cos\phi}{1-\cos\chi}.
\qe
The ``+'' under the radical goes with the ``+'' on the exterior
and vice versa. The derivatives at the zeros are 
\eq
f'(x_{\pm})=\frac{2\sqrt{1+z^2}}{x_{\pm}}.
\qe
Denoting the inner integrals over $\qt$ (and $q_0$) as $J_1$ and
$J_2$ corresponding to the two terms in Eq. (\ref{twoterm}) we write
the integral over the variable $x$ as
\eq
-\frac{g^4}{4k^2(2\pi)^4}\int d\phi \int_{-1}^1 dx 
\left[\frac{J_i(x,\phi)-
Q_i(x,\phi)}{f(x)+i\epsilon}+\frac{Q_i(x,\phi)}{f(x)+i\epsilon}\right],
\label{ji} 
\qe
where
\eq
Q_i(x,\phi)=f(x)\left[\frac{J_i(x_+,\phi)}{f'(x_+)(x-x_+)}
+\frac{J_i(x_-,\phi)}{f'(x_-)(x-x_-)}\right].
\qe

The second term in Eq. (\ref{ji}) can be evaluated by
$$
\int_{-1}^1dx\frac{Q_i(x,\phi)}{f(x)+i\epsilon}=
\frac{J_i(x_+,\phi)}{f'(x_+)}\ln \left(\frac{1-x_+}{1+x_+}\right)
-i\pi\frac{J_i(x_+,\phi)}{f'(x_+)}
$$
\eq
+
\frac{J_i(x_-,\phi)}{f'(x_-)}\ln \left(\frac{1-x_-}{1+x_-}\right)
-i\pi\frac{J_i(x_-,\phi)}{f'(x_-)},
\qe
with  
$$
J_1(x,\phi)=
\int_0^{\infty}d\qt \qt \frac{E+E_p}{\qt+2kx}
\int_{-\infty}^{\infty}dq_0
\frac{O(q_0,\bfq)\calf}{q_0^2-(E+E_p-i\epsilon)^2}\times
$$
\eq
\left[\frac{1}{q_0-E+E_p-i\epsilon}-\frac{1}{q_0+E-E_p+i\epsilon}\right]
\left[\frac{1}{q_0^2-\omega^2+i\epsilon}-
\frac{1}{q_0^2-\omega'^2+i\epsilon}\right]
\qe
and
$$
J_2(x,\phi)=
\int_0^{\infty}d\qt \qt \frac{E+E_p}{\qb-\qt}
\int_{-\infty}^{\infty}dq_0
\frac{O(q_0,\bfq)\calf}{q_0^2-(E+E_p+i\epsilon)^2}\times
$$
\eq
\left[\frac{1}{q_0-E+E_p-i\epsilon}-\frac{1}{q_0+E-E_p+i\epsilon}\right]
\left[\frac{1}{q_0^2-\omega^2+i\epsilon}-
\frac{1}{q_0^2-\omega'^2+i\epsilon}\right].
\qe

\section{Form Factors\label{form}}

The two protons are treated as finite-sized particles with their intrinsic
internal structure providing an extended source for the exchanged pions.  
In this case the range of the form factor is not a cut-off parameter to be
taken to infinity at the end of the calculation, as is often done in a
local field theory, but a physically meaningful quantity.  We consider
that the true interaction of the pion is with the partons within the
nucleons but we use an effective field theory (assumed to obey Feynman
rules) to describe the composite system.

\subsection{Basic Considerations \label{iva}}

In general, one may expect a finite distribution of the interacting
constituents of the nucleon to have two components: one due to the
confinement range of the valence quarks (and other non-color-singlet
partons) and one due to the meson cloud.  Because of G parity, the
exchange pions cannot couple directly to the pion part of the cloud
(likely to be the most important part) and the cloud can be expected to
have a larger extent (and hence be ``softer'' in momentum space).  For
these reasons we consider only that part of the density due to the
distribution of valence quarks. This argument is by no means new (see e.g.  
Maekawa and Robilotta \cite{maekawarobilotta} for the case of one-pion
exchange as well as Refs. \cite{bozoian,nozawa,haberzettl}). Form factors
of the mesonic dressing type have been extensively studied (see e.g. Ref.
\cite{gourdin}).

\begin{figure}[htb]
\epsfig{file=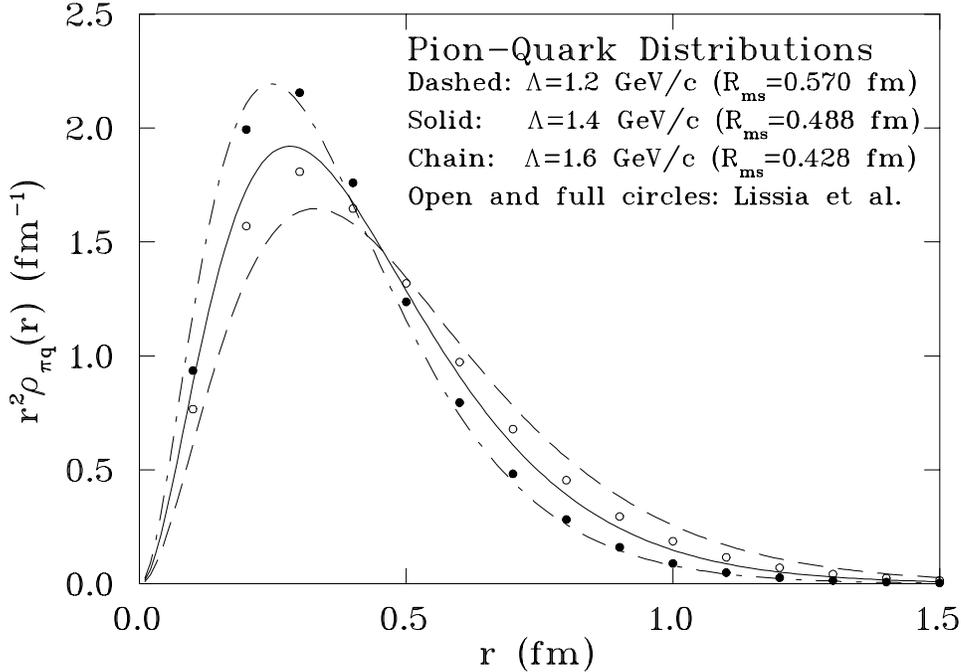,angle=90,height=3.5in}
\caption{Comparison with distributions arising from the dipole
form factor with those obtained from the lattice calculations
of Lissia et al. \protect{\cite{lissia}} by the procedures given
in the text.}
\label{quarkdist}
\end{figure}

The non-relativistic form factor can be obtained directly by calculating
the Fourier transform of the density (see e.~g. Ref. \cite{gl}).  While
more realistic forms are possible, we assume here an exponential density
in the center of mass of the nucleon corresponding to a dipole form factor
of the type
\eq 
\frac{(\Lambda^2-\mu^2)^2}{(\bfq^2+\Lambda^2)^2}=
\left(\frac{\Lambda^2-\mu^2}{\Lambda^2}\right)^2
\left(\frac{\Lambda^2}{\bfq^2+\Lambda^2}\right)^2 \label{dipole} 
\qe 
if we chose to normalize the form factor to unity at the nucleon pole and we have
neglected the small recoil correction, $\mu^4/(4m^2)$, to $-\mu^2$. The spatial
distribution corresponding to this form is proportional to $e^{-\Lambda r}$. It is
well known that there is a problem with this density since it has a finite
derivative with respect to $r$ at the origin but this presumably only causes
significant corrections at very high momenta. The value of $\Lambda=1.4$ GeV most
often used in our calculations corresponds to an rms core radius of the exponential
density of 0.49 fm.

An exponential shape is indeed suggested by lattice calculations. Lissia
et al. \cite{lissia} gave spatial distributions for the distance between
quarks, $\rho_{qq}(r_{qq})$, in the nucleon, and the pion and $\rho$
mesons.  To find the distribution about a fixed center, $\rho_q(r)$ needed
here, a conversion must be made. If we assume no correlations between
quarks and make no center-of-mass correction (in the same manner as Lissia
et al. did to compare with the MIT bag model) the conversion between these
two densities is given by the rule that the Fourier transform of
$\rho_{qq}$ is the square of that for $\rho_q$. The distributions given in
Ref. \cite{lissia} for $\rho_{qq}$ are well fit by an exponential,
$e^{-\alpha r_{qq}}$, except for the smallest values of $r_{qq}$, leading
to a dipole form factor. The square root of this form factor gives a
monopole form factor with the same value of $\alpha$.  Since the
pion-nucleon interaction will be governed by a convolution of the pion and
nucleon distributions, or a product of the form factors, the result of
this type of conversion is the product of two monopole form factors, or,
if the ranges are the same, a single dipole form factor. The ranges we
extract from the figures in Ref. \cite{lissia} are $\alpha=7.0\ {\rm
fm}^{-1}$ for the proton and $\beta=6.4\ {\rm fm}^{-1}$ for the pion. If
we were to take both ranges to be $7.0\ {\rm fm}^{-1}$ the result would be
a dipole form factor with a range very nearly 1.4 GeV/c which is what we
have used in most of the calculations shown. For two different values of
the ranges the density is given by

\eq 
\rho_{\pi q}(r)=
\frac{\alpha^2\beta^2}{\beta^2-\alpha^2}\frac{\left(e^{-\alpha r}
-e^{-\beta r}\right)}{r}. 
\qe 
This density is shown by the open circles in Fig. \ref{quarkdist} compared
with densities resulting from dipole form factors, i.e. pure exponentials.
The the root mean square radius, $R_{ms}$, corresponding to this
distribution is 0.52 fm.

Another possibility for the conversion of the pion distribution is to take 
the distance to the center of mass to be half the distance between quarks. 
This choice leads to a product form factor 
\eq
\frac{\alpha^2}{\alpha^2+\bfq^2}\left(\frac{\beta^2}{\beta^2+\bfq^2}
\right)^2,
\qe
where the $\beta$ is twice the value found in the fit to
$\rho_{qq}$. This form factor gives the density
\eq \rho_{\pi q}(r)=
\frac{\alpha^2\beta^4}{r(\beta^2-\alpha^2)^2}\left[
e^{-\alpha r}-e^{-\beta r}-\frac{(\beta^2-\alpha^2)r}{2\beta}
e^{-\beta r}\right],
\qe
which is shown as the solid dots in Fig. \ref{quarkdist}. The 
$R_{ms}$ corresponding to this density is 0.44 fm.

Alexandrou et al. \cite{alex} also made calculations of $\rho_{qq}$ and
commented that their distributions were very well fit with an exponential
form. In a variety of calculations they found $R_{rms}$ values of the
pion-proton system in the range from 0.55 to 0.60 fm.

Another estimate of this radius can be obtained directly from
form factors for pion-nucleon coupling. For a monopole form
factor, a value of $\Lambda$ of 800 MeV to 1 GeV was found
\cite{coon}. Choosing an equivalent rms radius leads to values 
in the range 1.13 to 1.41 GeV/c for a dipole form.

While the original MIT bag model used a radius near 1 fm. (for
all of the constituents of the nucleon), corrections due to the
inclusion of a pion cloud and the recoil degree of freedom made
by Bolsterli and Parmentola \cite{mark} led to values of the 
rms radius near 0.45 fm.

\subsection{Relativistic Generalization}

One often attempts to make a generalization of Eq.
(\ref{dipole})  to a relativistic dependence on momentum of
the type
\eq
\frac{(\Lambda^2-\mu^2)^2}{(q_0^2-\bfq^2-\Lambda^2)^2}, 
\label{mesonform}
\qe
i.e. $-\bfq^2\rightarrow q^2=q_0^2-\bfq^2$. While this
substitution makes the form factor invariant, it is not the only
way to achieve this objective. This procedure introduces
additional singularities on the real axis in $q_0$  which are often 
regarded as spurious.

We suppose that the distribution of quarks is known in the rest frame of
the nucleon and assume it to be spherically symmetric with its Fourier
transform being a function of the square of the conjugate momentum,
$\bft^2$. This distribution should be boosted into the center of mass
system of the two nucleons but a boost is only defined for on-shell
particles.
 
On general principle, to preserve Lorentz invariance, the form factor
should be a function of Lorentz scalars only.  In the case of a
pion-nucleon vertex (neglecting spin) there are three four-vectors to work
with, the initial and final nucleon momenta and the pion momentum.  Due to
4-momentum conservation only two are independent.  Let us choose the
initial nucleon momenta (k)  and the pion momentum (q).  From them we can
construct three scalars $q^2,\ k^2,\ \ {\rm and}\ \ k\cdot q $.

We expect the desired invariant to involve the nucleon momentum if it is
to represent a boost into the nucleon rest frame in the on-shell limit.  
Working with these invariants and taking into account that there should be
scaling with the dimensions of q, to evaluate to $\bft^2$, the expression
should be bilinear in q.  Since $k\cdot q$ is linear in q we must use its
square, which implies that, in order to be coherent in k as well, the only
two invariants to consider are $(k\cdot q)^2$ and $k^2q^2$. The linear
combination
\eq
 (k\cdot q)^2-k^2q^2\equiv m^2Z^2(k,q)\label{qdef}
\qe
indeed reduces to $m^2\bft^2=m^2\bfq^2$ when the nucleon 
four-momentum, k, is taken to be $(m,0)$.  We shall adopt the variable 
$Z^2(k,q)$, which provides an off-shell generalization of the boost to the 
nucleon rest frame, to replace $\bfq^2$ in Eq. (\ref{dipole}). For a 
discussion of the Lorentz tensor which represents the 4-dimensional cross 
product of k and q and can be contracted to form $Z^2$ see appendix C.

\subsection{Singularities in $q_0$}

The problem of the possible unwanted poles in $q_0$ on the 
real axis is at least partially resolved since the condition 
for a singularity $ (k\cdot q)^2-k^2q^2+m^2\Lambda^2=0 $
 gives as roots for $q_0$ 
\eq q_0=\frac{k_0\bfk\cdot\bfq \pm \sqrt{-\{
k^2[\bfk^2\bfq^2-(\bfk\cdot\bfq)^2]+\bfk^2m^2\Lambda^2\}}}
{\bfk^2}.  \label{polepos}
\qe
As long as the nucleon remains time-like ($k^2>0$) the quantity
under the radical is negative so that there is no zero on the
real axis.  Since, for the exchange of two pions, at least one of
the nucleons at each vertex is on shell, and there is no
zero on the real axis for $q_0$ for the nucleon momentum on
shell, the problem of unwanted additional singularities is
resolved.

Poles \underline{do} exist in the complex plane and lead a finite value of
the principal-value integral on $q_0$ (an integral over only poles on the
real axis will lead to a principal value contribution of zero).  Since, if
the pole is very far from the real axis its contribution will become
negligible, in the limit of $\Lambda\rightarrow\infty$ the principal-value
part of the $q_0$ integral vanishes as expected.  We can see from Eq.
(\ref{polepos}) that in the limit of $\bfk\rightarrow 0$\ the pole also
moves far from the real axis so at low energies the principal-value
contribution tends to zero.

Such poles in the upper half of the complex plane are often considered to
lead to a violation of causality.  In the present case we are considering
a pair of finite-size particles which are described in terms of the motion
of their centers.  Hence, they can start to interact when they are a
distance apart equal to twice their radii.  The usual considerations of
causality \cite{bjd,polyzou,nussenzveig} deal with two point particles and
the expectation that any interaction between them can not propagate faster
than the speed of light. This criterion does not apply to the present case
since the points which need to be connected by the speed of light are not
the centers but any co-moving points included in their structure. The
scattered wave can start to appear before the centers of the distributions
can be connected by a light ray. The appearance of these poles is natural
and expected.  Of course, their contribution tends to zero as the size of
the system goes to zero ($\Lambda \rightarrow\infty$).

Since the size of the system becomes unmeasurable at very low energy we
should expect that the nucleons could be treated as point particles in
this regime so that we expect that standard causality will be
valid in the low-energy limit. We see that, since the pole off of the real
axis gives a negligible contribution in this region being very far from
the real axis this expectation is realized.

\subsection{Properties of $Z^2$}

The function $Z^2$ has the property  
\eq
Z^2(p\pm q,q)=Z^2(p,q) \ {\rm or,\ in\ particular}\ 
Z^2(k',q)=Z^2(k,q).  \label{kpmq}
\qe
Since $Z^2$ is independent of the use of the initial or final
nucleon momentum, the vertex function is a property of the
vertex itself and not of the individual four-momenta.  If the
three 4-momenta which are connected to the vertex are $p_1,\
p_2$ and $p_3$ with a conservation law between them, say
$p_3=\pm p_1\pm p_2$, then any two of the momenta can be used
for the evaluation of the vertex function, i.e.
\eq
Z^2(p_1,p_2)=Z^2(p_1,p_3)=Z^2(p_3,p_2)=\dots
\qe
An equivalent way to express this function is
\eq
Z^2=\frac{[p_1^2]^2+[p_2^2]^2+[p_3^2]^2-2(p_1^2p_2^2
+p_1^2p_3^2+p_2^2p_3^2)}{4m^2}.\label{squares}
\qe

In a more general view, there are two invariants available;
$(k\cdot q)^2/m^2$, which evaluates to $q_0^2$ in the rest
frame of the nucleon and $\frac{(k\cdot q)^2-k^2q^2}{m^2}$
which evaluates to $\bfq^2$ in this rest frame. One could
choose any combination of $\bfq^2$ and $q_0^2$ desired for the
dependence in the rest frame. However, only $[(k\cdot
q)^2-k^2q^2]/m^2$ (and $q^2$ by dint of containing no
reference to the nucleon momentum at all) are independent of
which nucleon momentum ($k$ or $k'$) is used.

The use of the form factor obtained from the Fourier transform of the 
density in the nucleon center of mass without any dependence on $q_0$ 
corresponds to an interaction which is instantaneous in time. This is 
perhaps to be expected since the valence quarks are always present 
for interaction, whereas the interaction through the intermediate 
step of the emission of a meson requires a time $\hbar$/Mass for the 
meson to be formed and propagate.

Other authors \cite{pena,others} have used relativistic
generalizations of the form factor.  In particular Ramalho,
Arriaga and Pe\~na \cite{pena} have argued that the first
type of form factor dependence, Eq. (\ref{mesonform})  
corresponds to the dressing of the nucleon by mesons of mass
$\Lambda$ and suggest that the variable $\bfq^2$ should be
replaced by
\eq
{\cal Q}^2\equiv \frac{(P\cdot q)^2}{P^2}-q^2,
\qe
where $P$ is the center-of-mass momentum of the two nucleons.  
The value of ${\cal Q}^2$ evaluates to $\bfq^2$ in the center of 
mass system so that there is no dependence on $q_0$ if one
calculates in that reference frame, hence no possible
singularity in that variable. This may be a useful form but
the underlying basis in physics is not clear since the form factor
should be a property of the pion-nucleon vertex and not the
entire system.  How a pion interacts with a single nucleon
should not depend on the momentum of a second nucleon.

\subsection{Conditions on the Form Factor}

Reviewing the previous paragraphs, we may summarize the desired properties 
of an off-shell form factor arising from an extended source.

\begin{enumerate}
\item
There should be no poles of $q_0$ on the real axis.

\item
For the case of the nucleon on shell and at rest, the form factor
should be independent of $q_0$ so that the interaction is instantaneous.

\item
As the nucleon energy approaches zero, poles in $q_0$ should tend to $\pm
i\infty$. This is necessary since, in this limit, the interaction could be
regarded as being among point particles and, by causality, there can be no
contribution from poles off of the real axis. While the arguments of the
previous point and this one arise from different physical bases, they are
mathematically related.

For a typical example consider a form factor of the type
\eq
\frac{\Lambda^2}{aq_0^2+bq_0+\bfq^2+\Lambda^2},
\qe
where $a$ and $b$ are real functions of $\bfq$, $\bfk$ and $k_0$.
The poles in $q_0$ will be given by
\eq
q_0=\frac{-b\pm \sqrt{b^2-4a(\bfq^2+\Lambda^2)}}{2a}.
\qe
In order that $q_0\rightarrow \pm i\infty$ as $|\bfk|\rightarrow 0$
both $a$ and $b$ must tend to zero in this limit. This, in turn, 
means that the form factor becomes independent of $q_0$ in the limit 
$|\bfk|\rightarrow 0$.

\item
In the nucleon rest frame the form factor (which is a function of
$|\bfq|^2$ only) is the Fourier transform of the source density, a 
positive definite function.

\end{enumerate}

\subsection{Application to the present case.}

For the application of the form factor in the present calculation
we can always choose the nucleon to be one of the external 
lines and hence on shell with energy E.  If we define
\eq
f(E,\bfk ,q_0,\bfq)=f(k,q)=\left[\frac{\Lambda^2-\mu^2}{\frac{(k\cdot
q)^2}{m^2}-q^2+\Lambda^2}\right]^2, \label{fdef}
\qe
a product of four of these factors will appear, one for each 
vertex. For the box diagram we have

\eq \calf=
f(E,\bfk,q_0,\bfq)f(E,-\bfk,q_0,\bfq)f(E,\bfk',q_0,\bfq')
f(E,-\bfk',q_0,\bfq'), \label{ffbox}
\qe
while for the crossed diagram the product
\eq \calf=
f(E,\bfk,q_0,\bfq')f(E,-\bfk,q_0,\bfq)f(E,\bfk',q_0,\bfq)
f(E,-\bfk',q_0,\bfq') \label{ffcross}
\qe
enters. Since these functions are scalars and do not have any poles
on the real axis, the manipulation of the spin observables
and the treatment of the singularities is not affected by
their presence.

\section{Dispersion Relation Approach}

In order to cross-check our Feynman diagram results we have
calculated the two-pion exchange in the dispersion
relation approach following Ref. \cite{Cottingham73}. The
matrix element $\calM$ of Eq.~(1) in section II can be written
as,

\begin{equation}
        \calM=\sum_{i=1}^{5}\left[3p_{i}^{+}\left(w,t,\bar t\right)
+2p_{i}^{-}\left(w,t,\bar t\right)
\mbox{\boldmath $\tau$}_{1}\cdot\mbox{\boldmath $\tau$}_{2}\right]P_{i}.
        \label{M+-}
\end{equation}
Here, $w$, $t$, $\bar t$ are the Mandelstam invariants satisfying 
$w+t+\bar t=4m^2$, $\mbox{\boldmath $\tau$}_{1(2)}$ the 
isospin Pauli matrices for the nucleon 1(2) and $P_i$ the perturbative 
invariants. One has,

\begin{equation}
w  = \left(k_1+k_2\right)^2,
                \qquad
t  = \left(k_1-k_1'\right)^2,
                \qquad
\bar t  = \left(k_1-k_2'\right)^2
\label{wttb}
\end{equation}
and
 \begin{equation}
        \begin{array}{rcl}
                P_{1} & = & 1^{1}1^{2},  \\
                \ecart
                P_{2} & = & -\left[\left(\gamma^{1} \cdot N\right)1^{2}+\left(\gamma^{2} \cdot P\right)1^{1}\right],  \\
                \ecart
                P_{3} & = & \left(\gamma^{1} \cdot N\right)\left(\gamma^{2} \cdot P\right), \\
                \ecart
                P_{4} & = & \gamma^{1}\cdot \gamma^{2}, \\
                \ecart
                P_{5} & = & \gamma^{1}_{5} \gamma^{2}_{5},
        \end{array}
        \label{Pi}
  \end{equation}
with
  \begin{equation}
                N  =  \frac{1}{2}\left(k_{2}+k'_{2}\right),
                \qquad  
                P  =  \frac{1}{2}\left(k_{1}+k'_{1}\right).
                \label{NP}
   \end{equation}

Assuming the Mandelstam representation \cite{Mandelstam55}, the
two-pion exchange invariant amplitudes $p_{i}^{\pm}\left(w,t,\bar
t\right)$ of Eq.~(\ref{M+-}) satisfy the following double dispersion
relations,

 \begin{eqnarray}
    p_{i}^{\pm}\left(w,t,\bar{t}\right) & = &{\displaystyle
    \frac{1}{\pi^2}\int_{4\mu^{2}}^{+\infty}\frac{dt'}{t'-t-\imath\epsilon}\ 
\int_{4m^{2}}^{+\infty}    \frac{dw'}{w'-w-\imath\epsilon}\ y_{i}^{\pm}\left(w',t'\right)}
\nonumber\\
&&{\displaystyle\mp (-1)^i\frac{1}{\pi^2}\int_{4\mu^{2}}^{+\infty}\frac{dt'}{t'-t-\imath\epsilon}\
\int_{4m^2}^{+\infty}\frac{dw'}{w'-\bar t-\imath\epsilon}\ y_{i}^{\pm}\left(w',t'\right)},
\label{pi+-d}
\end{eqnarray}
where the $y_{i}^{\pm}(w,t)$ are the double spectral functions. In 
Eq.~(\ref{pi+-d}) the first double integrals represent the contributions 
of the two-pion-exchange box diagram (see Fig.~\ref{boite}).   The second 
double integrals  give the contribution of the crossed-pion diagram   
(Fig.~\ref{croise}). They are obtained from the contributions of the box  
diagram by the substitution $k_1 \leftrightarrow -k_1'$ which is 
equivalent  to that of $ w  \leftrightarrow \bar t$. This substitution  
changes the sign of the invariants $P_1$,  $P_3$, $P_5$ while  $P_2$ and  
$P_4$ are not modified. It also changes the isospin dependence. These  
changes result in the signs given here. We shall show below how one can  
calculate the two-pion-exchange double spectral functions 
$y_{i}^{\pm}(w,t)$.  Use of the following decomposition

\begin{eqnarray}
\label{twprime}
\frac{1}{t-t'-\imath\epsilon} \ \frac{1}{w'-(4m^2-w-t)-\imath\epsilon} &=&
\frac{1}{t-t'-\imath\epsilon}\ \frac{1}{w'-(4m^2-w-t')-\imath\epsilon}\nonumber\\
&&+\frac{1}{w'-(4m^2-w-t)-\imath\epsilon}\ \frac{1}{t'-(4m^2-w-w')-\imath\epsilon}
\end{eqnarray}
allows us to write Eq.~(\ref{pi+-d}) as

 \begin{eqnarray}
    p_{i}^{\pm}\left(w,t,\bar{t}\right) & = &{\displaystyle
    \frac{1}{\pi}\int_{4\mu^{2}}^{+\infty}\frac{dt'}{t'-t-\imath\epsilon}\ [\rho_{i}^{\pm}\left(w,t'\right)\mp (-1)^{i}\rho_{i}^{\pm}\left(4m^2-w-t',t'\right)]}\nonumber\\
&&{\displaystyle\mp (-1)^i\frac{1}{\pi}\int_{4m^2}^{+\infty}\frac{dw'}{w'-\bar t-\imath\epsilon}\ a_i^\pm\left(w',4m^2-w-w'\right)},
\label{pi+-}
\end{eqnarray}
with
\begin{equation}
     \rho_{i}^{\pm}(w,t)=\mbox{Im}_t\ p_{i}^{\pm}(w,t)=\frac{1}{\pi}
     \int_{4m^{2}}^{+\infty}    \frac{dw'}{w'-w-\imath\epsilon}\ y_{i}^{\pm}\left(w',t\right),
\label{rhoi+-}
\end{equation}
and
\begin{equation}
     a_{i}^{\pm}\left(w',4m^2-w-w'\right)=\frac{1}{\pi}
     \int_{4\mu^{2}}^{+\infty}    \frac{dt'}{t'-\left(4m^2-w-w'\right)-
\imath\epsilon}\ y_{i}^{\pm}\left(w',t'\right).
        \label{ai+-}
\end{equation}
Since the term with $a_{i}^{\pm}\left(w',4m^2-w-w'\right)$
in Eq.~(\ref{pi+-}) gives rise to a rather short range force
with mass exchanges larger than $2m$ it was neglected in
the Paris Potential \cite{Cottingham73, Richard75,
Lacombe81}.  In Eq.~(\ref{rhoi+-}) the notation
$\mbox{Im}_t$ comes from the use of the relation
\begin{equation}
        \frac{1}{t'-t-\imath\epsilon}=\frac{P}{t'-t}+\imath\pi\delta(t'-t),
        \label{delta}
\end{equation}
where $P$ indicates the principal value part of the 
integral.

\begin{figure}[htb]
\hspace*{0.9in}
{\large
\begin{picture}(500,180)
\put(5,170){$k_1=(\sqrt{t}/2,\bfk')$}
\put(275,170){$k_1'=(\sqrt{t}/2,-\bfk')$}
\put(120,170){$p=(q_0-\sqrt{t}/2,\bfq-\bfk')$}
\put(3,30){$k_2=(\sqrt{t}/2,\bfk)$}
\put(275,30){$k_2'=(\sqrt{t}/2,-\bfk)$}
\put(110,30){$p'=(\sqrt{t}/2-q_0,\bfk-\bfq)$}
\put(30,110){$q=(q_0,\bfq)$}
\put(265,110){$q'=(\sqrt{t}-q_0,-\bfq)$}
\put(5,160){\line(1,0){380}}
\put(5,161){\line(1,0){380}}
\put(5,50){\line(1,0){380}}
\put(5,49){\line(1,0){380}}
% first pion
\put(100,50){\line(0,1){10}}
\put(100,70){\line(0,1){10}}
\put(100,90){\line(0,1){10}}
\put(100,110){\line(0,1){10}}
\put(100,130){\line(0,1){10}}
\put(100,150){\line(0,1){10}}
% first arrow up
%\put(95,128){\line(1,2){5}}
\put(95,129){\line(1,2){5}}
\put(95,130){\line(1,2){5}}
%\put(105,129){\line(-1,2){5}}
\put(105,130){\line(-1,2){5}}
\put(105,131){\line(-1,2){5}}
% second arrow up
%\put(95,68){\line(1,2){5}}
\put(95,69){\line(1,2){5}}
\put(95,70){\line(1,2){5}}
%\put(105,69){\line(-1,2){5}}
\put(105,70){\line(-1,2){5}}
\put(105,71){\line(-1,2){5}}
% second pion
\put(260,50){\line(0,1){10}}
\put(260,70){\line(0,1){10}}
\put(260,90){\line(0,1){10}}
\put(260,110){\line(0,1){10}}
\put(260,130){\line(0,1){10}}
\put(260,150){\line(0,1){10}}
% first arrow up
%\put(255,78){\line(1,2){5}}
%\put(255,77){\line(1,2){5}}
%\put(255,76){\line(1,2){5}}
%\put(260,66){\line(-1,2){5}}
%\put(260,67){\line(-1,2){5}}
\put(255,69){\line(1,2){5}}
\put(255,70){\line(1,2){5}}
\put(265,70){\line(-1,2){5}}
\put(265,71){\line(-1,2){5}}
\put(255,129){\line(1,2){5}}
\put(255,130){\line(1,2){5}}
\put(265,130){\line(-1,2){5}}
\put(265,131){\line(-1,2){5}}
\end{picture}
}
\caption{Box diagram kinematics in the t-channel c.m.. 
}\label{tboite}
\end{figure}
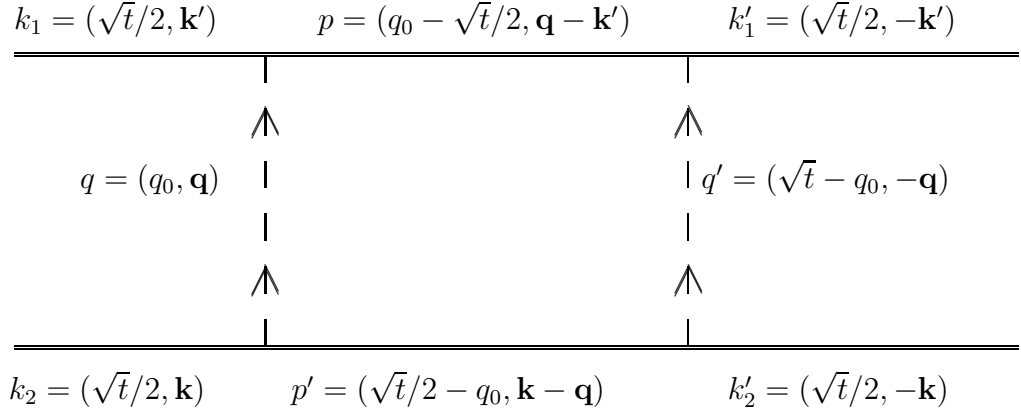
 
The unitarity condition in the t-channel for the $N \bar N \to
2\pi \to N \bar N$ process \cite{Richard75} leads to

\begin{equation}
  \begin{array}{l}
        \sum_{i}\lbrack \rho_{i}^{\pm}(w,t) \mp (-1)^{i}\rho_{i}^{\pm}\left(4m^{2}-w-t,t\right)
        \rbrack P_{i} =  \\
        \ecart
 {\displaystyle \frac{1}{128\pi^{2}}\sqrt{\frac{t-4\mu^{2}}{t}}       
       \int d \Omega_{Q}\left\lbrack A_{1}^{*\pm}\left(s_1,t,\bar s_1\right)1^1
        +\mbox{\boldmath $\gamma^1$}\cdot
        \mathbf{Q}B_{1}^{*\pm}\left(s_1,t,\bar s_1\right)\right\rbrack
        \left\lbrack A_{2}^{\pm}\left(s_2,t,\bar s_2\right)1^2
          +\mbox{\boldmath $\gamma^2$}\cdot
        \mathbf{Q}B_{2}^{\pm}\left(s_2,t,\bar s_2\right)\right\rbrack},
 \end{array}   
        \label{tchannelunitarity}
\end{equation}
where the four-momenta of the on-shell exchanged pions, in their t-channel center of mass (see the kinematics given in Fig.~\ref{tboite}), are,

\begin{equation}
  q = \left(\frac{\sqrt{t}}{2}, -{\bf Q}\right),
  \qquad 
  q' = \left(\frac{\sqrt{t}}{2},{\bf Q}\right).
\label{qq'tcms}
\end{equation}

The amplitudes $A(B)_{1(2)}^{\pm}\left(s_{1(2)},t,\bar
s_{1(2)}\right)$ are the usual invariants for the reaction $N
\bar N \to \pi \pi$ \cite{Hoehler83}. The Mandelstam invariants
$s_{1(2)}$, $t$ and $\bar s_{1(2)}$, satisfying, $s_{1(2)}+t+\bar
s_{1(2)}=2m^2+2\mu^2$, are

\begin{equation}
  \begin{array}{rcccl}
s_1      & = & \left(k_1-q'\right)^2   & = & m^2+\mu^2-\displaystyle\frac{t}{2}-2{\bf P}\cdot {\bf Q },\\
                \ecart
t        & = & \left(k_1-k_1'\right)^2 & = & 4\left({\bf Q}^2 + \mu^2\right), \\
                \ecart
\bar s_1 & = & \left(k_1+q\right)^2   & = & m^2+\mu^2-\displaystyle\frac{t}{2}+2{\bf P}\cdot {\bf Q} 
\end{array}
\label{s1ts1b}
\end{equation}
and

\begin{equation}
s_2       =  \left(k_2-q\right)^2    =  m^2+\mu^2-\frac{t}{2}-2{\bf N}\cdot {\bf Q} ,\\
                \qquad
\bar s_2  =  \left(k_2+q'\right)^2    =  m^2+\mu^2-\frac{t}{2}+2{\bf N}\cdot {\bf Q},
\label{s2s2b}
\end{equation}
with, $ q^2=q'^2=\mu^2=\frac{t}{4}-{\bf Q}^2 $.
Expressions~(\ref{s1ts1b}) and~(\ref{s2s2b}) follow from 

\begin{equation}
  \begin{array}{rcccl}
  k_1 & = & \left(-\displaystyle\frac{\sqrt{t}}{2}, {\bf k}_1\right),
\qquad
   k'_1 & = & \left(\displaystyle\frac{\sqrt{t}}{2}, {\bf k}_1\right),\\
                \ecart
    k_2  & = & \left(\displaystyle\frac{\sqrt{t}}{2}, {\bf k}_2\right),
\qquad
     k'_1  & = &  \left(-\displaystyle\frac{\sqrt{t}}{2}, {\bf k}_2\right),\\
                \ecart
     P & = & \left(0, {\bf k}_1\right),
\qquad
  \   N & = & \left(0, {\bf k}_2\right),
\end{array}
\label{ki}
\end{equation}
since $ {\bf k}_2- {\bf k'}_2= {\bf q}- {\bf q'}= {\bf k'}_1- {\bf k}_1=0 
$ (see Fig.~\ref{tboite} and Eq.(\ref{NP})).  Introducing the 
four-vector $ L_\lambda=\varepsilon_{\lambda \nu \rho \sigma} 
M^\nu K^\rho \Delta^\sigma $, where,

\begin{equation}
\begin{array}{rcccl}
M & = & N+P & =& (0,{\bf N} + {\bf P}),\\
                \ecart
K & = & N-P & =& (0,{\bf N} - {\bf P}),\\
                \ecart
\Delta & = & k'_1-k_1 & =& (\sqrt{t},0),
\end{array}
\label{MKD}
\end{equation}
$Q$ can be represented as
\begin{equation}
Q=\frac{Q\cdot M}{M^2}M+\frac{Q\cdot K}{K^2}K
+\frac{Q\cdot \Delta}{\Delta^2}\Delta+\frac{Q\cdot L}{L^2}L.
\label{Q}
\end{equation}
Eq.~(\ref{tchannelunitarity}) can then be decomposed in the invariants 
$P_i$ of Eq.~(\ref{Pi}) as \cite{Richard75,Brown76},
 \begin{eqnarray}
        &&\sum_{i}\lbrack \rho_{i}^{\pm}(w,t) \mp (-1)^{i}\rho_{i}^{\pm}\left(4m^{2}-w-t,t\right)
        \rbrack P_{i} = {\displaystyle \frac{1}{128\pi^{2}}\sqrt{\frac{t-4\mu^{2}}{t}}} \nonumber \\
        \ecart
                 && {\displaystyle \times \left[\left({\cal A}^{\pm}+2m {\cal C}^{\pm}+m^2 {\cal E}^{\pm} \right) P_1
+\left({\cal B}^{\pm}+m {\cal D}^{\pm}\right)P_2+{\cal E}^{\pm} P_3-{\cal F}^{\pm} P_4 \right]},
\label{rhoionPi}
\end{eqnarray}
with 

 \begin{equation}
        \begin{array}{rcl}
 {\cal A}^{\pm} & = &  {\displaystyle \int d \Omega_{Q} A_{1}^{*\pm} A_{2}^{\pm}}, \\
                \ecart
{\cal B}^{\pm} & = & {\displaystyle \int d \Omega_{Q} B_{1}^{*\pm} A_{2}^{\pm} 
\left(\frac{Q\cdot M}{M^2}+\frac{Q\cdot K}{K^2}\right)},\\
                \ecart
{\cal C}^{\pm} & = &  {\displaystyle \int d \Omega_{Q} B_{1}^{*\pm} A_{2}^{\pm}
\left(\frac{Q\cdot M}{M^2}-\frac{Q\cdot K}{K^2}\right)},\\
                \ecart
{\cal D}^{\pm} & = &  {\displaystyle\frac{1}{M^2 K^2}\int d \Omega_{Q} B_{1}^{*\pm} B_{2}^{\pm}
\left[\frac{(Q\cdot M)^2}{M^2}\left(2K^2+M^2\right)+\frac{(Q\cdot K)^2}{K^2}
\left(K^2+2M^2\right)-Q^2\left(K^2+M^2\right)\right]},\\
                \ecart

{\cal E}^{\pm} & = &  {\displaystyle\frac{1}{M^2 K^2}\int d \Omega_{Q} B_{1}^{*\pm} B_{2}^{\pm}
\left[\frac{(Q\cdot M)^2}{M^2}\left(2K^2-M^2\right)+\frac{(Q\cdot K)^2}{K^2}
\left(K^2-2M^2\right)-Q^2\left(K^2-M^2\right)\right]},\\
                \ecart
{\cal F}^{\pm} & = & {\displaystyle \int d \Omega_{Q} B_{1}^{*\pm} B_{2}^{\pm} 
\left[-Q^2+\frac{(Q\cdot M)^2}{M^2}+\frac{(Q\cdot K)^2}{K^2}\right]}.\\
\end{array}
\label{ABCDEF}
\end{equation}
Here $M^2=w$ and $K^2=\bar t = 4m^2-w-t$. Note that the 
Bjorken-Drell metric used here leads to some sign differences with
respect to references \cite{Richard75, Brown76} where the Pauli metric was 
used.

\subsection{Nucleon Pole Contribution for Pseudo-scalar Coupling}

If the invariant amplitudes $A(B)_{1(2)}^{\pm}$ are expressed as
fixed-t dispersion relations \cite{Hoehler83} one can write the
$\rho_i^{\pm}$ given by Eqs.~(\ref{rhoionPi}) and~(\ref{ABCDEF})
in the spectral form~(\ref{rhoi+-}). Following Ref. 
\cite{Nishijima69} one can use the Cutkosky rules 
\cite{Cutkosky54} to calculate the imaginary part of the 
$\rho_i^{\pm}$ from Eq.~(\ref{ABCDEF}). The expressions for the
$y_i^{\pm} (w,t)$ of Eq.~(\ref{rhoi+-}) in terms of double
integrals over $s_1$ and $s_2$ on quantities depending on elastic
pion-nucleon scattering absorptive parts, are given by Eqs.~(2.6)
to (2.8) of Ref. \cite{Cottingham73}.  The double spectral
functions $y_{i,N}^{\pm}(w,t)$ of the box diagram for
pseudo-scalar (PS) pion-nucleon coupling can then be obtained
from these equations. In this case the $A(B)_{1(2),PS}^{\pm}$ for
the $\pi N$ nucleon Born term are (see e.g. Eq.~(A.6.30) of
Ref. \cite{Hoehler83}),

\begin{equation}
A_{1(2),PS}^{\pm}=0,
\qquad
B_{1(2),PS}^{\pm}=g^2 \left(\frac{1}{m^2-s_{1(2)}-\imath\epsilon}
\mp\frac{1}{m^2-\bar s_{1(2)}-\imath\epsilon}\right).
\label{ABPS}
\end{equation}
The absorptive parts of $A_{1(2),PS}^{\pm}$ are zero and those of  
$B_{1(2),PS}^{\pm}$ are given by $\pi g^2\ \delta (s_{1(2)}-m^2)$ 
for  the direct term leading to,

  \begin{equation}
        \begin{array}{rcl}
         y_{1,N}^{\pm}(w,t) & = & {\displaystyle \pm 
         \frac{\pi^{2}}{2}\left(\frac{g^{2}}{4\pi}\right)^{2}
          \frac{m^{2}}{\sqrt{t}}\frac{\calK}{\left(4m^{2}-t-w\right)^{2}}
          \left[\frac{4m^{2}-2w-t}{w^{2}\calK^{2}}
          +\left(t-2\mu^{2}\right)^{2}\right]},\\
           \ecart    y_{2,N}^{\pm}(w,t)& = & {\displaystyle \pm 
           \frac{\pi^{2}}{2}\left(\frac{g^{2}}{4\pi}\right)^{2}
           \frac{m}{\sqrt{t}}\frac{\calK}{\left(4m^{2}-t-w\right)^{2}} 
  \left[-\frac{4m^{2}-t}{w^{2}\calK^{2}}+\left(t-2\mu^{2}\right)^{2} 
                \right]},\\
                \ecart
                y_{3,N}^{\pm}(w,t) & = & {\displaystyle\frac{ 
                y_{1}^{\pm}(w,t)}{m^{2}}},\\
                \ecart
                y_{4,N}^{\pm}(w,t) & = & {\displaystyle \pm 
                \frac{\pi^{2}}{2}\left(\frac{g^{2}}{4\pi}\right)^{2}
          \frac{1}{\sqrt{t}}\calK\left[-\frac{1}{w\left(4m^{2}-t-w\right) 
                \calK^{2}}\right]},\\
                \ecart
                y_{5,N}^{\pm}(w,t)& = & 0,  
        \end{array}
        \label{yiboite}
  \end{equation}
with
 
\begin{equation}
       \calK=w^{-1/2}\left\lbrack\left(t-4\mu^{2}\right)\left(w-4m^{2}\right)
        -4\mu^{4}\right\rbrack^{-1/2}.
        \label{Kwt}
\end{equation}
We define,

 \begin{equation}
        A(y,t')=\int_{x_{0}}^{+\infty}\frac{dw'}{w'-y}\ K(w',t'), 
\qquad
         C(y,t')=\int_{x_{0}}^{+\infty}\frac{dw'}{(w'-y)^{2}}\ K(w',t') 
        \label{DefdeC}
\end{equation}
where $x_{0}$ is the lower limit of the $w'$ integration such that
 $K(w',t')$ is defined, i. e.,
$(t'-4\mu^{2})(w'-4m^{2})-4\mu^{4}\ge 0$, which leads to, 
  \begin{equation}
        x_{0}(t')=4m^{2}+\frac{4\mu^{4}}{t'-4\mu^{2}} \ge 4m^2.
        \label{DefdeX0}
  \end{equation}
With the definition,

\begin{equation}
        \eta(w',t')  =  4\mu^{4}-(t'-4\mu^{2})(w'-4m^{2})
          = (x_{0}-w')(t'-4\mu^{2}),
        \label{DefdeAlpha}
  \end{equation}
we have  \cite{Zytnicki01},
 \begin{equation}
        \rho_{1}^{\pm}(w,t')=\rho_{11}^{\pm}(w,t')-\rho_{12}^{\pm}(w,t')
        \label{Ro1}
 \end{equation}
 with
 \begin{equation}
        \begin{array}{rcl}
                \rho_{11}^{\pm}(w,t') & = & {\displaystyle \pm \frac{\pi}{2} 
                \left(\frac{g^{2}}{4\pi}\right)^{2} \frac{m^{2}}{\sqrt{t'}}\ 
                \frac{(t'-2\mu^{2})^{2}}{4m^{2}-w-t'}\left\lbrack  
          \frac{A(w,t')-A(4m^{2}-t',t')}{4m^{2}-w-t'}
                  +C(4m^{2}-t',t') \right\rbrack}, \\
                \ecart
            \rho_{12}^{\pm}(w,t')& = & {\displaystyle \pm \frac{\pi}{2} 
                \left(\frac{g^{2}}{4\pi}\right)^{2} \frac{m^{2}}{\sqrt{t'}}
                \left\lbrack \mathcal{F}_{1}A(0,t')+\mathcal{F}_{2}A(w,t')
                +\mathcal{F}_{3}C(4m^{2}-t',t')
                 +\mathcal{F}_{4}A(4m^{2}-t',t')\right\rbrack}.
        \end{array}
        \label{Ro11etRo12}
 \end{equation}
 The ${\cal F}_i$ are given by,
\begin{equation}
        \begin{array}{rcccccl}
                \mathcal{F}_{1} & = & {\displaystyle 
                -\frac{4\left[m^{2}\left(t'-4\mu^{2}\right)+\mu^{4} \right]}{w\left(4m^{2}-t'\right)}},
                \qquad
                & \mathcal{F}_{2} & & = & {\displaystyle 
                -\frac{\eta\left(w,t'\right)\left(2w+t'-4m^{2}\right)}
                {w\left(w+t'-4m^{2}\right)^{2}}},  \\
                \ecart
                \mathcal{F}_{3} & = & {\displaystyle -\frac{\left(t'-2\mu^{2}\right)^{2}}
                {4m^{2}-w-t'}},
                \qquad
               & \mathcal{F}_{4}&  & = & -\mathcal{F}_{1} -\mathcal{F}_{2}.
        \end{array}
        \label{DefdeF}
\end{equation}
The expression for $\rho_{2}^{\pm}(w,t')$ is,
\begin{equation}
        \rho_{2}^{\pm}(w,t')=\rho_{21}^{\pm}(w,t')+\rho_{22}^{\pm}(w,t')
        \label{Ro2}
\end{equation}
with
\begin{equation}
        \begin{array}{rcl}
                \rho_{21}^{\pm}(w,t') & = & {\displaystyle \frac{\rho_{11}^{\pm}(w,t')}
                {m}},  \\
                \ecart
                \rho_{22}^{\pm}(w,t') & = & {\displaystyle \pm \frac{\pi}{2} 
                \left(\frac{g^{2}}{4\pi}\right)^{2} \frac{m}{\sqrt{t'}}}
(t'-4m^{2})
                \left[\mathcal{G}_{1}A(0,t')+ \mathcal{G}_{2}A(w,t')
                 +\mathcal{G}_{3}C(4m^{2}-t',t')
                +\mathcal{G}_{4}A(4m^{2}-t',t')
                 \right]
        \end{array}
        \label{Ro21etRo22}
\end{equation}
and,
\begin{equation}
        \begin{array}{rcccl}
                \ecart
                \mathcal{G}_{1} & = & {\displaystyle 
                -\frac{\mathcal{F}_{1}}{4m^{2}-t'}}, 
                \qquad
                \mathcal{G}_{2}& = & {\displaystyle \frac{\mathcal{F}_{2}}{2w+t'-4m^{2}}},  \\
                \ecart
               \mathcal{G}_{3} & = & {\displaystyle \frac{\mathcal{F}_{3}}
                {4m^{2}-t'}},
                \qquad
               \mathcal{G}_{4} & = & -\mathcal{G}_{1} -\mathcal{G}_{2}.
        \end{array}
        \label{DefdeG}
\end{equation}
Furthermore,
 
\begin{equation}
        \rho_{3}^{\pm}(w,t')  = {\displaystyle 
        \frac{\rho_{1}^{\pm}(w,t')}{m^{2}}},
        \label{Ro3}
\end{equation}
 
 \begin{equation}
        \rho_{4}^{\pm}(w,t')  =  \pm {\displaystyle 
\frac{\pi^{2}}{2} 
                \left(\frac{g^{2}}{4\pi}\right)^{2} 
\frac{1}{\sqrt{t'}}\ 
                \frac{1}{4m^{2}-w-t'}}
                \left[\eta(w,t')A(w,t')
                 -(t'-2\mu^{2})^{2}A(4m^{2}-t',t')\right].
        \label{Ro4}
 \end{equation} 

The functions $A(y,t')$ and  $C(y,t')$ can be calculated analytically. First notice that,

\begin{equation}
        C(y,t')=-\frac{\sqrt{t'-4\mu^{2}}}
        {(t'-4m^{2})(t'-2\mu^{2})^{2}}
        +\frac{(t'-4\mu^{2})(2t'-4m^{2})+4\mu^{4}}
        {(t'-4m^{2})(t'-2\mu^{2})^{2}}\ A(y,t'),
        \label{DefdeC2}
\end{equation}

\begin{equation}
     A(y,t')=\int_{x_{o}}^{+\infty}\frac{K(w',t')}{w'-y-\imath\epsilon}\ dw'
     =P\int_{x_{o}}^{+\infty}\frac{K(w',t')}{w'-y}\ dw'
     +i\pi K(y,t').                     
        \label{modifA}
\end{equation}
One obtains Eq.~(\ref{modifA}) using Eq.~(\ref{delta}). 
Depending on the value of $y$ one has the following results:
 
\begin{itemize}
        \item  if $y<0$, then $\eta(y,t')=(x_{0}-y)(t'-4\mu^{2})>0$
 \begin{equation}
        A(y,t')=\frac{1}{\sqrt{-\eta(y,t') y}}
                \ln{
                \frac{
                1+\sqrt{\frac{-y(t'-4\mu^{2})}{\eta(y,t')}}
                }
                {
                1-\sqrt{\frac{-y(t'-4\mu^{2}}{\eta(y,t')}}
                }}
               = \frac{1}{\sqrt{-\eta(y,t') y}}
                \ln{
                \frac{1+\sqrt{\frac{y}{y-x_{0}}}}
                {
                1-\sqrt{\frac{y}{y-x_{0}}}
                }
                },
        \label{A1}
 \end{equation} 

        \item  if $0<y<x_{0}$, then $\eta(y,t')>0$ and 
\begin{equation}
        A(y,t')=\frac{2}{\sqrt{\eta(y,t') y}}
                \arctan{\sqrt{\frac{y(t'-4\mu^{2})}{\eta(y,t')}}}
                =\frac{2}{\sqrt{\eta(y,t') y}}
                \arctan{\sqrt{\frac{y}{x_{0}-y}}},      
                \label{A2}
\end{equation}

        \item  if $y>x_{0}$, then $\eta(y,t')<0$, $A$ has an imaginary 
part:
\begin{equation}
     \begin{array}{rcl}
        \text{Re}A(y,t') & = & {\displaystyle 
        -\frac{1}{\sqrt{-\eta(y,t') y}}
        \ln{
        \frac{
        1+\sqrt{
        \frac{-y(t'-4\mu^{2})}{\eta(y,t')}
        }
        }
            {
            \sqrt{\frac{-y(t'-4\mu^{2})}{\eta(y,t')}}-1}
            }}
              =  {\displaystyle 
             -\frac{1}{\sqrt{-\eta(y,t') y}}
                \ln{
                \frac{
                1+\sqrt{\frac{y}{y-x_{0}}}
                }
                {
                \sqrt{\frac{y}{y-x_{0}}}-1
                }
                }
                },\\
            \ecart       
        \text{Im}A(y,t') & = & K(y,t').
     \end{array}
         \label{A3}
\end{equation}
\end{itemize}

It can be seen from Eqs.~(\ref{A1}) and~(\ref{A2}) that $A$ is continuous 
at y=0 with the value,

        \begin{equation}
                A(0,t')=\frac{2\sqrt{t'-4\mu^{2}}}{\eta(0,t')w}
                \label{Aen0}.
        \end{equation}

On the other hand, $A$ is discontinuous at $x_{0}$, since 
taking the limits in Eqs. ~(\ref{A2}) and~(\ref{A3}),

        \begin{equation}
                \begin{array}{l}
                        {\displaystyle A(y,t')\underset{y \to x_{0}^{-}}{\sim}\quad 
                        \frac{\pi}
                        {\sqrt{(x_{0}-y)(t'-4\mu^{2})y}}\,\longrightarrow  \, 
                        \infty}, \\
                        \ecart
                        {\displaystyle 
                        \lim_{y \to x_{0}^{+}}A(y,t')=-\frac{1}{y 
                        \sqrt{t'-4\mu^{2}}}}.
                \end{array}
                \label{limite}
        \end{equation}
This square root divergence will disappear after performing the 
integration over  $t'$.

The integration over $t'$ in Eq.~({\ref{ai+-}) is performed analytically. 
Let us define, 
 
\begin{equation}
        \widetilde K(w',t')=t'^{-1/2}
\left\lbrack\left(t'-4\mu^{2}\right)\left(w'-4m^{2}\right)-4\mu^{4}\right\rbrack^{-1/2},
        \label{tildaKwt}
\end{equation}

 \begin{equation}
        \widetilde A(w',\widetilde y)=\int_{\widetilde x_{0}}^{+\infty}\frac{dt'}{t'-\widetilde y}\ \widetilde K(w',t'), 
\qquad
         \widetilde C(w',\widetilde y)=
\int_{\widetilde x_{0}}^{+\infty}\frac{dt'}{(t'-\widetilde y)^{2}}\ 
\widetilde K(w',t'), 
        \label{DefdetildeC}
\end{equation}
where $\widetilde x_{0}$ is the lower limit of the $t'$ integration such that
 $\widetilde K(w',t')$ is defined, i. e.,
$(t'-4\mu^{2})(w'-4m^{2})-4\mu^{4}\ge 0$, which leads to, 
  \begin{equation}
        \widetilde x_{0}(t')=4\mu^{2}+\frac{4\mu^{4}}{w'-4m^{2}} \ge 4\mu^2.
        \label{DefdetildeX0}
  \end{equation}
One obtains, with $\widetilde y=4m^2-w-w'$ and $\widetilde w'=4m^2-w'$,
 \begin{equation}
        \begin{array}{rcl}
        a_{1}^{\pm}(w',\widetilde y) & = & {\displaystyle \pm \frac{\pi}{2} 
                \left(\frac{g^{2}}{4\pi}\right)^{2} 
 \left\lbrack a_{11}(w',\widetilde y)+a_{12}(w',\widetilde y)\right\rbrack},\\
	\ecart
        a_{2}^{\pm}(w',\widetilde y) & = & {\displaystyle \pm \frac{\pi}{2} 
                \left(\frac{g^{2}}{4\pi}\right)^{2} 
 \left\lbrack \frac{ a_{11}(w',\widetilde y)}{m}+ \widetilde a_{12}(w',\widetilde y)\right\rbrack},\\
	\ecart
        a_{3}^{\pm}(w',\widetilde y) & = & {\displaystyle 
	\frac{a_{1}^{\pm}(w',\widetilde y)}{m^2}},\\
	\ecart
        a_{4}^{\pm}(w',\widetilde y)& = & {\displaystyle  \pm \frac{\pi}{2} 
                \left(\frac{g^{2}}{4\pi}\right)^{2} 
             \frac{1}{w\sqrt{w'}}
                \left\lbrack \eta(w',\widetilde y)\widetilde A(w',\widetilde y)
                 - \eta(w',\widetilde w')\widetilde A(w',\widetilde w')
                \right\rbrack}.
        \end{array}
         \label{a1234+-}
 \end{equation}
 In Eq.~(\ref{a1234+-}),
 \begin{equation}
        \begin{array}{rcl}
a_{11}(w',\widetilde y) & = & {\displaystyle 
\frac{m^{2}}{w^2\sqrt{w'}}\ \left\lbrack
\left(\widetilde y -2\mu^2 \right )^2 \widetilde A(w',\widetilde w')
 -\left(\widetilde w'-2\mu^2\right)\left(\widetilde y-w-2\mu^2 
\right)\widetilde A(w',\widetilde w')
+w\left (\widetilde w'-2\mu^2 \right )^2 \widetilde 
C(w',\widetilde w') \right\rbrack}, \\
                \ecart
            a_{12}(w',\widetilde y)& = & {\displaystyle  
             \frac{m^{2}}{w^2w'\sqrt{w'}}
                \left\lbrack (w-w')\eta(w',\widetilde y)\widetilde A(w',\widetilde y)
                 - [w^2+(w'-w) \eta(w',\widetilde y)]\widetilde A(w',\widetilde w')
               +w w' \eta(w',\widetilde w') \widetilde C(w',\widetilde w')\right\rbrack},\\
	\ecart
            \widetilde a_{12}(w',\widetilde y)& = & {\displaystyle  
   \frac{m}{w^2w'\sqrt{w'}}
\left\lbrack (w+w')\eta(w',\widetilde y)\widetilde 
A(w',\widetilde y)
+[w^2-(w+w') \eta(w',\widetilde y)]\widetilde A(w',\widetilde w')
+ww'\eta(w',\widetilde w') \widetilde C(w',\widetilde 
w')\right\rbrack}.
        \end{array}
        \label{a11eta12}
 \end{equation}
The functions $\widetilde A(w',\widetilde y)$ and  
$\widetilde C(w',\widetilde y)$ with  $\widetilde y \le 0$  are, 
\begin{equation}
      \widetilde A(w',\widetilde y) = \frac{1}{\sqrt{-\eta(w',\widetilde y) \widetilde y}}
                \ln{
                \frac{1+\sqrt{\frac{\widetilde y}{\widetilde y-\widetilde x_{0}}}}
                {
                1-\sqrt{\frac{\widetilde y}{\widetilde y-\widetilde x_{0}}}
                }
                },
        \label{tildeA1}
 \end{equation} 
\begin{equation}
        \widetilde C(w',\widetilde y)=\frac{\sqrt{-\widetilde w'}}
        {\widetilde y \eta(w',\widetilde y)}
        -\frac{\widetilde w'(\widetilde y-2m^{2})+2\mu^{4}}
        {\widetilde y \eta(w',\widetilde y)}\ \widetilde A(w',\widetilde y).
        \label{DefdetildeC2}
\end{equation}

  We have calculated the discontinuities of the  
$p_{i}^{\pm}$ along $w$, i. e.,

 \begin{equation}
        \lim_{\epsilon \to 0} \left[
        p_{i}(w+\imath\epsilon,t')-p_{i}(w-\imath\epsilon,t')\right]
        =2i\text{Im}_w \  p_{i}(w,t')=2i 
                \frac{1}{\pi}\int_{4\mu^{2}}^{+\infty}\frac{dt'}{t'-t}
                \ \text{Im}_w \  \rho_{i}^\pm(w,t').    
      \label{rappel}
 \end{equation}
We introduce the following explicit notations
\begin{equation}
        \begin{array}{rcl}
           \rho_{i}^{\pm \, \text{box}}(w,t') & = & \text{Re}\rho_{i}^{\pm}(w,t'),  \\
           \ecart
           \rho_{i}^{\pm \, \text{cro}}(w,t') & = & \mp (-1)^{i}
           \rho_{i}^{\pm \, \text{box}}(w\leftrightarrow 4m^{2}-t'-w,t'),  \\
           \ecart
           \text{Im}_w\ \rho_{i}^{\pm \, \text{box}}(w,t') & = & 
           \text{Im}_w\ \rho_{i}^{\pm}(w,t')= y_{i,N}^{\pm}(w,t').
        \end{array}
        \label{notation1}
  \end{equation}
Here again, the last equality of Eq.~(\ref{notation1}) follows from
the application of Eq.~(\ref{delta}). In the expressions
(\ref{Ro11etRo12}), (\ref{Ro21etRo22}), (\ref{Ro4}) and (\ref{A3}),
the imaginary part of the $\rho_{i}^{\pm}(w,t')$ comes only from $A$.
There are only three $A$ terms:  $A(w,t')$, $A(0,t')$ and
$A(4m^{2}-t',t')$ in Eqs. (\ref{Ro11etRo12})  and (\ref{Ro21etRo22})
that have a non-zero imaginary part for $y>x_{0}$. However, from
(\ref{DefdeX0}) one cannot have $4m^{2}-t'>x_{0}$, so, \textit{a
fortiori} one cannot have $4m^{2}-t'-w>x_{0}$ and \textbf{only the box
diagram has an imaginary part} which justifies the superscript ``box''
in the expression for the imaginary part. The superscript "cro"
denotes the contribution from the crossed diagram. We write

\begin{eqnarray}
                p_{i}^{\pm \, \text{box}}(w,t) & = & {\displaystyle 
                \frac{1}{\pi}\int_{4\mu^{2}}^{+\infty}\frac{dt'}{t'-t}
                \ \rho_{i}^{\pm \, \text{box}}(w,t')},
                \label{Pech} \\
        \ecart 
                p_{i}^{\pm \, \text{cro}}(\bar{t},t) & = & {\displaystyle 
                \frac{1}{\pi}\int_{4\mu^2}^{+\infty}\frac{dt'}{t'-t}\ 
                \rho_{i}^{\pm \, \text{cro}}(\bar{t},t')}\nonumber\\&&
  {\displaystyle \mp (-1)^i\frac{1}{\pi}\int_{4m^2}^{+\infty}
 \frac{dw'}{w'-\bar t}\ a^{\pm}_i(w',4m^2-w-w')},  \label{Pcro} \\
                \ecart
                \text{Im}_w\ p_{i}^{\pm}(w,t) & = & {\displaystyle 
                \frac{1}{\pi}\int_{4\mu^{2}}^{+\infty}\frac{dt'}{t'-t}\ 
                y_{i,N}^{\pm}(w,t')
                }. \label{Pima}
\end{eqnarray}
It can be seen from studying the high $t'$ behavior of the functions
(\ref{notation1}) that they decrease like $1/t'$ or $\ln (t')/t'$ for
$t'\to +\infty$, so that the integrals~(\ref{Pech}), (\ref{Pcro}) and
(\ref{Pima}) on $t'$ converge. A similar convergence holds for the
integrals on $w'$. Note that the physical values of $t (\bar t)$ being
negative, there are no poles in the $t' (w')$ integration, so there is no
further imaginary part. In the present form, the numerical calculation of
Eqs.~(\ref{Pech}), (\ref{Pcro}) and (\ref{Pima}) requires one numerical
integration over $t'$ or $w'$.

We have seen (Eq. (\ref{limite})) that $A(y,t')$ has a divergence in $y \to
x_{o}^{-}$. In the expressions for integration on $t'$, $y$ takes the values
$4m^{2}-t'$, $0$, $4m^{2}-t'-w$ and $w$ (see Eqs. (\ref{Ro11etRo12}),
(\ref{Ro21etRo22}) and (\ref{notation1})). The divergence at $y=x_{0}$ is in
the expression for $\rho_{11}^{\pm}$ in Eq. (\ref{Ro11etRo12}) where the first
argument of $A$ is $w$, so that only the box diagram has this divergence.
Fixing $w$ and using Eq.~(\ref{DefdeX0}) the condition $w=x_{0}$ gives 
for $\rho_{i}^{\pm \, \text{box}}(w,t')$ a discontinuity at 
$t'=t'_{0}$ with

                \begin{equation}
                        t'_{0}=4\mu^{2}+\frac{4\mu^{4}}{w-4m^{2}}.
                        \label{defdeT0}
                \end{equation}
 The integral (\ref{Pech}) can be split into  two pieces,
 
                \begin{equation}
                        p_{i}^{\pm \, \text{box}}(w,t) = 
                \frac{1}{\pi}\int_{4\mu^{2}}^{t'_{0}}\frac{dt'}{t'-t}
               \ \rho_{i}^{\pm \, \text{box}}(w,t')+
                \frac{1}{\pi}\int_{t'_{0}}^{+\infty}\frac{dt'}{t'-t}
                \ \rho_{i}^{\pm \, \text{box}}(w,t').
                        \label{Pech2}
                \end{equation}
The first integral, of the type,

                \begin{equation}
                        \int_{4\mu^{2}}^{t'_{0}}\frac{dt'}{\sqrt{t'_{0}-t'}}\ F(t'),
                        \label{integrale1}
                \end{equation}
becomes, with $x=\sqrt{t'_{0}-t'}$, 

                \begin{equation}
                        2\int_{0}^{\sqrt{t'_{0}-4\mu^{2}}}dx\ F(t'_{0}-x^{2})
                        \label{integrale2}.
                \end{equation}
This transformation is applied to
$F(t')=\left[\sqrt{t'_{0}-t'}/(\pi(t'-t))\right]\ 
\rho_{i}^{\pm \, 
                \text{box}}(w,t')$.
The second integral can be recast into a finite domain by the following 
change of variable
                $$t'=\frac{\lambda (x+1)-2t'_{0}}{x-1},$$
where $\lambda$ is a free parameter. One then has to calculate, 
                \begin{equation}
                        \frac{1}{\pi}\int_{-1}^{1}\frac{2(t'_{0}-\lambda)dx}
                        {\left\lbrack\lambda (x+1)-2t'_{0}-t(x-1)\right\rbrack}
                        \ \rho_{i}^{\pm \, \text{box}}(w,t'(x)).
                        \label{integrale3}
                 \end{equation} 
        
For the calculation of $p_{i}^{\pm \,  \text{cro}}(\bar{t},t)$ and  
$\text{Im}_w\ p_{i}^{\pm}(w,t)$ there is no discontinuity in  $t'_{0}$. For 
$p_{i}^{\pm \, \text{cro}}(\bar{t},t)$, one can transform the first term of
(\ref{Pcro}) into an integral of the type (\ref{integrale3}). For the 
imaginary part the domain of definition of $K(w,t')$ restricts the 
integration interval to 
$[t'_{0},+\infty]$. There is a singularity at the lower limit in
                 $1/\sqrt{t'-t'_{0}}$ which we treat as above 
(see Eqs.~(\ref{integrale1}) and (\ref{integrale2})).

We have (see Eqs.~(\ref{yiboite}), (\ref{notation1}) and (\ref{Pcro})),
                \begin{equation}
                        p_{i}^{+ \, \text{box}}(w,t)=-p_{i}^{- \, \text{box}}(w,t),
                        \qquad
                        p_{i}^{+ \, \text{cro}}(\bar{t},t)=p_{i}^{- \, \text{cro}}
                        (\bar{t},t).
                        \label{isospin3}
                 \end{equation}

The proton-proton amplitude being  a pure isospin one state (I=1), 
one has (see Eq.~(\ref{M+-})), 

\begin{equation}
\calM^{I=1}=\sum_{i=1}^{4}\left[p_{i}^{+ \,
\text{box}}(w,t)+Imp_{i}^{+}(w,t)+5p_{i}^{+ \,
\text{cro}}(\bar{t},t)\right]P_{i}.
\label{isospin4}
                  \end{equation}

We give below the formulae which express, for a given isospin 
state, the 
nucleon-nucleon helicities amplitudes $\varphi_{i}$ in terms of the $p_i$ 
amplitudes. One has \cite{Richard75},
\begin{equation}
        \begin{array}{rcl}
                \varphi_{1} & = & {\displaystyle 
                \frac{1+z}{2}\left(p_{1}-2mDp_{2}+m^{2}D^{2}p_{3}\right)
                +\left(D- \frac{1-z}{2}\right)p_{4}
                 },\\
                \ecart
                \varphi_{2} & = & {\displaystyle -\frac{1-z}{2}\left(\frac{E^{2}}{m^{2}}p_{1}
                -\frac{2E^{2}}{m}p_{2}+E^{2}p_{3}+p_{4}+\frac{k^2}{m^2}p_{5}\right) }, \\
                \ecart
                \varphi_{3} & = & {\displaystyle \frac{1+z}{2}\left(p_{1}-2mDp_{2}  
                +m^{2}D^{2}p_{3}+Dp_{4}\right) },\\
                \ecart
                \varphi_{4} & = &  {\displaystyle \frac{1-z}{2}\left( 
                   \frac{E^{2}}{m^{2}}p_{1}-\frac{2E^{2}}{m}p_{2}+E^{2}p_{3}+p_{4}-\frac{k^2}{m^2}p_{5}\right) },\\
                \ecart
                 \varphi_{5}& = & {\displaystyle \frac{E\sqrt{1-z^2}}{2m}\left(-p_{1}
                 +\frac{2E^{2}}{m}p_{2}-m^{2}Dp_{3}-p_{4}\right)},
        \end{array}
        \label{invtohel2}
\end{equation}
where $D=(E^{2}+k^{2})/m^{2}$ and $z=\cos\theta$ (let us remember that in 
the center of mass of the nucleon-nucleon w-channel, $w=4E^2=4(k^2+m^2)$ 
and $t=-2k^2(1-z)$). The Saclay amplitudes $a,\ b,\ c,\ d$ and $e$ are 
given in terms of the $\varphi_i$ amplitudes by the following 
expressions  \cite{Bystricky78},

 \begin{equation}
        \begin{array}{rcl}
                a & = &  {\displaystyle \frac{1}{2}\left[ \left(\varphi_{1}   
                 +\varphi_{2}+\varphi_{3} -\varphi_{4}\right)z 
-4\varphi_{5}\sqrt{1-z^2}\right]},\\
                \ecart
                b & = &  {\displaystyle \frac{1}{2} \left(  \varphi_{1}   
                -\varphi_{2} +\varphi_{3} +\varphi_{4}
                \right)}, \\
                \ecart
                c & = & {\displaystyle \frac{1}{2} \left( - \varphi_{1}   
                +\varphi_{2} +\varphi_{3} +\varphi_{4}
                \right)},  \\
                \ecart
                d & = &  {\displaystyle \frac{1}{2} \left(  \varphi_{1}   
                +\varphi_{2}- \varphi_{3} +\varphi_{4}
                \right)}, \\
                e & = & {\displaystyle -\frac{i}{2} \left[ \left( \varphi_{1}   
                +\varphi_{2} +\varphi_{3} -\varphi_{4}
                \right)\sqrt{1-z^2}+4\varphi_{5}z\right]}.
        \end{array}
        \label{heltobis}
 \end{equation}

\subsection{Nucleon pole contribution for Pseudo-vector coupling}

For pseudo-vector pion-nucleon coupling, the $A(B)_{1(2),PV}^\pm$ 
for the $\pi N$ nucleon Born term are (see e.g. Eqs.~(A.8.1) and (A.8.2) 
of Ref. \cite{Hoehler83}),

\begin{equation}
 \begin{array}{rcl}
\label{ABPV}
A_{1(2),PV}^+ & = & \displaystyle\frac{g^2}{m},\quad A_{1(2),PV}^-=0, \\
\ecart
B_{1(2),PV}^+  & = & B_{1(2),PS}^+,\  B_{1(2),PV}^-= 
B_{1(2),PS}^--\displaystyle\frac{g^2}{2m^2}.
 \end{array}
\end{equation}
One can then use Eqs.~(\ref{rhoionPi}) and (\ref{ABCDEF}) to calculate 
the corresponding $\rho^\pm_i$.
One recovers first the contribution for pseudo-scalar coupling from 
$B_{1,PS}^\pm\ B_{2,PS}^\pm$ given in the above section.
There are further bubble ($bub$) diagram contributions arising from 
$A_{1,PV}^+\ A_{2,PV}^+$ and proportional to 
$\left(\displaystyle\frac{g^2}{m}\right)^2$ and from $B_{1,PV}^-\ 
B_{2,PV}^-$ and proportional to $\left(\displaystyle\frac{g^2}{2m^2}\right)^2$.
There are also triangle ($tri$) diagram contributions arising from $ 
B_{1,PV}^-\ B_{2,PV}^-$ and proportional to 
$\displaystyle\frac{g^2}{2m^2}\ B_{1(2),PS}^-$. 
One finds
\begin{equation}
\label{rhoibubtri}
\begin{array}{rcl}
 \rho_{1}^{+bub}(t')& = &\displaystyle\frac{g^4}{32\pi 
m^2}\sqrt{\displaystyle\frac{t'-4\mu^2}{t'}},\\
\ecart
\rho_{1}^{+tri}(t')& = &\displaystyle\frac{g^4}{16\pi}\sqrt{\displaystyle\frac{t'-4\mu^2}{t'}}\ 
\displaystyle\frac{t'-4\mu^2}{t'-2\mu^2}\ I_2(t'), \\
\ecart
\rho_{2}^{-tri}(t')& = &-\displaystyle\frac{g^4}{128\pi 
m}\sqrt{\displaystyle\frac{t'-4\mu^2}{t'}}\ 
\displaystyle\frac{t'-4\mu^2}{t'-4m^2}\ [I_0(t')-3I_2(t')], \\
\ecart
\rho_{4}^{-bub}(t')& = &-\displaystyle\frac{t'-4\mu^2}{48m^2}\ 
\rho_{1}^{+bub}(t'), \\
\ecart
\rho_{4}^{-tri}(t')& = &\displaystyle\frac{g^4}{256\pi 
m^2}\sqrt{\displaystyle\frac{t'-4\mu^2}{t'}}\ 
(t'-4\mu^2)\ [I_0(t')-I_2(t')].
 \end{array}
\end{equation}
In the above formulae~(\ref{rhoibubtri}) 
\begin{equation}
\label{I2n}
I_{2n}(t')=\int_{-1}^{+1}\displaystyle\frac{u^{2n}du}{\alpha+\bar\beta u},\ 
I_2(t')=\displaystyle\frac{\alpha}{\bar\beta^2}
[-2+\alpha I_0(t')],
\end{equation}
with $\alpha=\mu^2-t'/2$. For $t'<4m^2,\ 
\bar\beta=i\beta=i2[(m^2-t'/4)(t'/4-\mu^2)]^{1/2}$ and
\begin{equation}
\label{I0petit}
I_0(t')=\displaystyle\frac{2}{\beta}\arctan\displaystyle\frac{\beta}{\alpha}.
\end{equation}
For $t'>4m^2,\ \bar\beta=\beta=2[(t'/4-m^2)(t'/4-\mu^2)]^{1/2}$ and
\begin{equation}
\label{I0grand}
 I_0(t')=\displaystyle\frac{1}{\beta}\ln\left\vert\displaystyle\frac{\alpha+\beta}{\alpha-\beta}\right\vert.
\end{equation}
The functions $I_0(t')$ and $I_2(t')$ are continuous at $t'=4m^2$ and
\begin{equation}
\label{I0I2t4m2}
I_0(4m^2)=\displaystyle\frac{2}{\alpha},\ 
I_2(4m^2)=\displaystyle\frac{2}{3\alpha}.
\end{equation}
Furthermore $\rho_2^{-tri}(t')$ is also continuous at $t'=4m^2$, with,
\begin{equation}
\label{rho2tri4m2}
\lim_{t'\to 4m^2}\displaystyle\frac{I_0(t')-3I_2(t')}{t'-4m^2}=
\displaystyle\frac{2(m^2-\mu^2)}{(\mu^2-2m^2)^3}.
\end{equation}
 For $t'\to\infty$ the functions $I_{0,2}(t')\to 
 (2/t')\ln(t'/m^2)$ and $(t'/4-\mu^2)[I_0(t')-I_2(t')]\to 
\textrm{constant}$.
It can then be seen from Eqs.~(\ref{pi+-}) and (\ref{rhoibubtri}) that 
the amplitudes $p_1^{+tri}(t)$ and $p_2^{-tri}(t)$ are convergent, 
$p_1^{+bub}(t)$ and $p_4^{-tri}(t)$ are logarithmically divergent and 
$p_4^{-bub}(t)$ is linearly divergent. Introducing an upper limit of 
integration, $t_M$, in the integrals~(\ref{pi+-}), one finds
\begin{equation}
\label{p1+bub}
p_1^{+bub}(t,t_M)=\displaystyle\frac{g^4}{32\pi^2m^2}\ I_1(t,t_M),
\end{equation}
\begin{equation}
\label{p4-bub}
p_4^{-bub}(t,t_M)=\displaystyle\frac{g^4}{1536\pi^2m^4}\ I_4(t,t_M),
\end{equation}
with
\begin{equation}
\label{I1ttm}
I_1(t,t_M)=I_1(0,t_M)+J_1(t,t_M),
\end{equation}
\begin{equation}
\label{I10ttm}
I_1(0,t_M)=-2X(t_M)-I_\textrm{ln}(X(t_M),1),
\end{equation}
\begin{equation}
\label{J1ttm}
J_1(t,t_M)=2X(t_m)+X(t)I_\textrm{ln}(X(t_M),X(t)),
\end{equation}
\begin{equation}
\label{XIln}
X(t)=\sqrt{1-\displaystyle\frac{4\mu^2}{t}},\ I_\textrm{ln}(Y,Z)=\ln
\left\vert
\frac{Y-Z}{Y+Z}
\right\vert,
\end{equation}
\begin{equation}
\label{I4ttm}
I_4(t,t_M)=I_4(0,t_M)+J_4^1(t,t_M)+J_4^2(t,t_M),
\end{equation}
\begin{equation}
\label{I40tm}
I_4(0,t_M)=t_MX^5(t_M)-6(\mu^2/t_M)J_4^1(t_M,t_M),
\end{equation}
\begin{equation}
\label{J4ttm}
J_4^1(t,t_M)=-t[(2/3)X^3(t_M)-I_1(0,t_M)],
\end{equation}
\begin{equation}
\label{J42ttm}
J_4^2(t,t_M)=t
\left[
(2/3)X^3(t_M)+X^2(t)J_1(t,t_M)
\right].
\end{equation}
The logarithmic divergence of $p_1^{+bub}(t)$ is seen in
Eq.~(\ref{J1ttm}) and the linear divergence of $p_4^{-bub}(t)$ in
Eq.~(\ref{I40tm}).

The $\rho_1^{+bub}(t'),\ \rho_1^{+tri}(t')$ contributions and those of
$\rho_2^{-tri}(t'),\ \rho_4^{-bub(tri)}(t')$ can be interpreted as
coming from the correlated $2\pi$-exchange in the S-wave ($\sigma$
or $f_0(600)$ exchange) and in the P-wave ($\rho$ exchange),
respectively (see Eqs.~(2.29), (2.30), (2.31) of Ref. \cite{Richard75}
and Eqs.~(2.12), (2.13) of Ref. \cite{Cottingham73}). However, here there
is no contribution to $\rho_1^-$ as $A^-=0$.

It is interesting to examine the box- and crossed-diagram
contributions to the bubble and triangle diagrams arising from the
pseudo-vector coupling case. Performing a similar reduction to that of
$P_{PV}$ (see Eq.~(26) in section II.B) one can write the direct and
crossed, $T_{s_1}^{\alpha\beta}\pi N$ amplitudes as

\begin{equation}
\label{Ts1ab}
T_{s_1}^{\alpha\beta}=\tau_\alpha^1\tau_\beta^1
[A1^1+\mbox{\boldmath{$\gamma^1$}}\cdot \textbf{Q}(B_{s_1}-B_r)],
\end{equation}
\begin{equation}
\label{Ts1bab}
T_{\bar s_1}^{\alpha\beta}=\tau_\beta^1\tau_\alpha^1
[A1^1-\mbox{\boldmath{$\gamma^1$}}\cdot \textbf{Q}(B_{\bar s_1}-B_r)].
\end{equation}
Here the Pauli matrices $\tau_{\alpha(\beta)}^1$ represent the isospin 
coupling of the pion field $\pi_{\alpha(\beta)}$ to the nucleon 1.
In Eqs.~(\ref{Ts1ab}) and (\ref{Ts1bab}),
\begin{equation}
\label{ABrBs}
A=\displaystyle\frac{g^2}{2m},\ B_r=
\displaystyle\frac{g^2}{4m},\ B_s=\displaystyle\frac{g^2}{m^2-s}.
\end{equation}
Note that
\begin{equation}
\label{taubox}
\sum_{\alpha,\beta=1,2,3}\tau_\alpha^1\tau_\beta^1\tau_\alpha^2\tau_\beta^2=
\sum_{\alpha,\beta}
\left(\delta_{\alpha\beta}+\imath\epsilon_{\alpha\beta\gamma} 
\tau_\gamma^1\right)
\left(\delta_{\alpha\beta}+\imath\epsilon_{\alpha\beta\gamma} 
\tau_\gamma^2\right)
=3-2\mbox{\boldmath{$\tau^1$}}\cdot\mbox{\boldmath{$\tau^2$}}.
\end{equation}
The $2\pi$-exchange box diagram contribution arises from
 \begin{eqnarray}
\label{Tbox}
  T_{box}    &  = & T_{s_1}^{\alpha\beta} T_{s_2}^{\alpha\beta} +
  T_{\bar s_1}^{\alpha\beta} T_{\bar s_2}^{\alpha\beta}  \nonumber \\
      \ecart
      &   = & (3-2\mbox{\boldmath{$\tau^1$}}\cdot\mbox{\boldmath{$\tau^2$}})
      \left\{
      2A^21^11^2+(B_{s_1}-B_{\bar s_1})A(1^2\mbox{\boldmath{$\gamma^1$}}\cdot \textbf{Q}
      +1^1\mbox{\boldmath{$\gamma^2$}}\cdot \textbf{Q})\right.\nonumber\\
      \ecart
    &&  \left. +2
    \left[B_{s_1}B_{s_2}-B_r(B_{s_1}+B_{s_2})+B_r^2\right]
   \mbox{\boldmath{$\gamma^1$}}\cdot \textbf{Q}\mbox{\boldmath{$\gamma^2$}}\cdot \textbf{Q}\right\}.
\end{eqnarray}
The crossed diagram contribution is,
 \begin{eqnarray}
\label{Tcro}
  T_{cro}    &  = & T_{s_1}^{\alpha\beta} T_{\bar s_2}^{\alpha\beta} +
  T_{\bar s_1}^{\alpha\beta} T_{s_2}^{\alpha\beta}  \nonumber\\
  \ecart
      &   = & (3+2\mbox{\boldmath{$\tau^1$}}\cdot\mbox{\boldmath{$\tau^2$}})
      \left\{
      2A^21^11^2+(B_{s_1}-B_{\bar s_1})A(1^2\mbox{\boldmath{$\gamma^1$}}\cdot \textbf{Q}
      +1^1\mbox{\boldmath{$\gamma^2$}}\cdot \textbf{Q})\right.\nonumber\\
      \ecart
    &&  \left. -2
    \left[B_{s_1}B_{\bar s_2}-B_r(B_{s_1}+B_{\bar s_2})+B_r^2\right]
   \mbox{\boldmath{$\gamma^1$}}\cdot Q \gamma^2\cdot \textbf{Q}\right\}.
\end{eqnarray}
In Eqs.~(\ref{Tbox}) and (\ref{Tcro}) the terms $B_{s_1}B_{s_2}$ and $B_{s_1}B_{\bar s_2}$ correspond to the $2\pi$-exchange nucleon pole contribution for pseudo-scalar coupling. For the bubble and triangle diagram contributions,
use of Eqs.~(\ref{rhoionPi}) and (\ref{ABCDEF}) leads to
\begin{equation}
\label{rhoi+-boxcro}
\begin{array}{rcl}
\rho_{1\ box}^{\pm bub}(t')   &= & \pm\ \rho_{1\ cro}^{\pm bub}(t')=\pm 
   \displaystyle\frac{1}{2}\   \rho_1^{+ bub}(t'), \\
   \ecart
\rho_{1\ box}^{\pm tri}(t')   &= & \pm\ \rho_{1\ cro}^{\pm tri}(t') = 
  \pm \displaystyle\frac{1}{2}\   \rho_1^{+ tri}(t'), \\
   \ecart
\rho_{2\ box}^{\pm tri}(t') &  = & \mp\ \rho_{2\ cro}^{\pm tri}(t') = 
  \mp \displaystyle\frac{1}{2}\   \rho_2^{- tri}(t'), \\
   \ecart
\rho_{4\ box}^{\pm bub}(t')   &  = & \mp\ \rho_{4\ cro}^{\pm bub}(t')= 
 \mp  \displaystyle\frac{1}{2}\   \rho_4^{- bub}(t'), \\
  \ecart
\rho_{4\ box}^{\pm tri}(t')   &  = & \mp\ \rho_{4\ cro}^{\pm tri}(t') = 
  \mp \displaystyle\frac{1}{2}\   \rho_4^{- tri}(t').
 \end{array}
\end{equation}
It can be seen that the total $\rho_{i\ box}^{\pm 
bub(tri)}(t')+\rho_{i\ cro}^{\pm bub(tri)}(t')$ from the above 
formulae~(\ref{rhoi+-boxcro}) are in agreement with the 
formulae~(\ref{rhoibubtri}) calculated from Eqs.~(\ref{ABPV}). 
Formulae~(\ref{rhoi+-boxcro}) satisfy the relations of 
Eq.~(\ref{notation1}) between the box and crossed $\rho_i$.

\subsection{Comparison with Feynman diagram calculation}

We have checked, at different energies and angles, that the Saclay
amplitudes given by Eq. (\ref{heltobis}) for both box and crossed diagrams
for pseudo-scalar coupling calculated in Section II-A from the Feynman
diagram expressions with no form factor ($\Lambda\rightarrow\infty$)  are
in agreement with the corresponding dispersion relation
results Eq.~(\ref{Pech}), Eq.~(\ref{Pcro}) and Eq.~(\ref{Pima}). We 
found that the
short range contribution of the crossed diagram coming from the term
containing $a_{i}^{\pm}\left(w',4m^2-w-w'\right)$ in Eq.~(\ref{pi+-}) is
crucial for the agreement. Numerical comparisons will be shown in the next
section.

For the pseudo-vector coupling, the comparison of the bubble and triangle
contributions Eq.~(\ref{rhoibubtri}) is less straightforward due to the
divergence of some of these terms when integrated on $t'$ in the
dispersion relation or of the corresponding terms in the limit of
$\Lambda\rightarrow\infty$ in the Feynman method. The results of both
approaches for the convergent invariant amplitudes $p_1^{+tri}(t)$ and
$p_2^{-tri}(t)$ (see Eqs.~(\ref{rhoibubtri})) compare well. The
logarithmically divergent $p_1^{+bub}(t)$ and $p_4^{-tri}(t)$ require one
subtraction and a comparison can be made with the $dp_1^{+bub}(t)/dt$ and
$dp_4^{-tri}(t)/dt$.  For the linearly divergent $p_4^{-bub}(t)$ the
second derivative, $d^2p_4^{-bub}(t)/dt^2$, converges and can be compared
with the Feynman calculation.

To obtain these amplitudes in the Feynman calculation the amplitudes $p_4$
and $p_1$ are extracted by solving for them from Eqs. (\ref{invtohel2})
and (\ref{heltobis}) thus removing the explicit dependence on the cosine
of the scattering angle, $z$. Then the derivatives with respect to $z$ are
essentially the same as the derivatives with respect to $t$ in the case of
the dispersion relations allowing a direct numerical comparison. Since
this process requires numerical derivatives and a calculation for large
$\Lambda$ it is more difficult, but a good agreement is obtained.

\begin{figure}[htb]
\epsfig{file=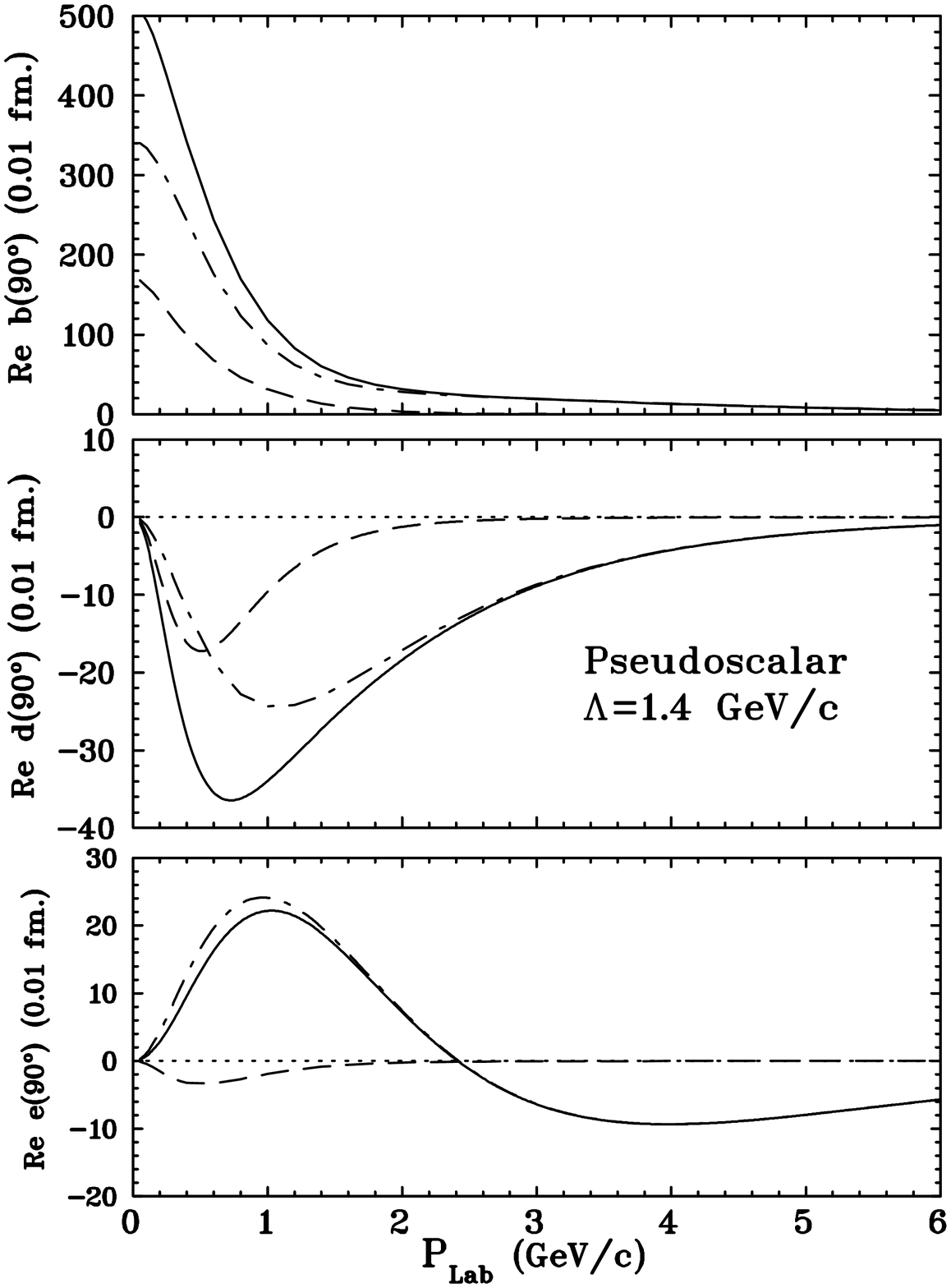,height=4.5in}
\epsfig{file=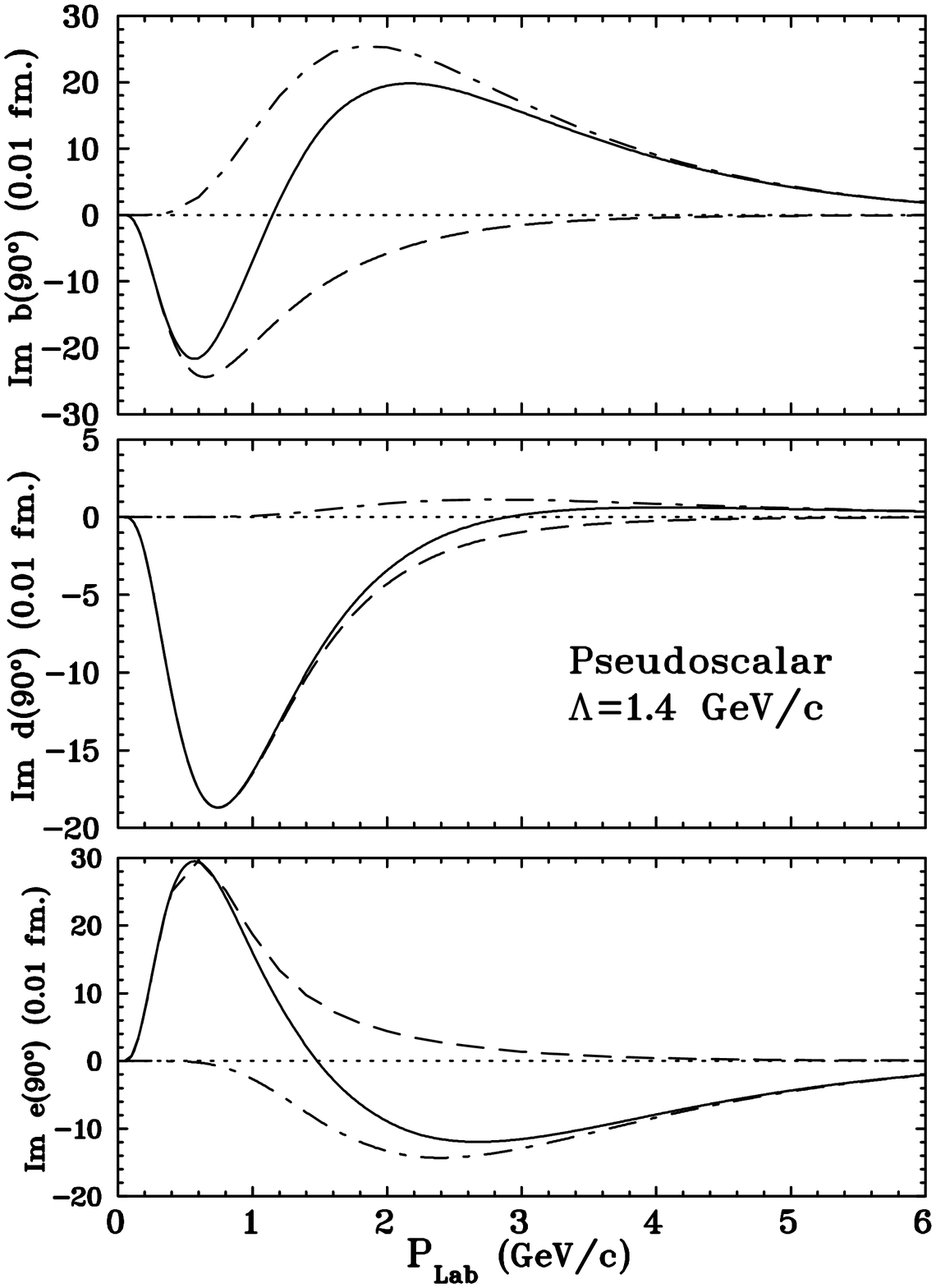,height=4.5in}
\caption{Contributions to the real (left) and imaginary (right) 
parts of the amplitudes $b$, $d$ and $e$ as a function of energy for  
pseudo-scalar coupling. The dashed line corresponds to the box, 
the dashed-dot line to the crossed-pion diagram and the solid 
line to their sum.}
\label{amrps}   
\end{figure}

\begin{figure}[htb]
\epsfig{file=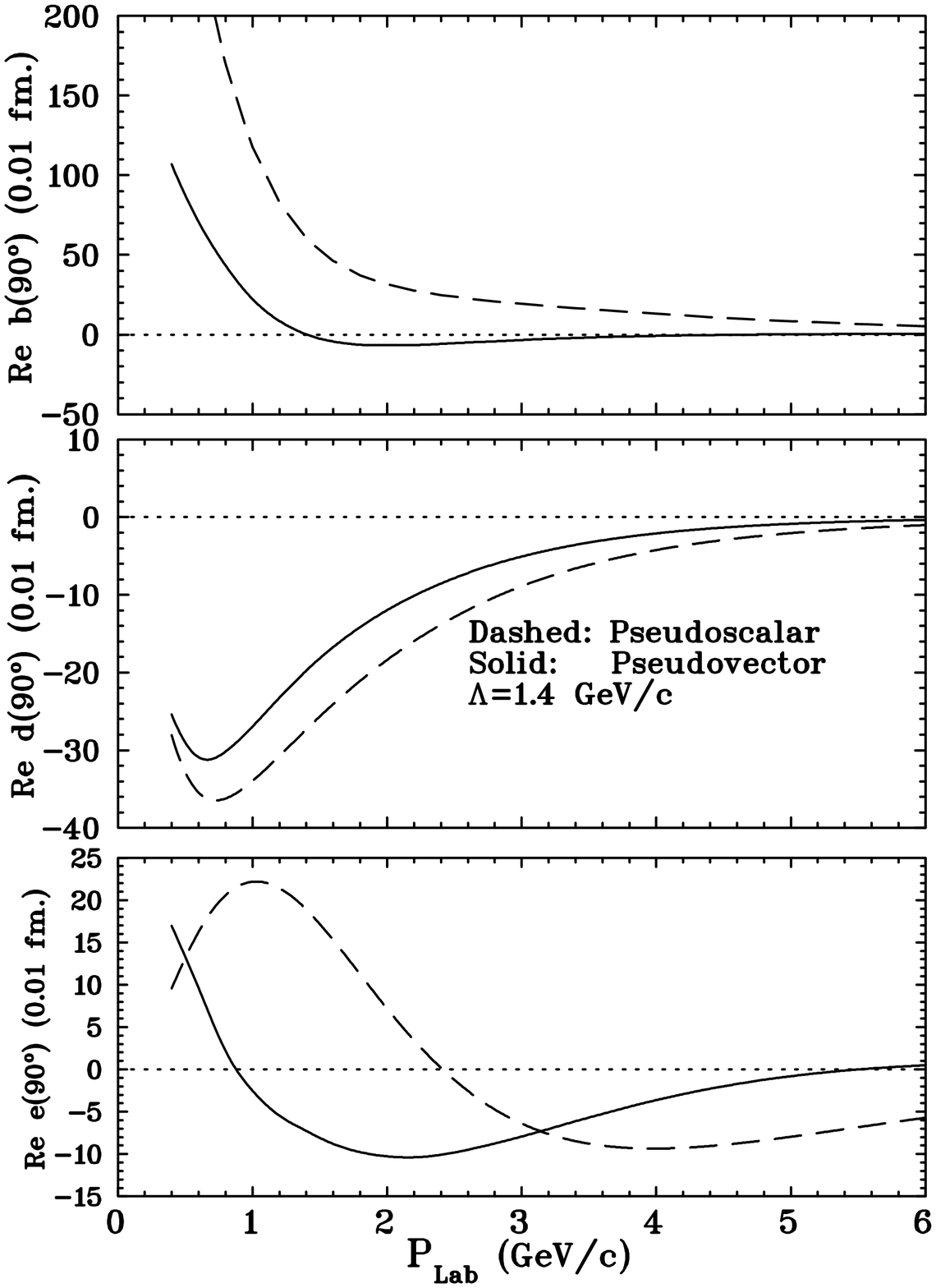,height=4.4in}
\epsfig{file=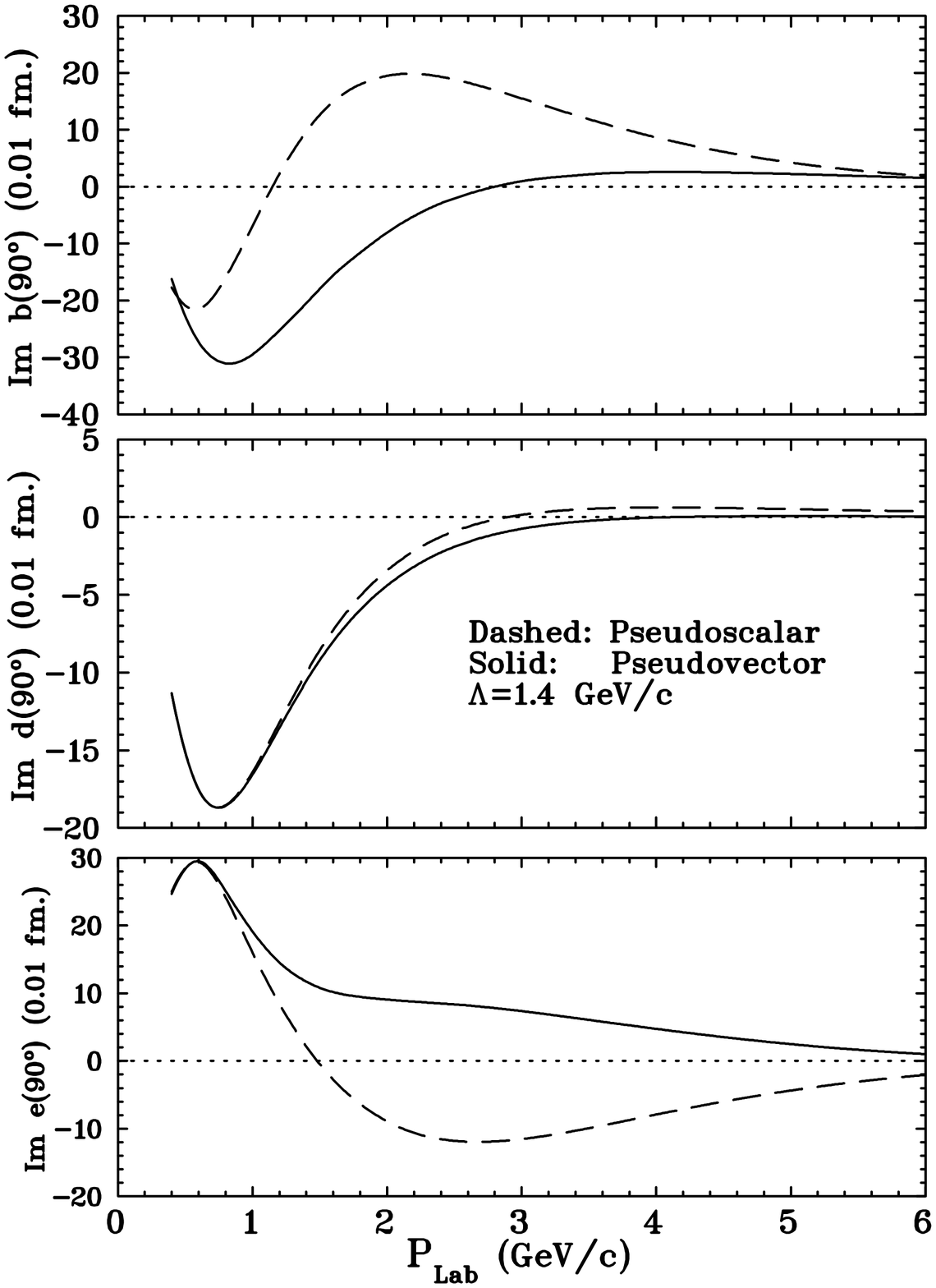,height=4.4in}
\caption{Comparison of the real (left) and imaginary (right)  
parts of the  amplitudes $b$, $d$ and $e$ corresponding to the 
pseudo-scalar and pseudo-vector couplings. The dashed curve 
represents the PS and the solid curve the PV.}\label{amr}   
\end{figure}

\begin{figure}[htb]
\epsfig{file=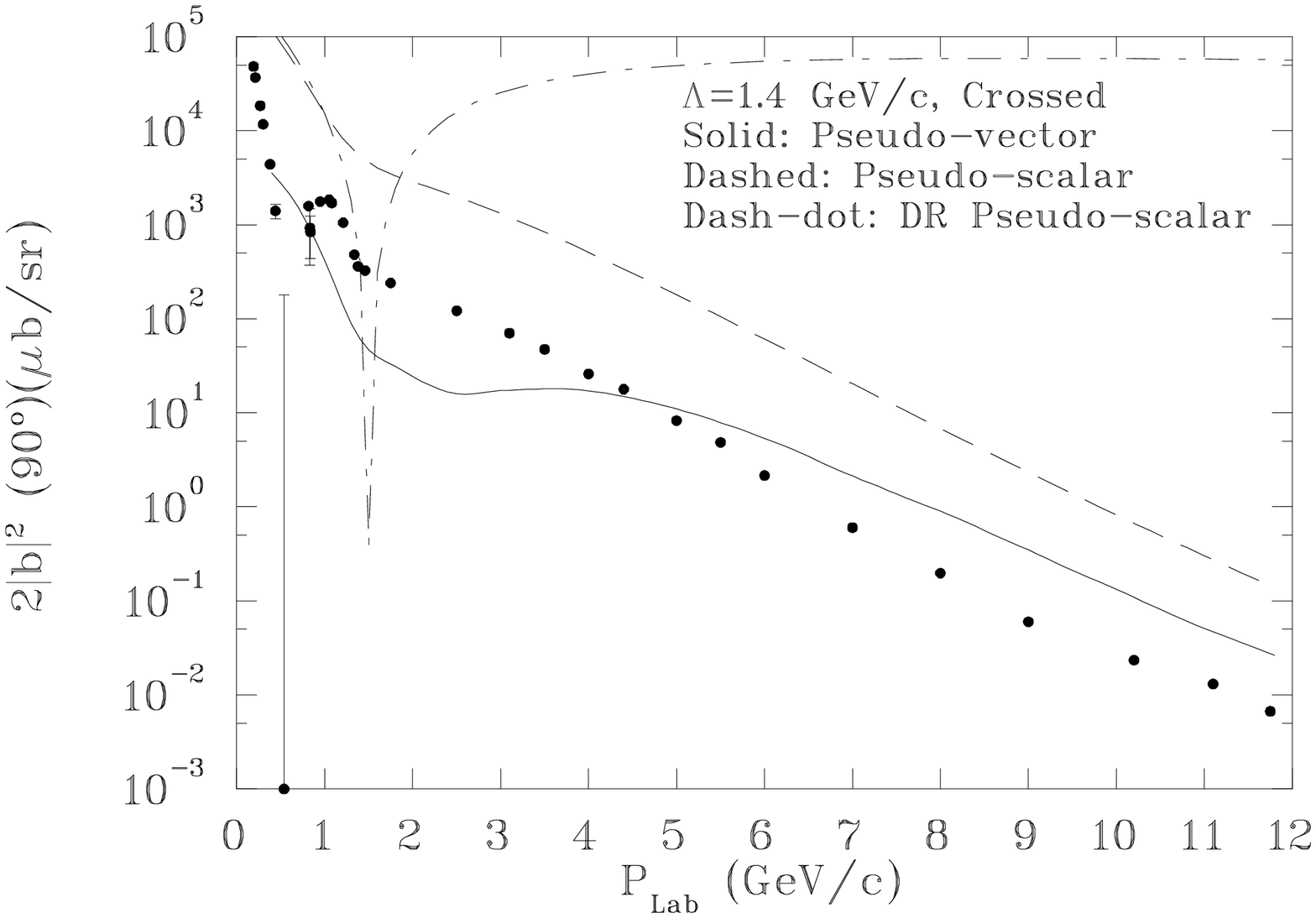,angle=90,height=2.2in}
\epsfig{file=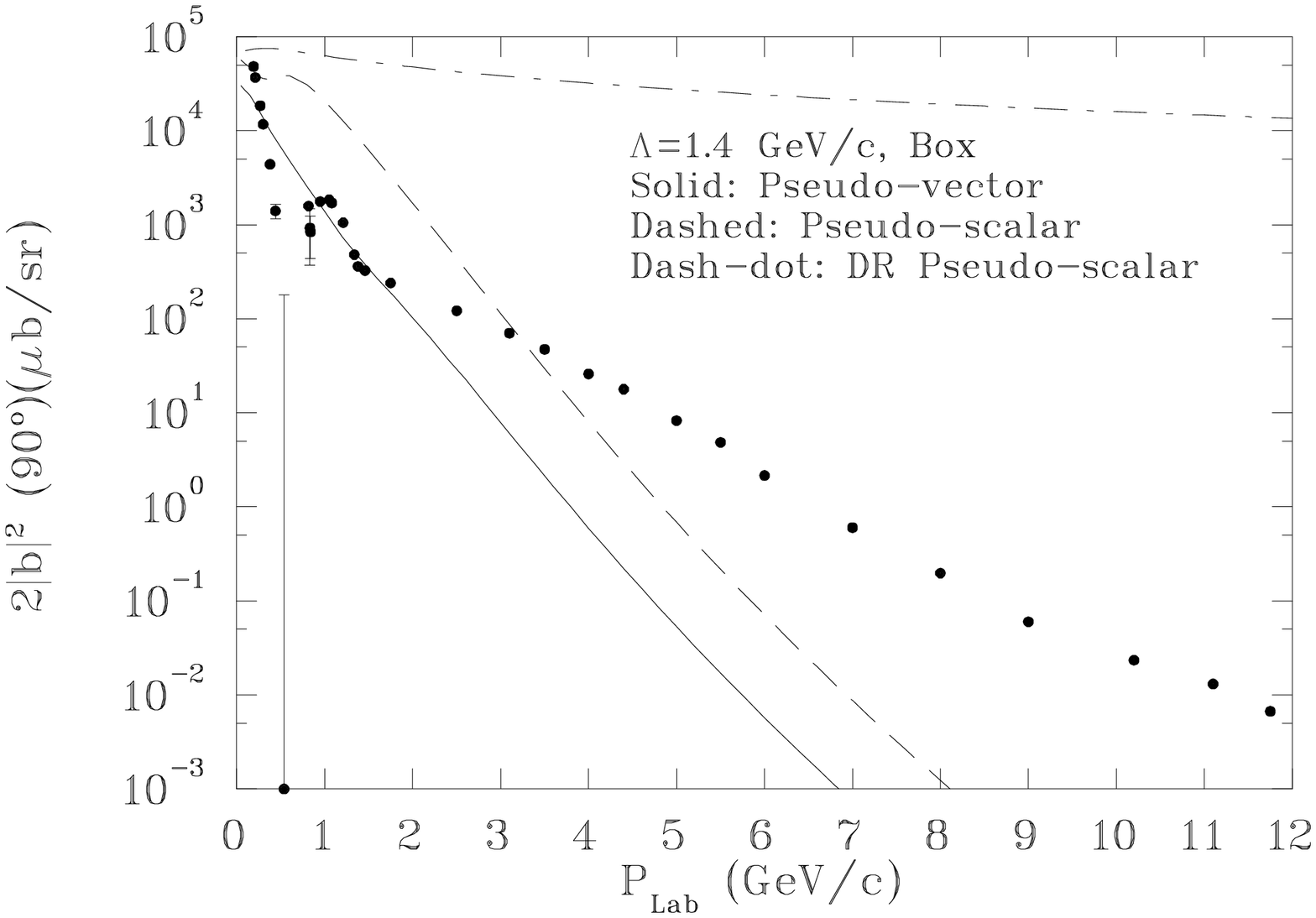,angle=90,height=2.2in}
\caption{Results for the crossed (left) and box diagram (right) for 
the $b$ amplitude. The dash-dot curves show the results of the 
dispersion relation calculation, equivalent to the Feynman 
calculation for $\Lambda\rightarrow\infty$.}
\label{b2c1400}
\end{figure}

\begin{figure}[htb]
\epsfysize=165mm
\epsfig{file=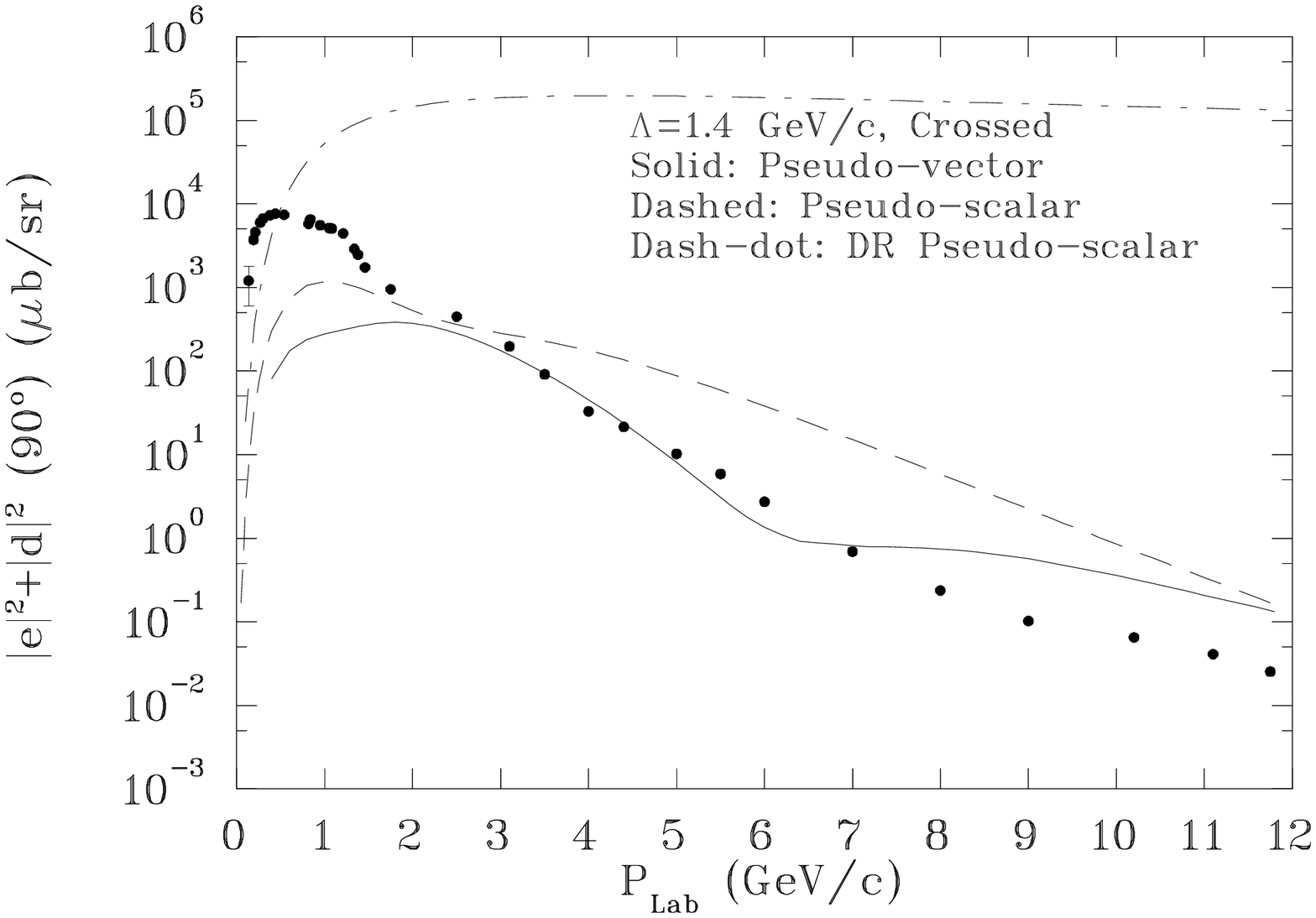,angle=90,height=2.2in}
\epsfig{file=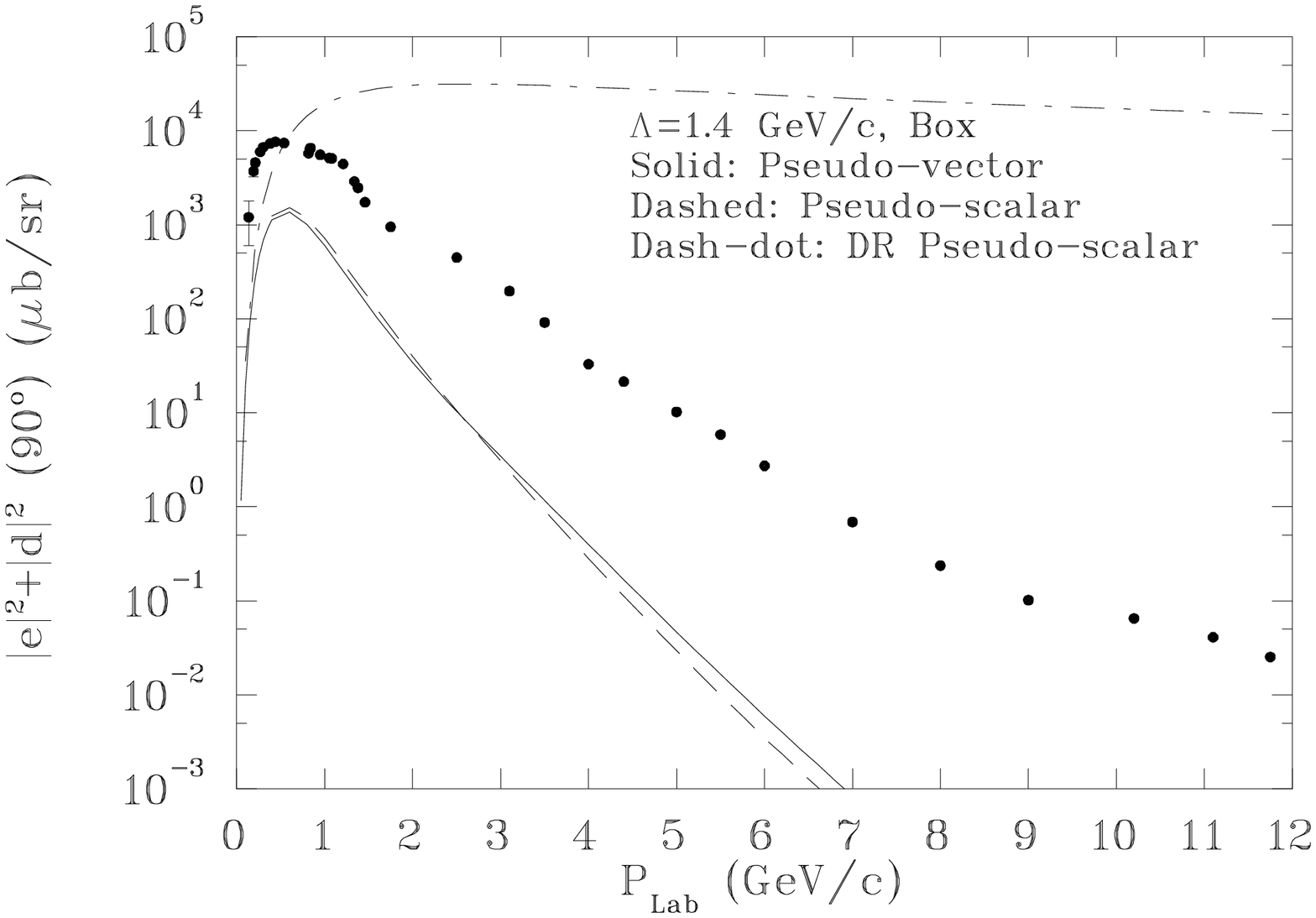,angle=90,height=2.2in}
\caption{Results for the crossed (left) and box diagram (right) for the $d$ 
and $e$ amplitudes. The curves have the same meaning as in Fig. 
\protect{\ref{b2c1400}}.}\label{d2e2c1400}
\end{figure}

\begin{figure}[htb]
\epsfig{file=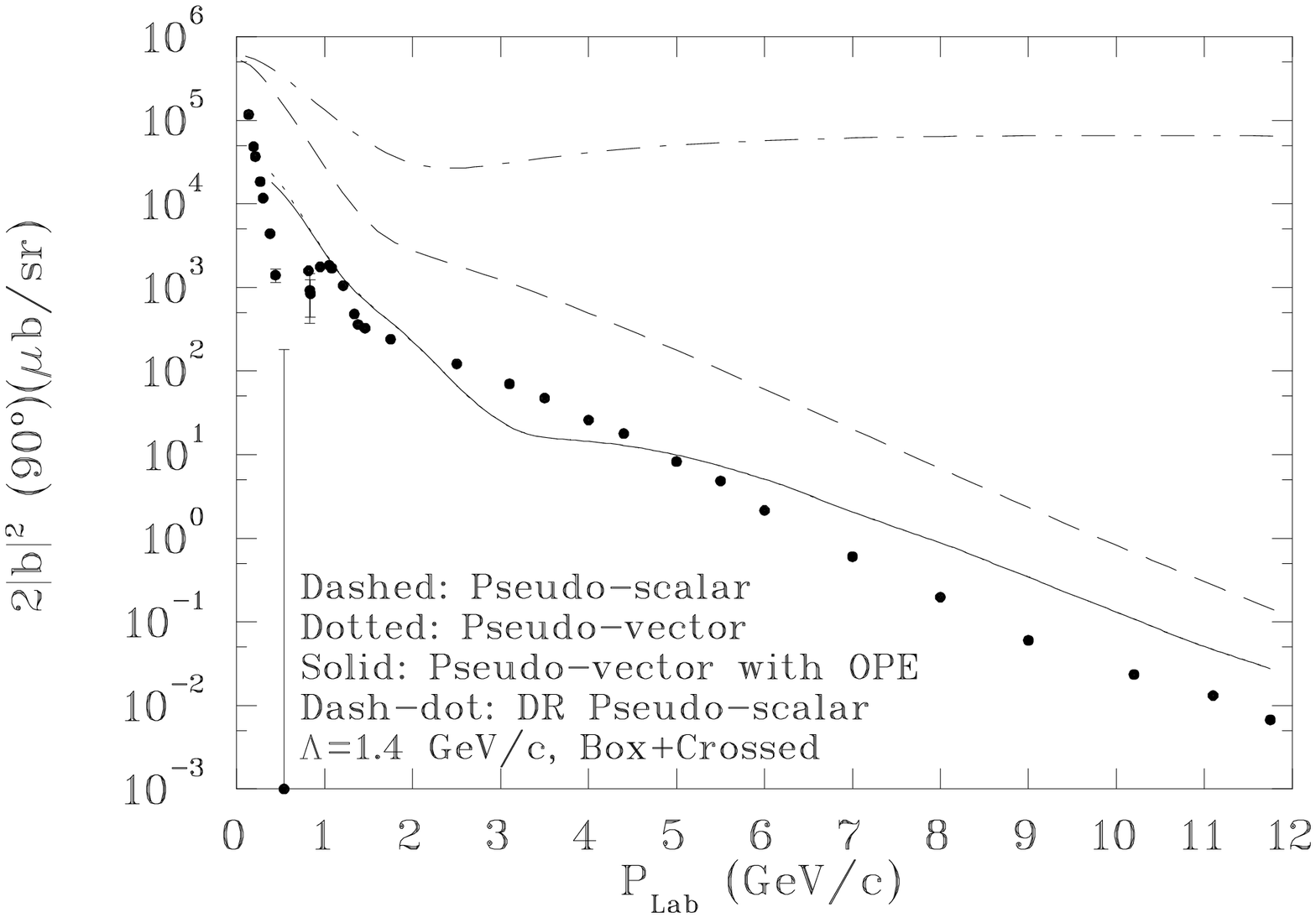,angle=90,height=2.2in}
\epsfig{file=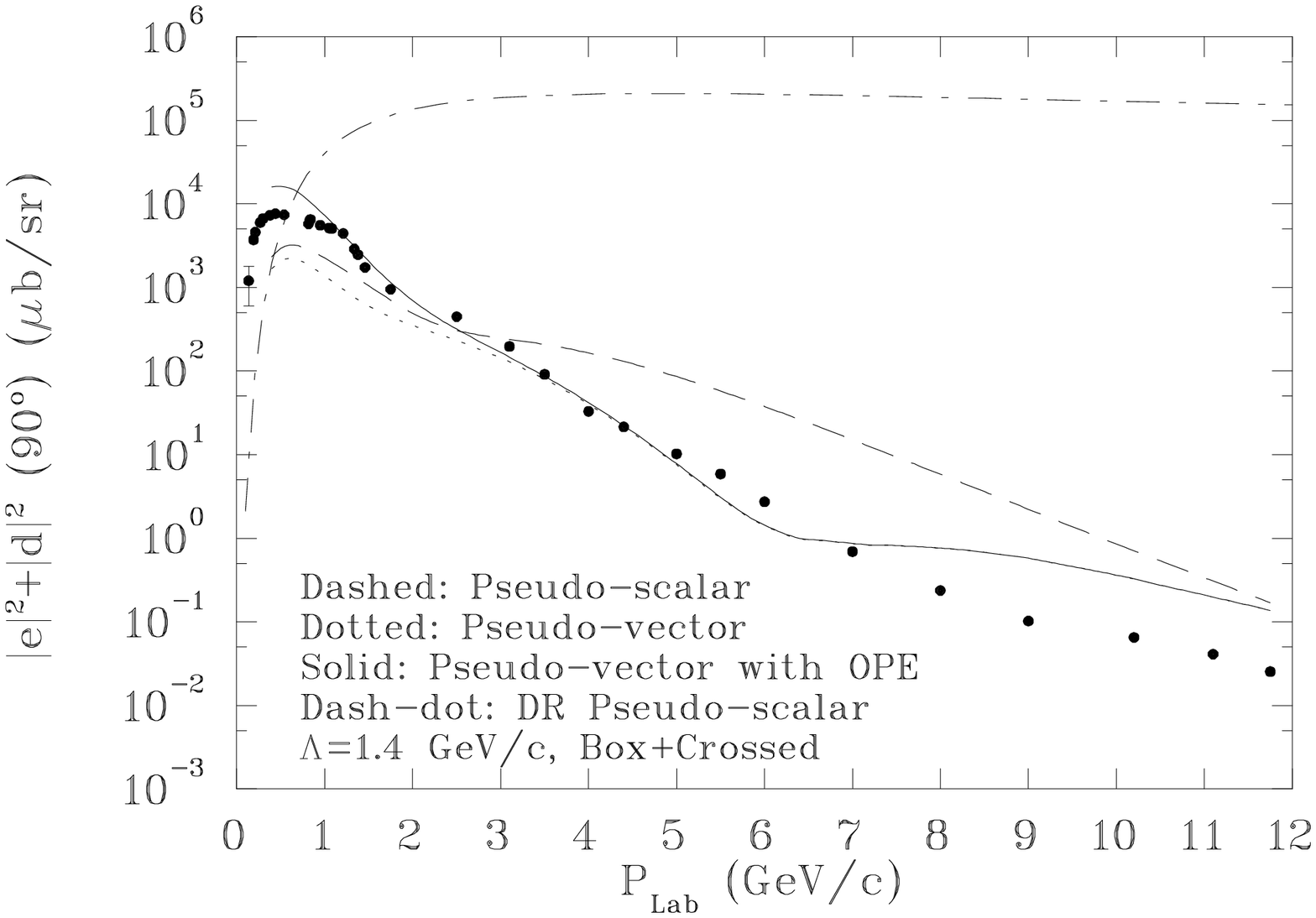,angle=90,height=2.2in}
\caption{Results of the sum of the crossed and box diagrams for 2$|b|^2$ (left) and
$|d|^2+|e|^2$ (right). The dash-dot curves show the results of the
dispersion relation calculation, equivalent to the Feynman calculation
for $\Lambda\rightarrow\infty$.}\label{b2t1400}
\end{figure}

\begin{figure}[htb]
\epsfig{file=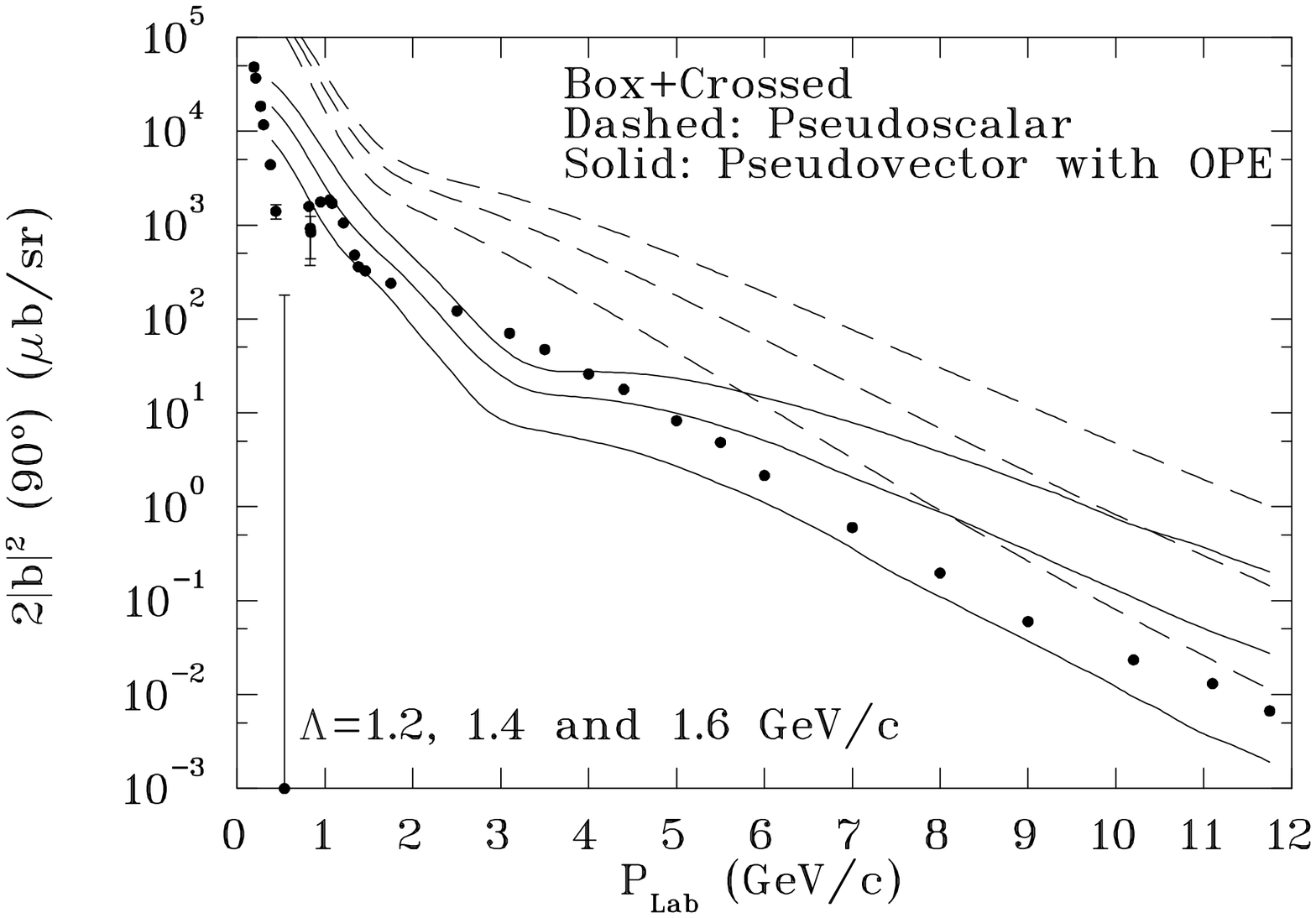,angle=90,height=2.2in}
\epsfig{file=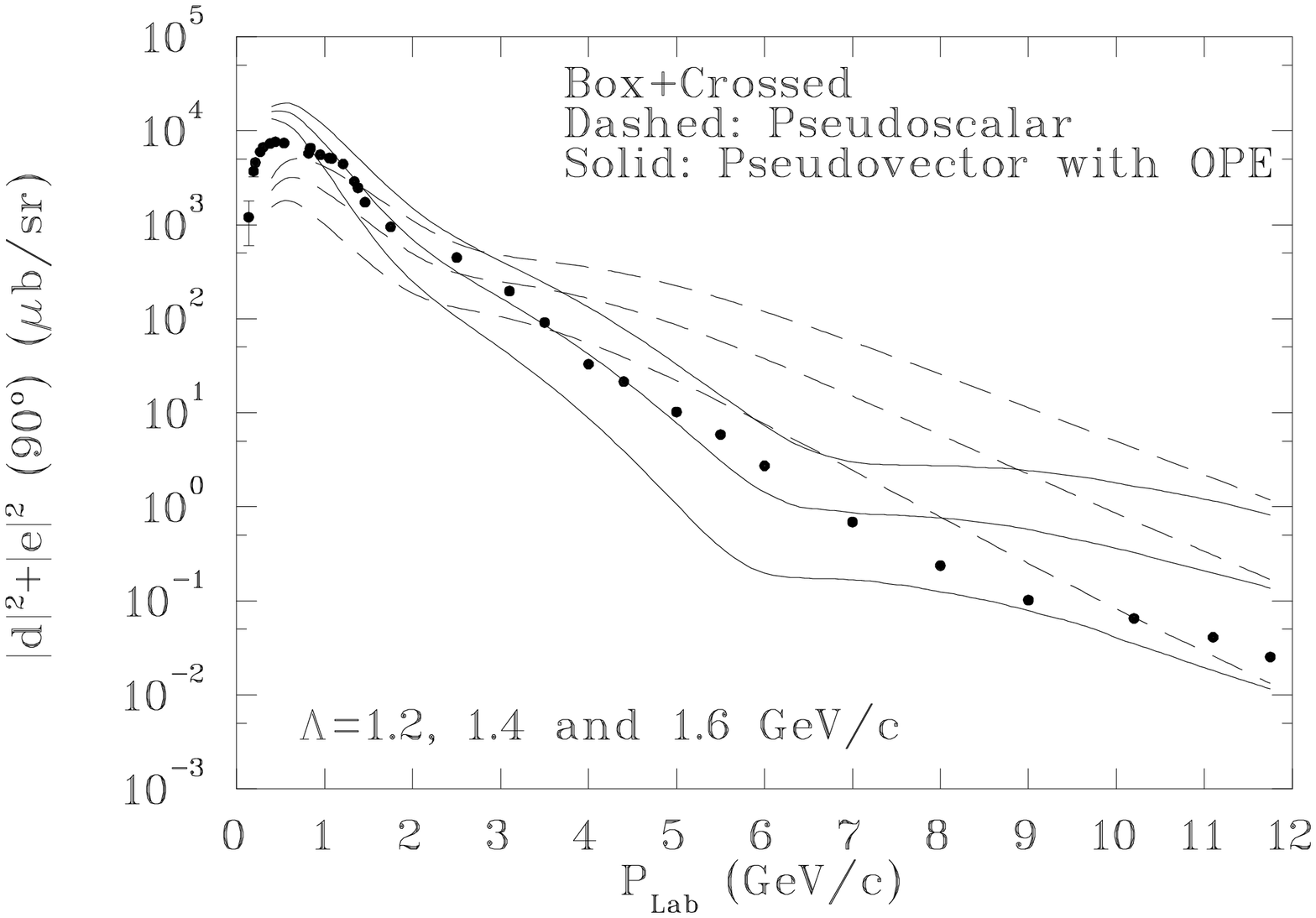,angle=90,height=2.2in}
\caption{Comparison of the data for 2$|b|^2$ (left) and $|d|^2+|e|^2$ (right) 
with the results of the sum of the crossed and box diagrams for different values 
of $\Lambda$. Increasing cross sections correspond to increasing values of $\Lambda$.}
\label{b2tcmp}
\end{figure}

\begin{figure}[htb]
\epsfig{file=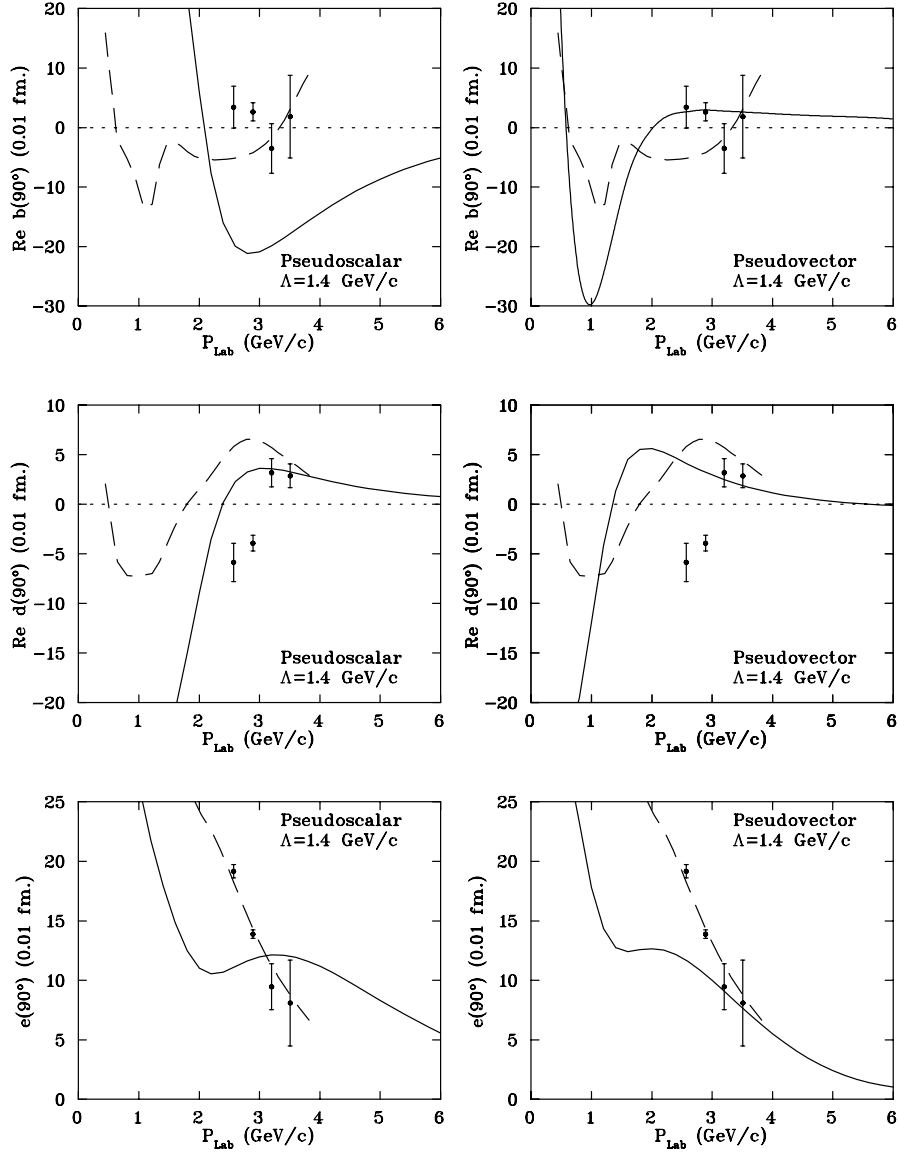,height=6in}
\caption{Comparison of the real parts of the calculated PS (left) and PV (right) 
$b$, $d$ and $e$ amplitudes (solid lines) with the results of SAID \cite{arndt} 
(dashed lines) and those obtained by Bystricky et al.  (points) \cite{bystdata}. 
The phase has been chosen such that the amplitude $e$ is real.}
\label{ampsr}   
\end{figure}

\begin{figure}[htb]
\epsfig{file=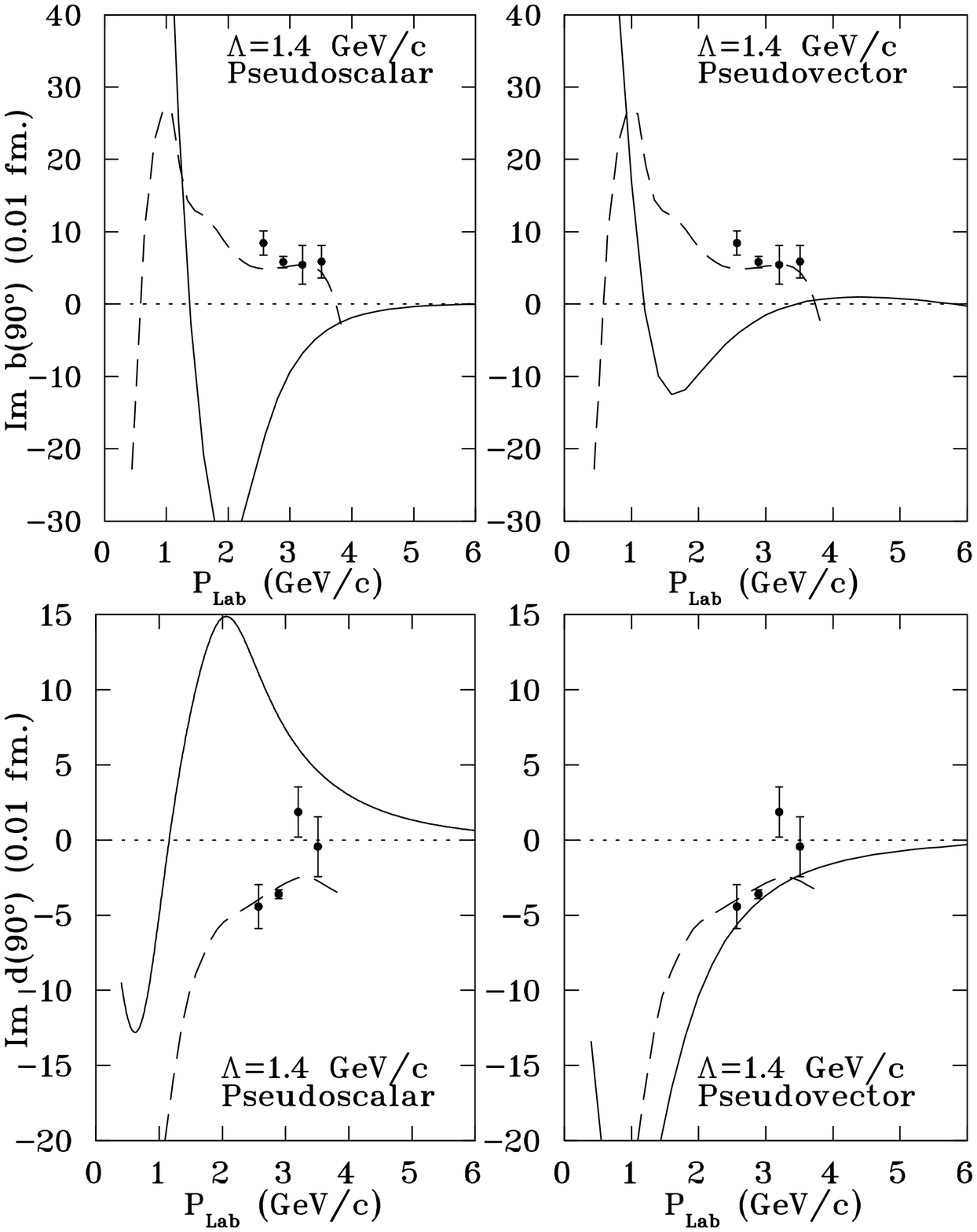,height=5in}
\caption{As in Fig. \protect{\ref{ampsr}} but for the imaginary part of 
the $b$ and $d$ amplitudes.}
\label{ampsi}   
\end{figure}

\section{Results and Discussion}}

Test calculations were made with and without the antiproton poles 
($p_0=-E$). At low values of P$_{Lab}$ or for large $\Lambda$ they give 
significant contributions and are essential to get agreement with the 
dispersion relation results.  However, for realistic values of $\Lambda$ 
at higher energies they become unimportant and can be neglected.

Figure \ref{amrps} shows the behavior of the real and imaginary parts 
of the amplitudes as a function of incident momentum for the 
pseudo-scalar coupling. For the crossed-pion diagram only terms with 
a single nucleon or pion on shell and the principal value integral on 
$q_0$ survive.  That is to say, only the contributions from the $q_0$ 
integral need to be treated specially, the rest of the indicated 
integrals have no singularities. For the box diagram, in addition to 
the two contributions just mentioned, the diagram in which two 
nucleons are on shell also gives a real contribution (in the form we 
are looking at now, it becomes an imaginary contribution to the 
amplitude after application of the factor of $i$ to be consistent with the 
dispersion relation approach).  That is, the 
integral on the magnitude of the three-momentum also has a 
$\delta$-function part which contributes.

In figure \ref{amrps} (imaginary part) we see that the principal-value
part of the integral goes rapidly to zero for the crossed diagram as the
momentum drops below 500 MeV/c.  This behavior can be traced to the form
factor and its dependence on $q_0$.  For low energy the form factor
becomes independent of $q_0$, since the pole moves far from the real axis.  
Since it is the only pole off of the real axis the integrand limits to a
function with only poles on the real axis and hence contains only the
$\delta$-function parts.

The imaginary part of the box contains both the principal value and double
pole contribution.  At low energies the real parts of the amplitudes are
dominated by the term with two nucleons on shell. By 2 GeV/c that
situation has been reversed with the crossed-pion principal value
dominating.

Figure \ref{amr} compares the results for the pseudo-scalar and
pseudo-vector couplings. One sees that the real part of the Saclay
amplitude $b$ is very large for the PS coupling but is greatly reduced
for PV coupling. The $d$ amplitude shows only a modest difference
between PS and PV coupling while the $e$ amplitude is again
significantly modified. We note that the $b$ amplitude is a mixture of
central and spin-spin character, the $d$ amplitude is of tensor
character and the $e$ amplitude is of spin-orbit character \cite{gl}.

Figures \ref{b2c1400} and \ref{d2e2c1400} show a comparison of the box and
crossed-pion contributions.  It is readily seen that the crossed-pion
diagram dominates at high energy.  This result can be traced back to the
relative ability of the two diagrams to transfer energy and momentum.  
Because factors of momentum transfer in the numerator cancel the decrease
coming from the propagators, the integrand without form factors does not
fall off with increasing momentum transfer so that the decrease in cross
section is mainly due to the form factors.  Since the form factor falls
off with increasing pion momentum in the center of mass of the nucleon,
one can see that for the box diagram with the four-vector $q$ connecting
vertices with $(E,\bfk)$ and $(E,-\bfk)$ both momenta in the respective
centers of momentum cannot be zero (or even small) at the same time. The
crossed diagram does not suffer from this contradiction since $\bfq$
connects $-\bfk$ and $\bfk'$ and, for some value of the 4-vector, $q$,
there is a chance to minimize the momenta at both vertices. We now look at
the behavior of the form factor chosen for these calculations.

The crucial element for the understanding of the behavior of
the box and crossed diagrams is the argument $Z^2$ in Eq. 
(\ref{fdef}). Since one nucleon is always on-shell we can write 
$Z^2$ as
\eq
Z^2=\frac{(a\cdot b)^2}{m^2}-b^2,
\qe
where $a=\pm k\ {\rm or}\ \pm k'$ and $b=q\ {\rm or}\ q'$.

The order of magnitude of $\qt=|\bfq|$ will be the same as $|\bfk|$ since
$\qt$ cannot be taken as small to maximize the integrand because of the
powers of $\qt$ in the numerator.  For values of momenta greater than the
mass of the nucleon, the first term, $(a\cdot b)^2$, will dominate so we
need only consider the size of that term to obtain some sense of the
behavior of the form factor. Out-of-plane values of $\bfq$ (in the y
direction in the system we are using) only increase $Z^2$ so consider
$q_y=0$. Let $x$ be the cosine between the incident direction $\bfk$ and
the vector $\bfq$ and consider the pole at $q_0=\omega$ in an extreme
relativistic limit $E\rightarrow \kt$ ($\kt=|\bfk|$) and
$\omega\rightarrow \qt$.

It is useful to look for the minimum in the four values of $Z^2$ in order
to maximize the form factor.  Corresponding to the four factors in Eq.
(\ref{ffbox}) we have (neglecting the second term in $Z^2$)  the following
four values for $a\cdot b$ for the box diagram;

\newpage

$$ 1)\ E\omega-\bfk\cdot
\bfq \rightarrow \kt\qt(1-x);\ \ 2)\ E\omega+\bfk\cdot \bfq
\rightarrow \kt\qt(1+x);
$$ 
\eq 3)\ E\omega-\bfk'\cdot \bfq'
\rightarrow \kt[\qt(1-\sqrt{1-x^2})+\kt];\ \ 4)\
E\omega+\bfk'\cdot \bfq' \rightarrow
\kt[\qt(1+\sqrt{1-x^2})-\kt] .
\qe 
Since the first and second vertices involve the sum and difference of the
same quantities, they cannot both be small at the same time. The form factor
being an even function of $x$, there is an extremum at $x=0$ and, in fact, it
is a maximum. If we choose $x=0$ and the fourth value of $a\cdot b$ to be
zero with $\qt=\h \kt$ then, in this limit

\eq \calf_{box}^{\rm max}\rightarrow \frac{1}{\left(
1+\frac{\kt^4}{4m^2\Lambda^2}
\right)^4}\frac{1}{\left(1+\frac{\kt^4}{m^2\Lambda^2}\right)^2}.
\qe
We can apply the same considerations for the crossed-pion graph from
Eq (\ref{ffcross}). 
The values of $a\cdot b$ at the four vertices are
$$
1)\ E\omega-\bfk\cdot\bfq' \rightarrow \kt\qt(1-x)-\kt^2;\ \ 
2)\ E\omega+\bfk\cdot\bfq \rightarrow \kt\qt(1+x);
$$
\eq
3)\ E\omega-\bfk'\cdot \bfq \rightarrow \kt\qt(1- \sqrt{1-x^2});\ \ 
4)\ E\omega+\bfk'\cdot \bfq' \rightarrow \kt\qt(1+\sqrt{1-x^2})-\kt^2.
\qe
It is possible to find values of $\qt$ and $x$ such that any
of the values of $Z^2$ is zeroed, unlike the case of the box
diagram where the third value can never be zero. For example,
one could take $x=-1$ and $\qt=\h \kt$ to set the first and
second values of $Z^2$ to zero.
The  maximum comes about, however, if we take the first and
fourth values of $a\cdot b$ to be zero which leads to
\eq
\qt=(2-\sqrt{2})\kt;\ \ x=-1/\sqrt{2} \label{peakvalues}.
\qe
The second and third values of $a\cdot b$ become
\eq
(3-2\sqrt{2})\kt^2\approx 0.1716\kt^2,
\qe 
which is much smaller than what one was able to achieve 
in the case of the box diagram. The form factor is then
\eq
\calf_{crossed}^{\rm max}\rightarrow 
\frac{1}{\left[1+\left(\frac{0.1716\kt^2}{m\Lambda}\right)^2\right]^4}.
\qe

Thus, not only is it possible to make two of the arguments zero at the
same time but the other two are relatively small also. Numerical studies
of the relative sizes of the form factors alone confirm that the values of
the maximum of the crossed form factor occurs at the values given by Eq.
(\ref{peakvalues}).  While it is necessary to go to very high energy to
justify the ultra-relativistic limit used above for illustration, at
P$_{Lab}$=5.5 GeV/c typical values of the crossed diagram form factor are
more than two orders of magnitude larger than that for the box diagram.

Since the crossed-pion form factor peaks at definite values of
$\bfq$ and $x$ at high energies, analytical predictions of the
relative size of the different amplitudes can be made. From
these values we find for the Saclay amplitudes, in the very high-energy 
limit
\eq
e=-2b;\ \ \ \ d=0,
\qe
which was verified by a calculation at P$_{Lab}= 46$ GeV/c. 
These values lead to a value of $C_{NN}$ which is again 1/3.
In the energy range where we compare with data we are far
from this limit, however.

\begin{figure}[htb]
\epsfig{file=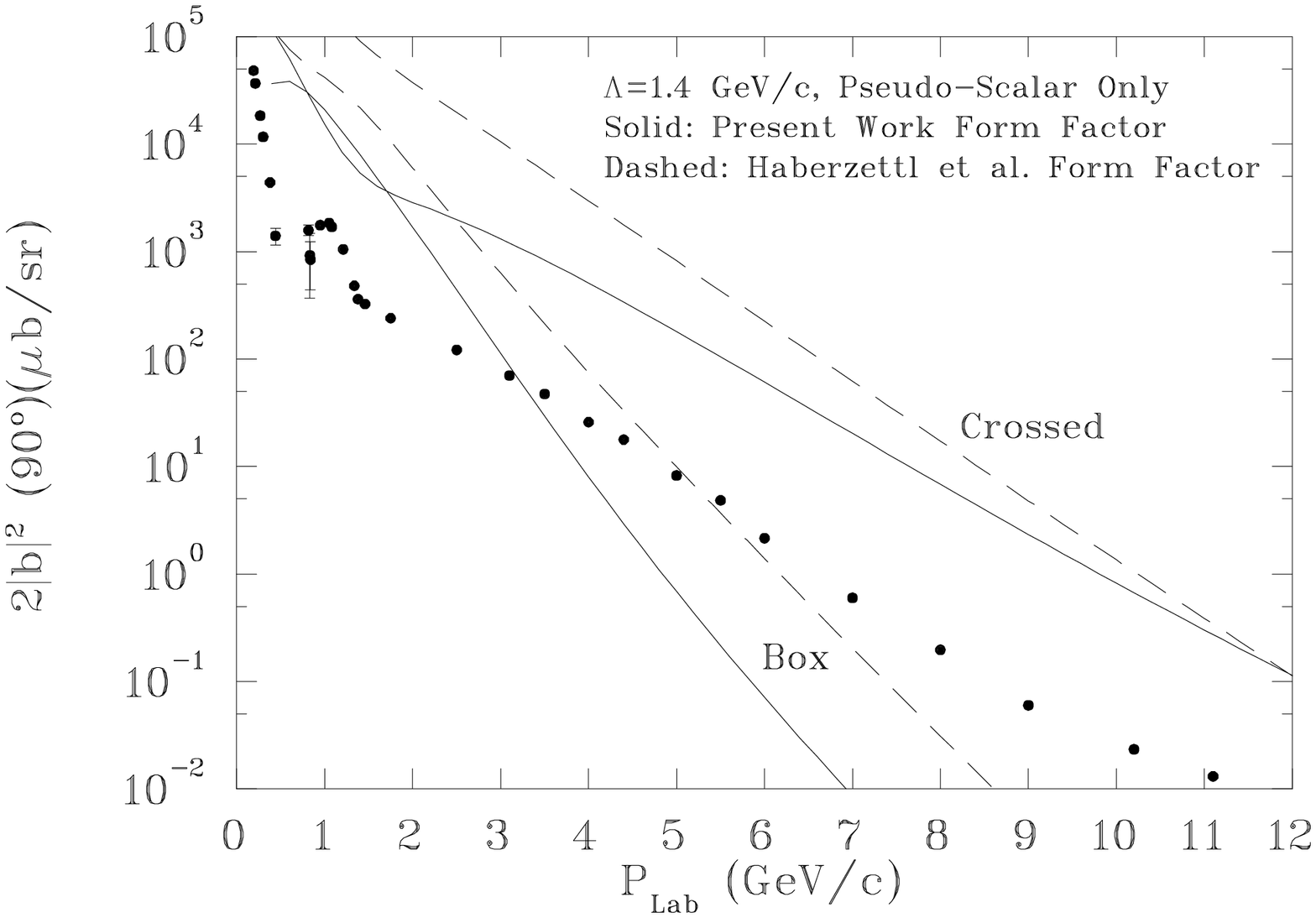,height=3.5in,angle=90}
\epsfig{file=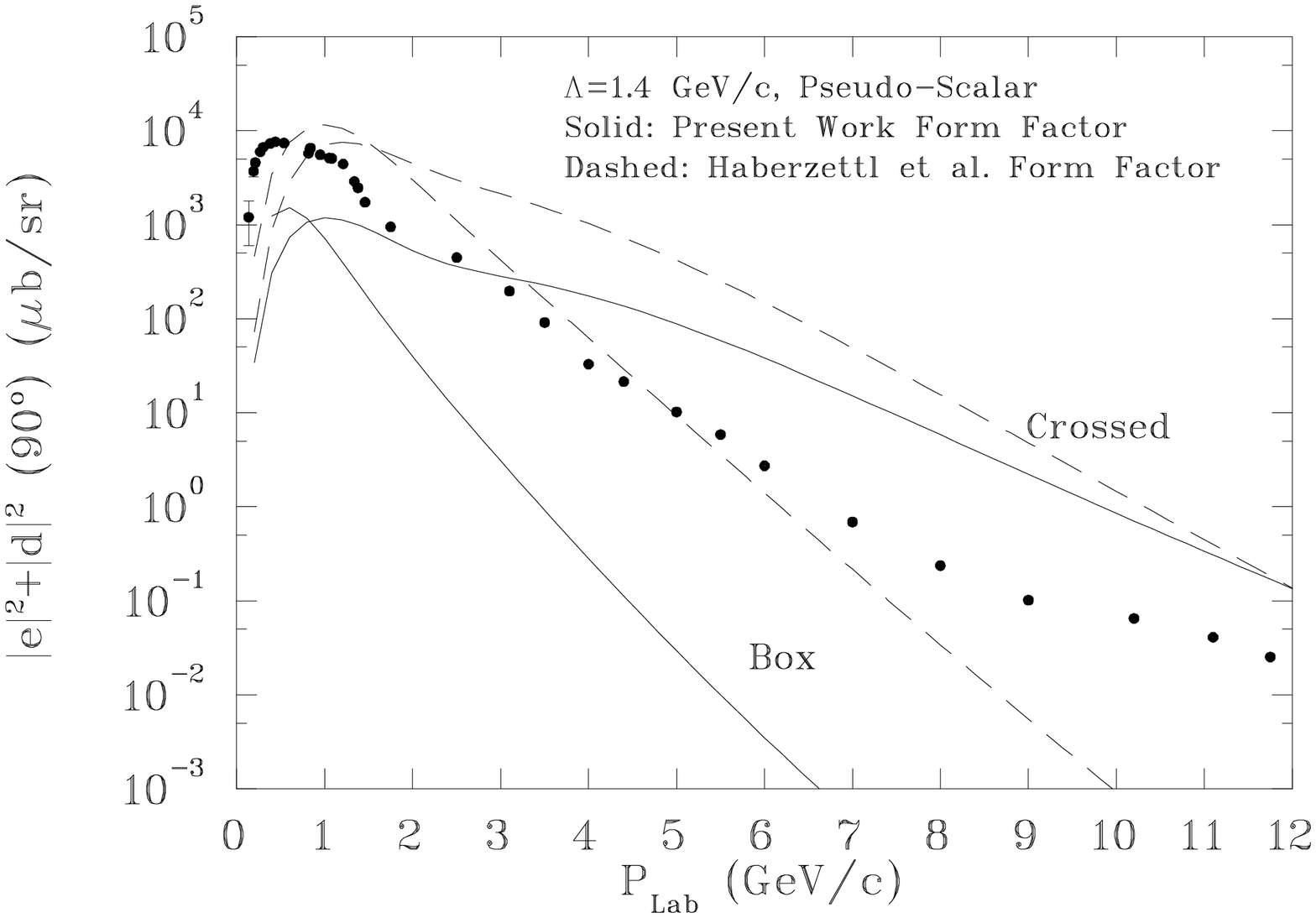,height=3.5in,angle=90}
\caption{Comparison of calculations using $Z^2$ as the 
variable in a dipole form factor with that of Eq. 
(\ref{hhvariable}) from Haberzettl et al. \cite{haberzettl}.
No attempt has been made to make the
range the same in the two calculations and this figure gives
only the PS contribution. The graph shows that the slopes of 
the calculations with the different form factors are similar 
and in both cases the box diagram contribution falls off much 
more rapidly than that of the crossed diagram.}\label{hh}   
\end{figure}

We see that the values of the spin separated cross sections shown in Fig.
\ref{b2tcmp} indicate a preference for the PV coupling and values of
$\Lambda$ in the range 1.2 to 1.4.  Since $\Lambda$ has been limited from
other sources (see Section \ref{iva}), the agreement is obtained in a 
nearly parameter free manner.

Some experimental values of the Saclay amplitudes are known at high
enough energy that unitarity corrections can be expected to be small
enough that they can be compared with the present calculation. Arndt et
al. \cite{arndt} give amplitudes up to P$_{Lab}=3.82$ GeV/c and
Bystricky et al. \cite{bystdata} present several single energy values.  
Ghahramany and Forozani \cite{ghah} found values of the amplitudes at
two energies in reasonable agreement with Refs. \cite{arndt} and
\cite{bystdata}. Figure \ref{ampsr} shows a comparison of the present
calculation with the SAID values (broken line) and the Bystricky et al.
points. The agreement for the pseudo-vector coupling is seen to be
considerably better than the pseudo-scalar one and gives a reasonable
representation of the data except for the imaginary part of $b$.

One may ask if the difference in fall-off of the crossed and box diagrams
is a general result of the kinematics or if it depends on the particular
form of the variable $Z^2$ used in the form factor.  To attempt to give a
partial answer to this question we have calculated the pseudo-scalar
result for the crossed diagram with a different, though superficially
similar, variable. Haberzettl et al.  \cite{haberzettl} use a variable
which is of the form

\eq
W^2(p_1,p_2,p_3)=\frac{(p_1^2-m_1^2)^2+(p_2^2-m_2^2)^2+
(p_3^2-m_3^2)^2}{4m^2}\label{hhvariable}
\qe
where we have chosen the normalization such that the limit
for large values of the invariant masses the limit matches that
of Eq. (\ref{squares}). Figure \ref{hh} shows the comparison
of the two calculations and one sees that the fall-off for this
variable is very similar to the presently used form factor.
The rise at small momenta is at least partially due to the
contribution of the principal value integral since the pole
in the complex plane does not move far from the real axis as
the energy goes to zero so that causality is not respected
in that limit.

It is useful to examine the decrease of the cross section with increasing
energy.  The prediction of slope of Brodsky and Farrar \cite{bf} (see Fig.
\ref{b2ope}) based on counting of internal propagators and our calculation of
$2|b|^2$ are very similar.  The high energy limit of a form factor has been
compared with the picture of the propagators between interacting constituents
by the authors of Ref. \cite{amado} for the case of scattering from nuclear
constituents. The two-pion-exchange diagram may be able to provide a bridge
between the low and high-energy points of view.

The present result bears directly on the question of the energy regime
where the transition from a color singlet hadronic exchange to a
quark-gluon basis might reasonably take place.  This, in turn, may impact
the question of color transparency (see, for example Jain et al.  
\cite{jain}). If the exchange of color singlets (pions in this case)
continues to be important through 12 GeV/c it might negate the basis for
color transparency (dominance of quark-gluon exchange) in the moderate
energy range P$_{Lab}\le$ 12 GeV/c.  This is the entire range covered in
the color transparency experiments of Carroll et al. \cite{carroll}.
However, it may be that, even if color singlet exchange continues to be
very important for elastic scattering, quark-gluon exchange might dominate
the inelastic processes.

\begin{figure}[htb]
\epsfig{file=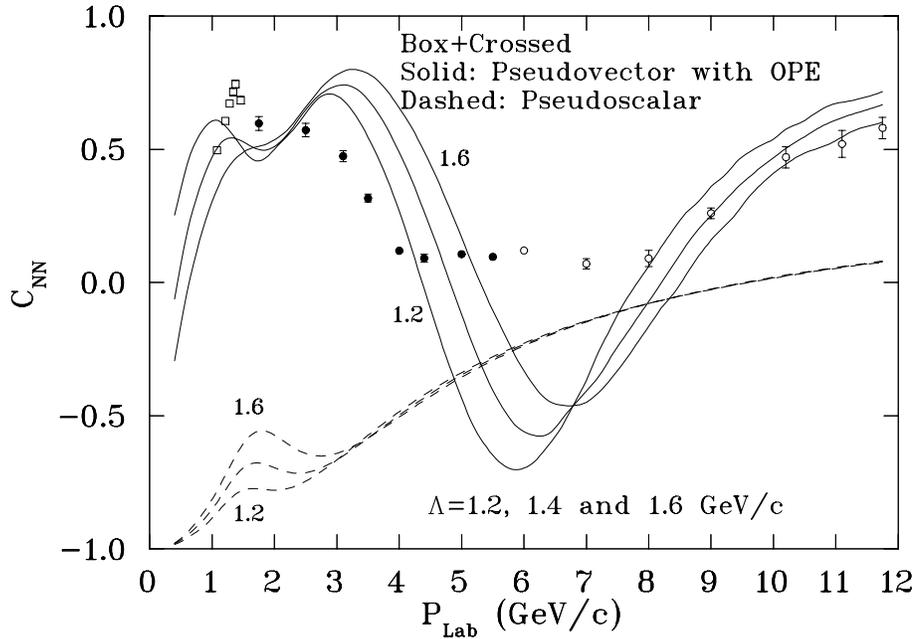,angle=90,height=3.3in}
\caption{Comparison of the result for the spin-correlation observable
$C_{NN}$ for the PS and PV couplings for three values of 
$\Lambda$. Data are: Squares, Bhatia et al. 
\protect{\cite{bhatia}}, Filled circles, Lin et al. 
\protect{\cite{lin}}, Open circles, Crosbie 
et al. \protect{\cite{crosbie}}.}\label{cpi}   
\end{figure}

The comparison with the spin transfer observable $C_{NN}$ is given in Fig.
\ref{cpi} for several values of $\Lambda$. For the pseudo-vector choice of
coupling it is seen that the qualitative features are reproduced. One, in
fact, needs a mechanism (or several mechanisms) with considerable
structure in order to reproduce the data. The plateau in the data from 4
to 8 GeV/c is not found but there is a minimum in the calculation in this
region.

\section{Summary and Conclusions}

We have calculated the contribution of the box and crossed
two-pion-exchange diagrams to proton-proton scattering at
90$^{\circ}_{c.m.}$ for laboratory momenta up to 12 GeV/c.  The cases of
both pure pseudo-scalar and pure pseudo-vector pion-nucleon coupling are
treated. 

We assume that the interaction of the pion is with the valence quarks
within the nucleons and use an effective field theory obeying Feynman rules
to describe the composite system.  At each pion-nucleon-interaction vertex
we introduce a relativistic Lorentz-invariant form factor. These form
factors are related to the convolution of the nucleon and pion sizes and
represent the pion source distribution based on the quark structure of the
hadronic cores and explicitly introduce an interaction range of this size.
These form factors are functions of the four-momenta of the exchanged pions
and scattering nucleons. One can use any two of the three momenta which
converge at a given vertex to calculate the function so that the dependence
on the nucleon and pion momenta displays a high degree of symmetry.  This
behavior can be traced to the fact that the basic scalar on which the form
factors depend can be written as the square of the four-dimensional cross
product of any two of the three four-vectors.

While we believe the form factors that we introduce are quite reasonable,
we make no claim as to their uniqueness. However, there exist a set of
conditions, based on short reaction time, causality etc., that any
relativistic Lorentz invariant form factor should satisfy.  Our form
factors obey these rules, but other form factors used by some authors do
not.

As a check of the calculation, we compare the two-pion-exchange
Feynman-diagram amplitudes for both pseudo-scalar and pseudo-vector
coupling calculated with point-like nucleons with those obtained from the
dispersion relation approach and find agreement.  The antiproton poles in
the Feynman calculations are essential for the comparison but for a more
realistic calculation with a finite value of the form-factor range,
$\Lambda$, they are unimportant. While performing this study, we found it
essential to include a contribution of the crossed diagram which was
neglected in the dispersion-relation calculations of the two-pion-exchange
Paris potential because of its very short range.

The numerical technique for calculating the 4-dimensional Feynman
integrals, taking into account multiple poles is presented. Standard
transformation methods are not always applicable in the presence of  
form factors.

In the Feynman diagram calculations, using the form factors constrained
by the valence quark distribution, comparison with experimental data
favors the pseudo-vector coupling over the pseudo-scalar one.  This is
seen in the comparison with the magnitude of the spin-separated cross
sections, in the agreement with amplitudes extracted up to $~4$ GeV/c
and in the spin transfer parameter $C_{NN}$, although in this later case
it is only the general behavior which is correctly given.

The exchange of one pion is important only at momentum less than
P$_{Lab}$=1 GeV/c. While the strengths of the box and crossed diagrams for
the exchange of two pions are comparable for laboratory momenta below 2
GeV/c, for larger momenta the crossed diagram dominates, mainly due to the
mathematical structure of the form factors and to the fact that the
kinematics of the crossed diagram allows a repartition of momenta at the
vertices in a favorable manner. An important contribution arises from the
principal-value contribution of the integrals which is non-zero when form
factors are included.

While here we compare only with the extracted spin-separated cross
sections at 90$^{\circ}_{c.m.}$, our calculation can be extended to all
scattering angles. Further studies such as unitarity corrections can be
considered.

To conclude, we have shown that the two-pion exchange plays a
significant role in proton-proton scattering even up to P$_{Lab}$=12
GeV/c.  This result suggests that the importance of the exchange of
color singlets in elastic scattering may extend to higher energies than
expected.

\begin{center}Acknowledgments\end{center}

We thank Helmut Haberzettl for a useful conversation regarding causality,
Henri Zytnicki for providing support for this project with his work and Jean-Pierre 
Dedonder and R. Vinh Mau for helpful comments on the manuscript.

This work was supported by the National Science Foundation under
contract PHY-0099729.

\cen{\bf \Large Appendices}

\begin{appendix}

\section{Projection onto the Saclay Amplitudes}

The Saclay amplitudes \cite{Bystricky78} are defined by the equation

$$
M(\bfk_f,\bfk_i)=\frac{1}{2}[(a+b)+(a-b)
\bfs_1\cdot\bfn \bfs_2\cdot\bfn
+(c+d)\bfs_1\cdot\bfm \bfs_2\cdot\bfm, $$

\eq
+(c-d)\bfs_1\cdot\bfl \bfs_2\cdot\bfl
+e(\bfs_1+\bfs_2)\cdot\bfn]
\qe
where the center-of-mass basis vectors are:

\eq
\bfl=\frac{\bfk_f+\bfk_i}{|\bfk_f+\bfk_i|},\ \ 
\bfm=\frac{\bfk_f-\bfk_i}{|\bfk_f-\bfk_i|},\ \
\bfn=\frac{\bfk_i\times\bfk_f}{|\bfk_i\times\bfk_f|}.\ \
\qe

Observables at $90^{\circ}$ are

\eq \sigma=\frac{1}{2}(2|b|^2+|d|^2+|e|^2);\ \ \ 
\sigma C_{NN}=\frac{1}{2}(-2|b|^2+|d|^2+|e|^2); \qe

\eq \sigma D_{NN}=\frac{1}{2}(-|d|^2+|e|^2) =\sigma K_{NN};\ \ 
\sigma C_{MMMM}=\frac{1}{2}(2|b|^2+|d|^2-|e|^2). \qe

The expressions for the amplitudes in terms of the 
singlet-triplet matrix elements are:

\eq
a=\frac{1}{2}(M_{11}+M_{00}-M_{1-1});\ \ \ 
b=\frac{1}{2}(M_{11}+M_{ss}+M_{1-1});
\qe

\eq
c=\frac{1}{2}(M_{11}-M_{ss}+M_{1-1});\ \ \ 
d=-\frac{1}{\sqrt{2}\sin\theta}(M_{10}+M_{01});\ \ \ 
e=\frac{i}{\sqrt{2}}(M_{10}-M_{01}).
\qe

The two-pion exchange box and crossed amplitudes can be written in the form
\eq
M(\bfk_f,\bfk_i)=<<(G^1+i\sigma_1\cdot\bfH^1)(G^2+i\sigma_2\cdot\bfH^2)>>
\qe
where the double brackets indicate integrations over the propagators and the
G and $\bfH$ here are related to the corresponding quantities in Section II by
the factor $\frac{m^2}{4\pi E}$ ($G^1=\frac{m^2}{4\pi E}G$,
$G^2=\frac{m^2}{4\pi E}G'$, etc.).   With our choice of coordinate system $G$ and 
$H_y$ are even in
$q_y$ while $H_x$ and $H_z$ are odd.  Any totally odd quantity will integrate
to zero and those terms have been dropped in the following expressions. In
the remainder of this appendix it is to be understood that each term bilinear
in G and/or $\bfH$ is surrounded by double brackets. One has
\eq
M_{1,1}=G^1G^2-H^1_zH^2_z;\ \ M_{s,s}=G^1G^2+\bfH^1\cdot\bfH^2;\ \ 
M_{1,-1}=-H_x^1H_x^2+H_y^1H_y^2;
\qe
\eq
M_{0,0}=G^1G^2+H^1_zH^2_z-H_x^1H_x^2-H_y^1H_y^2;
\qe
\eq
M_{0,1}=\frac{1}{\sqrt{2}}[-H_z^1H_x^2-H_x^1H_z^2
-G^1H_y^2-G^2H_y^1];
\qe
\eq
M_{1,0}=\frac{1}{\sqrt{2}}[-H_z^1H_x^2-H_x^1H_z^2
+G^1H_y^2+G^2H_y^1].
\qe

From these expressions we can construct the amplitudes as
\eq
a=G^1G^2-H_y^1H_y^2;\ \ b=G^1G^2+H_y^2H_y^2;\ \ \ c=-H_x^1H_x^2-H_z^1H_z^2;
\qe
\eq
d=\frac{1}{\sin\theta}(H_z^1H_x^2+H_x^1H_z^2);\ \ \ 
e=i(G^1H_y^2+G^2H_y^1).
\qe

With the symmetry due to the identical particles
\eq
a(\theta)\rightarrow a(\theta)-a(\pi-\theta),
\qe
\eq
b(\theta)\rightarrow b(\theta)-c(\pi-\theta)=G^1(\theta)G^2(\theta)+
\bfH^1(\pi-\theta)\cdot\bfH^2(\pi-\theta),
\qe
\eq 
c(\theta)=-b(\pi-\theta),
\qe
\eq
d(\theta)\rightarrow \frac{1}{\sin\theta}
[H_z^1(\theta)H_x^2(\theta)+H_x^1(\theta)H_z^2(\theta)
H_z^1(\pi-\theta)H_x^2(\pi-\theta)+H_x^1(\pi-\theta)H_z^2(\pi-\theta)],
\qe
\eq
e(\theta)\rightarrow G^1(\theta)H_y^2(\theta)+G^2(\theta)H_y^1(\theta)
+G^1(\pi-\theta)H_y^2(\pi-\theta)+G^2(\pi-\theta)H_y^1(\pi-\theta).
\qe
At 90$^{\circ}$
\eq
a=0;\ \ \ b=G^1G^2+\bfH^1\cdot\bfH^2=-c;\ \ \ d=2(H_z^1H_x^2+H_x^1H_z^2);
;\ \ \ e=2(G^1H_y^2+G^2H_y^1),
\qe
where the quantities are evaluated at 90$^{\circ}$.

\section{Calculation Technique}

Consider the interior integrals over $q$ and $q_0$ with the
integrals over the solid angle to be done at the exterior.
We apply two formulae depending on the case. 
In the integral of the type
\eq
\int dy \int dx \left\{ \frac{f(x,y)}{[x-x_1(y)][x-x_2(y)]}
\right\}
\qe
there are two possibilities.  
If $x_1$ and $x_2$ are distinct
for all values of $y$ then we can write the integral as
\eq
\int dy \int dx \frac{f(x,y)-g(x,y)}{[x-x_1(y)][x-x_2(y)]}
+\int dy \int dx \frac{g(x,y)}{[x-x_1(y)][x-x_2(y)]},
\qe
where

\eq
g(x,y)=\frac{x-x_2(y)}{x_1(y)-x_2(y)}f(x_1(y),y)+
\frac{x-x_1(y)}{x_2(y)-x_1(y)}f(x_2(y),y).
\qe
The combination $f(x,y)-g(x,y)$ vanishes at each of the poles
so that there is no singularity in the first term. For the
second integral we have
\eq \int_{-x_0}^{x_0}  
\frac{g(x,y)dx}{[x-x_1(y)][x-x_2(y)]}=
\frac{f(x_1(y),y)}{x_1(y)-x_2(y)}\int_{-x_0}^{x_0} \frac{1}{x-x_1(y)}
+\frac{f(x_2(y),y)}{x_2(y)-x_1(y)}\int_{-x_0}^{x_0} \frac{1}{x-x_2(y)}
\qe

\eq
=\frac{f(x_1(y),y)}{x_1(y)-x_2(y)}\left[\pm i\pi 
+ln\left(\frac{x_0-x_1(y)}{x_0+x_1(y)}\right)\right]
+ \frac{f(x_2(y),y)}{x_2(y)-x_1(y)}\left[\pm i\pi 
+ln\left(\frac{x_0-x_2(y)}{x_0+x_2(y)}\right)\right].
\qe

For the general case of $n$ distinct poles
\eq
\int dy \int \frac{dx f(x,y)}{\prod_{i=1}^n[x-x_i(y)]}
=\int dy \int \frac{dx [f(x,y)-g(x,y)]}{\prod_{i=1}^n[x-x_i(y)]}
+\int dy \int \frac{dx g(x,y)}{\prod_{i=1}^n[x-x_i(y)]},
\qe
where
\eq
g(x,y)=\sum_{i=1}^n \frac{\prod_{j\ne i}[x-x_j(y)]}
{\prod_{j\ne i}[x_i(y)-x_j(y)]}f(x_i(y),y)
\qe
and the second integral is given by
\eq
\sum_{i=1}^n \frac{f(x_i(y),y)}{\prod_{j\ne i}
[x_i(y)-x_j(y)]}\left[\pm i \pi 
+ln\left(\frac{x_0-x_i(y)}{x_0+x_i(y)}\right)\right].
\qe

If two poles are not distinct (for some value of $y$) then
this method does not work.  Instead we may write

\eq
\int dy \int \frac{dxf(x,y)}{[x-x_1(y)][x-x_2(y)]}=
\int dy \frac{1}{x_1(y)-x_2(y)}\int dxf(x,y)
\left[\frac{1}{x-x_1(y)}-\frac{1}{x-x_2(y)}\right].
\qe
Each of these integrals can be done separately and
a second singularity has been pushed into the $y$ integral.
In the cases where the pole occurs in quadratic expressions
it may be more efficient to take that into account as
\eq
\int dy \int dx  \frac{f(x,y)}{[x^2-x_1^2(y)]
[x^2-x_2^2(y)]}=
\int dy \frac{1}{x_1^2(y)-x_2^2(y)}\int dxf(x,y)
\left[\frac{1}{x^2-x_1^2(y)}-\frac{1}{x^2-x_2^2(y)}
\right].
\qe

\section{Interpretation of $Z^2$}

One can be led to the form of Eq. (\ref{qdef}) from the condition given in  
Eq. (\ref{kpmq}).
Since $k'=k+q$ (for example) we observe the analogous condition in three
dimensions is indicative of the vector cross product and are thus led to
consider the four-dimensional version of the cross product.  If we define
this cross product by the use of a totally antisymmetric 4-component
tensor, $\epsilon_{ijk\ell}$, in analogy with the 3-dimensional case,
for two vectors (say a and b) the result is a tensor

\eq
T_{ij}=\sum_{0,1,2,3}\epsilon_{ijk\ell}a_kb_{\ell}.
\qe
Since this tensor has 6 independent components, it cannot be expressed as
an ordinary 4-vector.  It is useful to separate the components into two
classes:  one in which the zero index is free and one in which it is
contained in the sum. 

\eq
T_{0j}=\sum_{1,2,3}\epsilon_{0jk\ell}a_kb_{\ell}=[\bfa\times\bfb]_j
;\ \ j=1,2,3,
\qe

\eq
T_{ij}=a_0b_k-b_0a_k=[a_0\bfb -b_0\bfa]_k,\ \ \ i,j,k=1,2,3
\ \  {\rm and\ cyclic}.
\qe
Contracting this tensor with itself with the standard metric tensor
$g_{i,j}=g_i\delta_{i,j};\ \ g_0=1;\ \ g_i=-1,\ i=1,2,3$
we find

\eq
\h\sum g_{ii'}g_{jj'}T_{ij}T_{i'j'}=
\h \sum g_{ii}g_{jj}T_{ij}T_{ij}=(a_0\bfb-b_0\bfa)^2- 
(\bfa\times\bfb)^2
\equiv (a\cdot b)^2-a^2b^2.
\qe
The last identity may be verified by direct evaluation and has the
form used for $Z^2$ showing that it corresponds to the contraction of
a four-dimensional cross product.

\end{appendix}

\end{document}